\documentclass[reprint,amsmath,amssymb,aps,prd,onecolumn,nofootinbib,superscriptaddress]{revtex4-1}
\usepackage{graphicx}
\graphicspath{{fig/}} 
\usepackage[letterpaper,margin=1in]{geometry}
\usepackage{booktabs}
\usepackage{multirow}
\usepackage{pifont}
\usepackage{todonotes}
\usepackage{amssymb}
\usepackage{amsmath}
\usepackage{xcolor, colortbl}
\usepackage{array}
\usepackage[colorlinks = true, bookmarksopenlevel=4,
            linkcolor = blue,
            urlcolor  = blue,
            citecolor = blue,
            anchorcolor = blue]{hyperref}

\newcommand\pubnumber{arXiv: }
\newcommand\pubdate{\today}

\def\Title#1{\begin{center} {\LARGE #1 } \end{center}}

\newcommand\pubblock{\rightline{\begin{tabular}{l} \pubnumber\\
         \pubdate \end{tabular}}}
\newenvironment{Abstract}{\begin{quotation} \begin{center}
                       ABSTRACT
     \end{center}\bigskip  }{\end{quotation}}

\newcommand\snowmass{\begin{center}\rule[-0.2in]{\hsize}{0.01in}\\\rule{\hsize}{0.01in}\\
\vskip 0.1in Submitted to the Proceedings of the US Community Study\\
on the Future of Particle Physics (Snowmass 2021)\\
\rule{\hsize}{0.01in}\\\rule[+0.2in]{\hsize}{0.01in} \end{center}}

\begin{document}

\pubblock

\Title{The COHERENT Experimental Program}

\bigskip

\newcommand{\Mephidesc}{\affiliation{National Research Nuclear University MEPhI (Moscow Engineering Physics Institute), Moscow, 115409, Russian Federation}}
\newcommand{\NCSUnucengdesc}{\affiliation{Department of Nuclear Engineering, North Carolina State University, Raleigh, NC, 27695, USA}}
\newcommand{\Dukedesc}{\affiliation{Department of Physics, Duke University, Durham, NC, 27708, USA}}
\newcommand{\TUNLdesc}{\affiliation{Triangle Universities Nuclear Laboratory, Durham, NC, 27708, USA}}
\newcommand{\UTKdesc}{\affiliation{Department of Physics and Astronomy, University of Tennessee, Knoxville, TN, 37996, USA}}
\newcommand{\ITEPdesc}{\affiliation{National Research Center  ``Kurchatov Institute'' Kurchatov Complex for Theoretical and Experimental Physics, Moscow, 117218, Russian Federation}}
\newcommand{\USDdesc}{\affiliation{Physics Department, University of South Dakota, Vermillion, SD, 57069, USA}}
\newcommand{\NCSUdesc}{\affiliation{Department of Physics, North Carolina State University, Raleigh, NC, 27695, USA}}
\newcommand{\Sandiadesc}{\affiliation{Sandia National Laboratories, Livermore, CA, 94550, USA}}
\newcommand{\ORNLdesc}{\affiliation{Oak Ridge National Laboratory, Oak Ridge, TN, 37831, USA}}
\newcommand{\UWdesc}{\affiliation{Center for Experimental Nuclear Physics and Astrophysics \& Department of Physics, University of Washington, Seattle, WA, 98195, USA}}
\newcommand{\LANLdesc}{\affiliation{Los Alamos National Laboratory, Los Alamos, NM, 87545, USA}}
\newcommand{\Laurentiandesc}{\affiliation{Department of Physics, Laurentian University, Sudbury, Ontario, P3E 2C6, Canada}}
\newcommand{\Tuftsdesc}{\affiliation{Department of Physics and Astronomy, Tufts University, Medford, MA, 02155, USA}}
\newcommand{\IUdesc}{\affiliation{Department of Physics, Indiana University, Bloomington, IN, 47405, USA}}
\newcommand{\VTdesc}{\affiliation{Center for Neutrino Physics, Virginia Tech, Blacksburg, VA, 24061, USA}}
\newcommand{\NCCUdesc}{\affiliation{Department of Mathematics and Physics, North Carolina Central University, Durham, NC, 27707, USA}}
\newcommand{\CMUdesc}{\affiliation{Department of Physics, Carnegie Mellon University, Pittsburgh, PA, 15213, USA}}
\newcommand{\UFdesc}{\affiliation{Department of Physics, University of Florida, Gainesville, FL, 32611, USA}}
\newcommand{\SLACdesc}{\affiliation{SLAC National Accelerator Laboratory, Menlo Park, CA, 94025, USA}}
\newcommand{\SNUdesc}{\affiliation{Department of Physics and Astronomy, Seoul National University, Seoul, 08826, Korea}}
\author{D.~Akimov}\Mephidesc
\author{S.~Alawabdeh}\NCSUnucengdesc
\author{P.~An}\Dukedesc\TUNLdesc
\author{C.~Awe}\Dukedesc\TUNLdesc
\author{P.S.~Barbeau}\Dukedesc\TUNLdesc
\author{C.~Barry}\Dukedesc
\author{B.~Becker}\UTKdesc
\author{V.~Belov }\ITEPdesc\Mephidesc
\author{I.~Bernardi}\UTKdesc
\author{C.~Bock}\USDdesc
\author{B.~Bodur}\Dukedesc
\author{A.~Bolozdynya}\Mephidesc
\author{R.~Bouabid}\Dukedesc\TUNLdesc
\author{A.~Bracho}\Dukedesc\TUNLdesc
\author{J.~Browning}\NCSUdesc
\author{B.~Cabrera-Palmer}\Sandiadesc
\author{E.~Conley}\Dukedesc
\author{J.~Daughhetee}\ORNLdesc
\author{J.~Detwiler}\UWdesc
\author{K.~Ding}\USDdesc
\author{M.R.~Durand}\UWdesc
\author{Y.~Efremenko}\UTKdesc\ORNLdesc
\author{S.R.~Elliott}\LANLdesc
\author{L.~Fabris}\ORNLdesc
\author{M.~Febbraro}\ORNLdesc
\author{A.~Gallo Rosso}\Laurentiandesc
\author{A.~Galindo-Uribarri}\ORNLdesc\UTKdesc
\author{M.P.~Green }\TUNLdesc\ORNLdesc\NCSUdesc
\author{B.~Harris}\Tuftsdesc
\author{M.R.~Heath}\ORNLdesc
\author{S.~Hedges}\Dukedesc\TUNLdesc
\author{R.~Henderson}\Dukedesc
\author{M.~Hughes}\IUdesc
\author{P.~Jairam}\Dukedesc
\author{B.A.~Johnson}\IUdesc
\author{T.~Johnson}\Dukedesc\TUNLdesc
\author{A.~Khromov}\Mephidesc
\author{A.~Konovalov}\Mephidesc\ITEPdesc
\author{E.~Kozlova}\Mephidesc\ITEPdesc
\author{A.~Kumpan}\Mephidesc
\author{L.~Li}\Dukedesc\TUNLdesc
\author{J.M.~Link}\VTdesc
\author{J.~Liu}\USDdesc
\author{A.~Major}\Dukedesc
\author{K.~Mann}\NCSUdesc
\author{D.M.~Markoff}\NCCUdesc\TUNLdesc
\author{J.~Mastroberti}\IUdesc
\author{J.~Mattingly}\NCSUnucengdesc
\author{M.~Mishra}\NCSUnucengdesc
\author{P.E.~Mueller}\ORNLdesc
\author{J.~Newby}\ORNLdesc
\author{D.S.~Parno}\CMUdesc
\author{S.I.~Penttila}\ORNLdesc
\author{D.~Pershey}\Dukedesc
\author{C.~Prior}\Dukedesc\TUNLdesc
\author{F.~Rahman}\NCSUnucengdesc
\author{R.~Rapp}\CMUdesc
\author{H.~Ray}\UFdesc
\author{O.~Razuvaeva}\Mephidesc\ITEPdesc
\author{D.~Reyna}\Sandiadesc
\author{G.C.~Rich}\TUNLdesc
\author{A.~Rouzky}\Dukedesc
\author{D.~Rudik}\Mephidesc
\author{J.~Runge}\Dukedesc\TUNLdesc
\author{D.J.~Salvat}\IUdesc
\author{A.M.~Salyapongse}\CMUdesc
\author{J.~Sander}\USDdesc
\author{K.~Scholberg}\email{kate.scholberg@duke.edu}\Dukedesc
\author{P.~Siehien}\USDdesc
\author{A.~Shakirov}\Mephidesc
\author{G.~Simakov}\Mephidesc\ITEPdesc
\author{W.M.~Snow}\IUdesc
\author{V.~Sosnovstsev}\Mephidesc
\author{J.~Steele}\Dukedesc\TUNLdesc
\author{A.~Strasbaugh Hjelmstad}\Dukedesc
\author{T.~Subedi}\VTdesc
\author{B.~Suh}\IUdesc
\author{R.~Tayloe}\IUdesc
\author{K.~Tellez-Giron-Flores}\VTdesc
\author{F.~Tsai}\UWdesc
\author{Y.-T.~Tsai}\SLACdesc
\author{E.~Ujah}\NCCUdesc\TUNLdesc
\author{E.~van Nieuwenhuizen}\Dukedesc\TUNLdesc
\author{R.L.~Varner}\ORNLdesc
\author{C.J.~Virtue}\Laurentiandesc
\author{G.~Visser}\IUdesc
\author{K.~Walkup}\VTdesc
\author{J.~Wang}\Dukedesc
\author{E.M.~Ward}\UTKdesc
\author{T.~Wongjirad}\Tuftsdesc
\author{Y.~Yang}\USDdesc
\author{J.~Yoo}\SNUdesc
\author{C.-H.~Yu}\ORNLdesc
\author{J.~Zettlemoyer}\altaffiliation{Now at: Fermi National Accelerator Laboratory, Batavia, IL, 60510, USA}\IUdesc

\maketitle
\medskip

\medskip

\begin{Abstract}
The COHERENT experiment located in Neutrino Alley at the Spallation Neutron Source (SNS),  Oak Ridge National Laboratory (ORNL), has made the world's first two measurements of coherent elastic neutrino-nucleus scattering (CEvNS), on CsI and argon, using neutrinos produced at the SNS. The COHERENT collaboration continues to pursue CEvNS measurements on various targets as well as additional studies of inelastic neutrino-nucleus interactions, searches for accelerator-produced dark matter (DM) and physics beyond the Standard Model, using the uniquely high-quality and high-intensity neutrino source available at the SNS. This white paper describes primarily COHERENT's ongoing and near-future program at the SNS First Target Station (FTS). Opportunities enabled by the SNS Second Target Station (STS) for the study of neutrino physics and development of novel detector technologies are elaborated in a separate white paper.
\end{Abstract}


\snowmass

\def\thefootnote{\fnsymbol{footnote}}
\setcounter{footnote}{0}

\clearpage
\tableofcontents

\clearpage
\section{Executive Summary}

\begin{itemize}
    \item The SNS at ORNL in Tennessee provides very high-quality neutrinos from stopped pions.  The neutrinos are produced with high intensity and sharp timing beneficial for background reduction, and have decay-at-rest (DAR) purity greater than 99\%~\cite{Akimov:2021geg}.
    \item CEvNS is the neutral-current scattering of a neutrino off an entire nucleus; its measurement offers a broad physics program, including multiple probes of BSM physics~\cite{COHERENT:2018gft}.  The COHERENT experiment in Neutrino Alley at the SNS exploited the SNS neutrino source~\cite{Akimov:2021geg} to make the first-light measurement of CEvNS on CsI~\cite{COHERENT:2017ipa} in 2017, and followed this with the first measurement of CEvNS on argon in 2020~\cite{COHERENT:2020iec}. 
 
    \item Beyond CEvNS, COHERENT can explore many other physics signals.  These include DM-induced recoils~\cite{COHERENT:2021pvd} and inelastic neutrino-nucleus interactions~\cite{COHERENT:2018gft} on several nuclear targets of particular interest for the core-collapse-supernova and solar-neutrino energy regime. 
 
    \item The COHERENT collaboration has additional existing and planned near-future deployments in Neutrino Alley at the SNS with exciting physics potential.  Deployments already underway include 18 kg of Ge and 2 tonnes of NaI, as well as a heavy-water detector for flux normalization~\cite{COHERENT:2021xhx}.  The experimental program under development includes tonne-scale argon, cryogenic inorganic scintillator and a liquid argon time-projection chamber.  These detectors will further broaden and deepen the physics reach of the COHERENT experiment.

    \item  An upgrade to the SNS proton beam will bring the power to 2~MW by 2025, and a second target station is planned in the 2030s, for a final power of 2.8~MW with protons divided between the two stations.  This offers many possibilities for the future, which are detailed in another white paper.
    
    \item The COHERENT collaboration is committed to constructively addressing issues of diversity, equity and inclusion within the physics community, as well as open sharing of data and tools.
    
\end{itemize}

\begin{figure}[htbp]\centering
\includegraphics[width=0.7\linewidth]{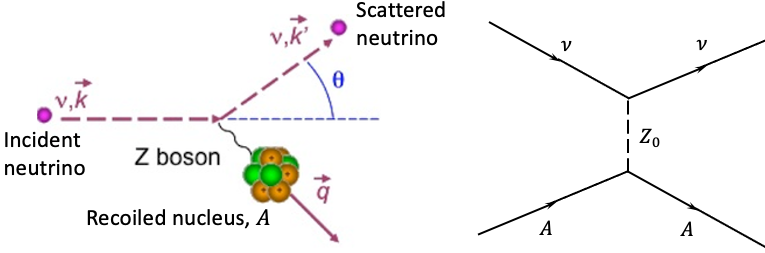}
\caption{Sketch of CEvNS interaction (left) and its corresponding Feynman diagram (right).}
\label{f:cevns}
\end{figure}

\section{COHERENT Program Overview}

\subsection{Coherent elastic neutrino-nucleus scattering (CEvNS)}

\begin{figure}[htbp]\centering
\includegraphics[width=0.7\linewidth]{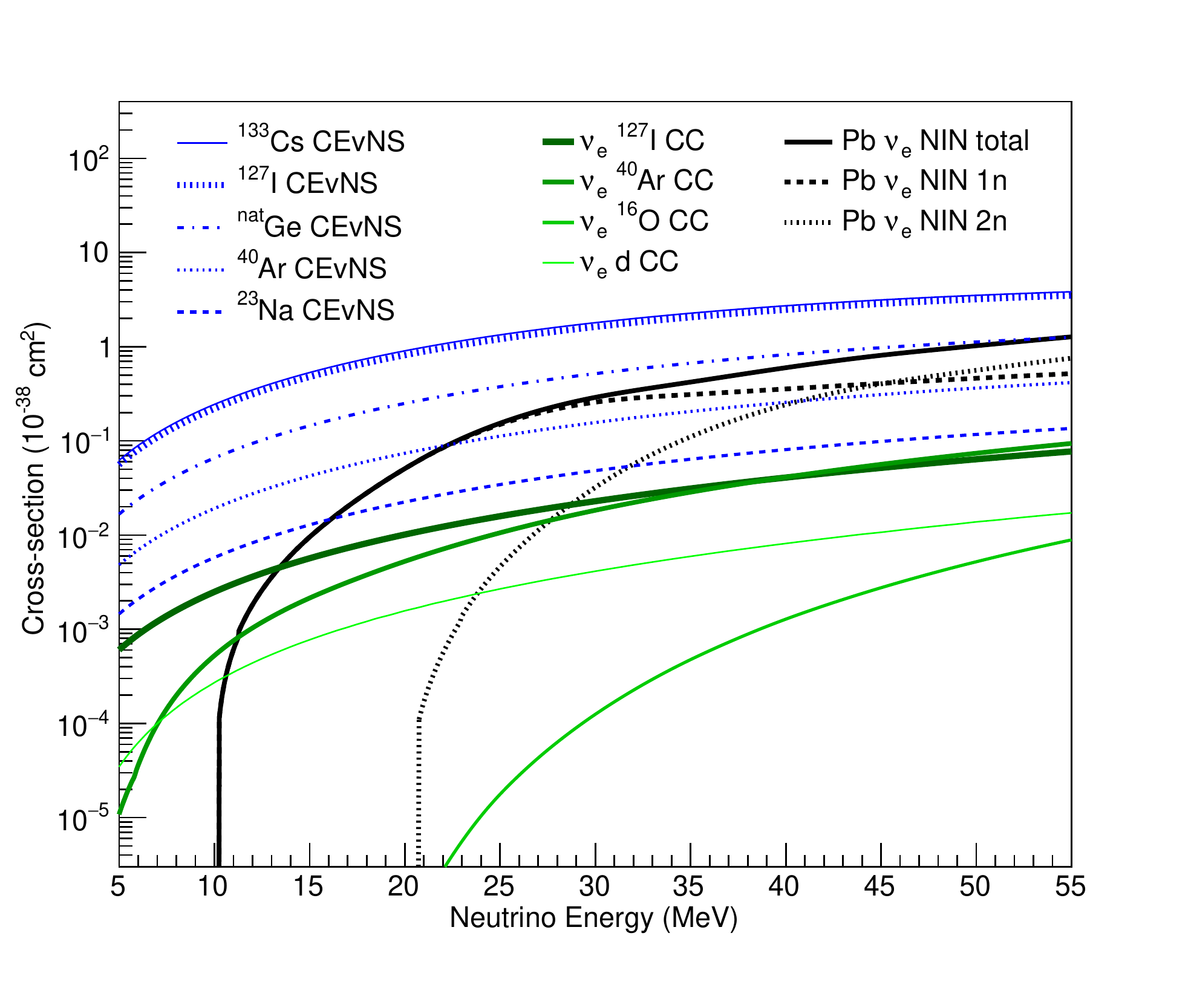}
\caption{Neutrino cross sections for neutrino energies up to 100~MeV relevant for COHERENT, from Ref.~\cite{Barbeau:2021exu}.}
\label{f:nuCrossSections}
\end{figure}

First proposed in 1974~\cite{Freedman:1973yd, Kopeliovich:1974mv}, CEvNS is a weak neutral-current (NC) process in which a low-energy neutrino, $\nu$, elastically scatters from a target nucleus, denoted by its atomic number $A$, through the exchange of a $Z_0$ boson as depicted in Fig.~\ref{f:cevns}. The coherence condition, in which the neutrino scatters off all nucleons of the nucleus in phase with each other, is satisfied when the neutrino energy is in the tens of MeV range and its momentum transfer to the recoiling nucleus is small (sub-keV to tens-of-keV range). Because the neutrino interacts with a nucleus as a whole, the CEvNS cross section for a given nuclear target are much larger than that of inelastic charged-current (CC) or neutrino interactions, for which a neutrino interacts with an individual electron or nucleon within a nucleus.  Cross sections of relevance to COHERENT are shown in Fig.~\ref{f:nuCrossSections}.

The differential cross section of CEvNS predicted by the Standard Model is given by~\cite{Barranco:2005yy}
\begin{equation}\label{eq:CEvNS_crossSection}
    \dfrac{d\sigma}{dT}(T, E_\nu) = \dfrac{G_{F}^{2}M}{2\pi}\left[(G_{V}+G_{A})^{2}+(G_{V}-G_{A})^{2}\left(1-\dfrac{T}{E_{\nu}}\right)^{2}-(G_{V}^{2}-G_{A}^{2})\dfrac{MT}{E_{\nu}^{2}}\right],
\end{equation}
where $T$ is the recoil energy, $E_{\nu}$ is the incident neutrino energy, $G_{F}$ is the Fermi constant, $M$ is the target nuclear mass,
\begin{equation}\label{eq:crossSection_vectorCouplingConstant}
    G_{V}=(g^{p}_{V}Z+g_{V}^{n}N)F^{V}_{nucl}(Q^{2}),
\end{equation}
\begin{equation}\label{eq:crossSection_axialCouplingConstant}
    G_{A}=(g^{p}_{A}(Z_{+}-Z_{-})+g^{n}_{A}(N_{+}-N_{-}))F^{A}_{nucl}(Q^{2}),
\end{equation}
$g_{V}^{n,p}$ and $g_{A}^{n,p}$ are vector and axial-vector coupling factors, respectively, for protons and neutrons, $Z$ and $N$ are the proton and neutron numbers, $Z_{\pm}$ and $N_{\pm}$ refer to the number of spin up or down nucleons, $F^{V,A}_{nucl}$ are vector and axial nuclear form factors, and $Q$ is the momentum transfer. As the numbers of spin up and down nucleons in a nucleus are either precisely zero or much smaller than the number of nucleons, the axial-vector contribution $G_A$ is small. The couplings are subject to percent-level $Q$-dependent radiative corrections~\cite{Tomalak:2020zfh}, with values of $g_V^n\sim -0.511$ and $g_V^p\sim 0.03$.

\begin{figure}[htbp]\centering
\includegraphics[width=0.7\linewidth]{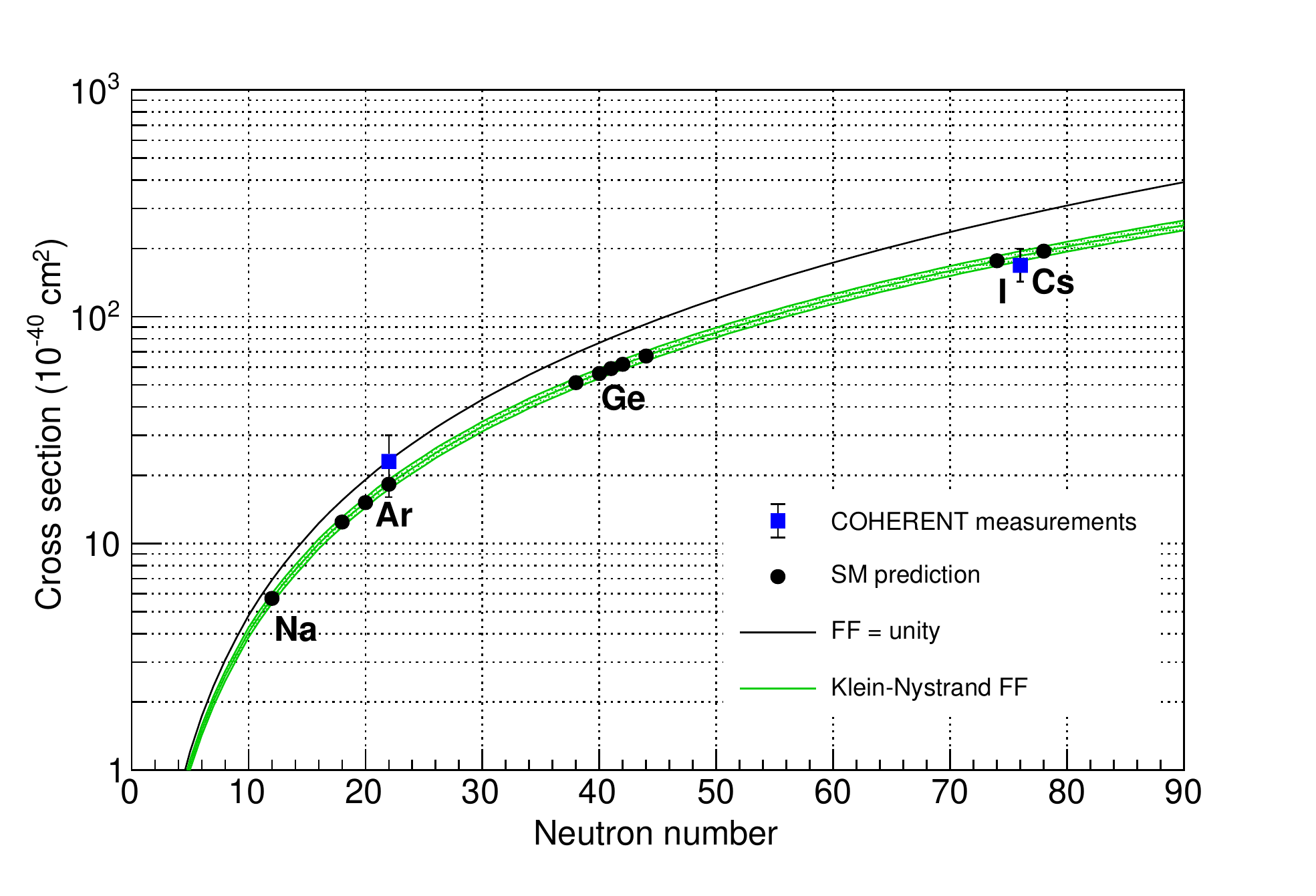}
\caption{Flux-averaged CEvNS cross sections as a function of neutron number, from Ref.~\cite{Barbeau:2021exu}.  The black line assumes unity form factor.  The two COHERENT measurements~\cite{COHERENT:2017ipa, COHERENT:2020iec} are indicated.}
\label{f:nuxs}
\end{figure}

The vector contribution $G_V$ is hence mainly determined by the total number of neutrons in the target nucleus. The cross section as a function of the number of neutrons is shown in Fig.~\ref{f:nuxs}.

Even though CEvNS has a relatively large cross section, it was not until recently that this process was observed~\cite{COHERENT:2017ipa, COHERENT:2020iec} due to the low recoil energy of the nucleus. A pulsed neutrino source optimized for generating high-flux neutrinos in the tens of MeV range and detectors that are sensitive to nuclear recoils down to keV or even sub-keV range enabled the first measurements, and this setup provides further opportunities for detailed study of CEvNS.

\subsection{Neutrinos at the SNS}
\begin{figure}[htbp]\centering
\includegraphics[width=0.9\linewidth]{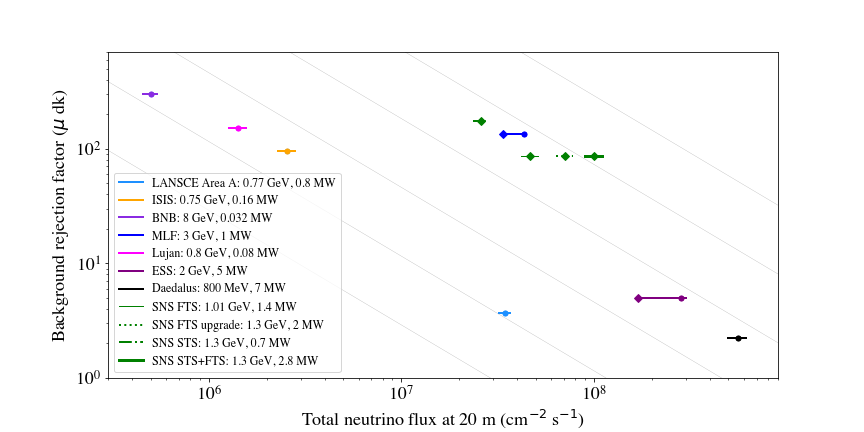}
\includegraphics[width=0.9\linewidth]{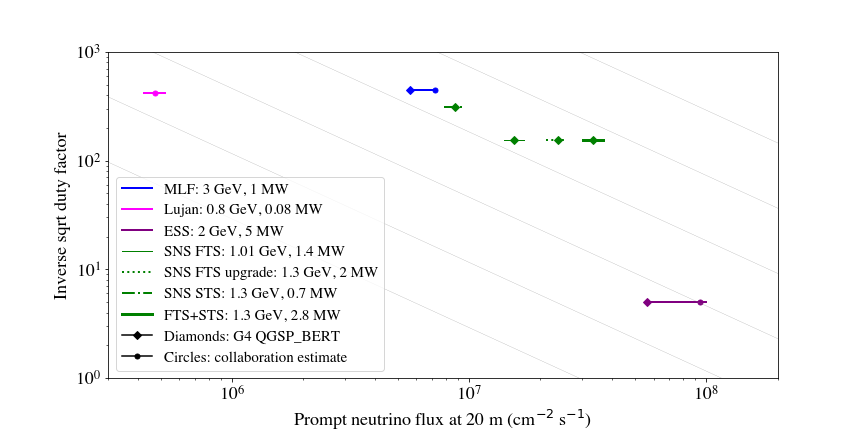}
\caption{Figures of merit for stopped-pion neutrino sources worldwide (past, current and future).  Top plot: The x-axis shows the total (all-flavor) neutrino flux, which depends on the beam power and proton energy, as well as details of the target geometry.  The uncertainty is shown as the range of fluxes.  The diamonds show flux estimates based on Geant4 QGSP\_BERT; the circles indicate the relevant collaboration's chosen baseline flux estimate. The y-axis shows the inverse square root of the duty factor, a measure of impact of background rejection due to beam pulsing, where duty factor is determined from the maximum of the muon decay lifetime, 2.2~$\mu$s, and the beam pulse window.   A larger value corresponds to improved steady-state background rejection.    The diagonal lines represent contours of equal flux over inverse square root of the duty factor.  Bottom plot: The x-axis shows $\nu_\mu$ flux.  This plot demonstrates the impact of sharp pulsing on flavor separation.  The y-axis here represents the duty factor computed using the time window that can be used for prompt $\nu_\mu$ selection.  For the MLF, for which there are two pulses separated by 540~ns, only the first one is considered.  This plot considers only current and near-future sources. }
\label{f:worldNuSources}
\end{figure}

The Spallation Neutron Source (SNS), a User Facility of the Department of Energy Office of Science operated by Oak Ridge National Laboratory (ORNL) offers a high-flux, short-pulsed, stopped-pion neutrino source.  The SNS is best suited for CEvNS detection compared to other similar facilities in the world as shown in Fig.~\ref{f:worldNuSources}.   The plots show figures of merit for total neutrino flux (for which the relevant time window is the maximum of the beam window and the muon-decay time scale) and for prompt neutrino flux (for which the relevant time window is the proton pulse width).  The prompt-flux figure of merit is especially relevant for physics for which flavor separation matters, or for which one expects specific prompt new physics signal (see, e.g., Sec.~\ref{s:dm}.)

As currently operating, the SNS utilizes a superconducting linear accelerator to bring hydrogen ions to a kinetic energy of 1 GeV. The ions are stripped of their electrons with a thin foil and the protons are injected into a 1 microsecond period storage ring.
In just over a millisecond, 1200 of these proton pulses are accumulated before the beam is extracted and directed onto a massive liquid mercury target with a longitudinal profile of only 350 ns full-width-half-maximum.
This entire process is repeated 60 times a second to deliver a total of $10^{16}$ protons on target (POT) per second at 1.4 MW, producing 20-30 neutrons per proton-Hg collision.
  
The neutrons are heavily moderated and sent down beam lines to neutron scattering instruments as well as a fundamental-neutron-physics experiment hall.
The SNS operates 5000 hours per year with very consistent beam conditions, enabling reliable, scheduled operations for the user program.

\begin{figure}[htbp]\centering
\includegraphics[width=\linewidth]{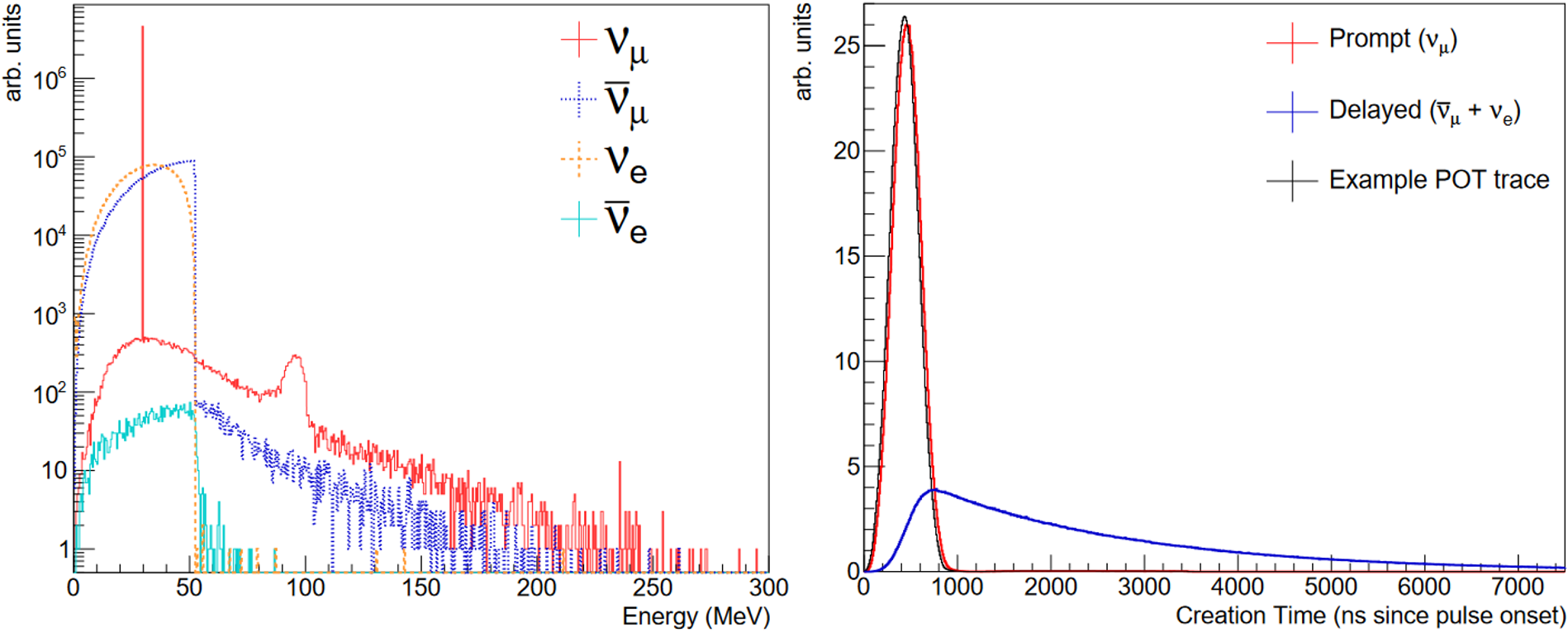}
\caption{Distributions of neutrino energy (left) and creation time (right) produced at the SNS predicted by our Geant4 simulation~\cite{Akimov:2021geg}.}
\label{f:tnu}
\end{figure}

As a by-product of the spallation, charged and neutral pions are also produced. About 99$\% $ of $\pi^{-}$ produced are captured within the thick and dense mercury target, while the majority of $\pi^{+}$ stop and decay at rest with a lifetime of 26~ns according to Eq.~\ref{eq:pionDecay}.  The majority of $\mu^{+}$ also stop inside the target and decay at rest, but with a longer lifetime of 2.2~$\mu s$ according to Eq.~\ref{eq:muonDecay}. This produces three distinct neutrino flavours, prompt $\nu_{\mu}$, and delayed $\nu_{e}$ and $\bar{\nu}_{\mu}$, with the kinetically well-defined energy spectra as shown in Fig.~\ref{f:tnu}.  

\begin{equation}\label{eq:pionDecay}
    \pi^{+}\rightarrow \mu^{+}+\nu_{\mu}
\end{equation}

\begin{equation}\label{eq:muonDecay}
    \mu^{+}\rightarrow e^{+}+\bar{\nu}_{\mu}+\nu_{e}
\end{equation}

Using a Geant4 simulation of the SNS~\cite{Akimov:2021geg}, we calculate a total luminosity of $2.36 \times 10^{15}$ neutrinos produced per second for an incident 1~GeV proton beam, or $4.25 \times 10^{22}$ neutrinos per year assuming typical SNS operations of 7.0 GWhr/yr.  The decay-at-rest production results in an intense, pulsed, and isotropic source of neutrinos with energies up to 52~MeV.

\subsubsection{Accelerator upgrade}
\begin{figure}[htbp]\centering
  \includegraphics[width = 0.45\textwidth]{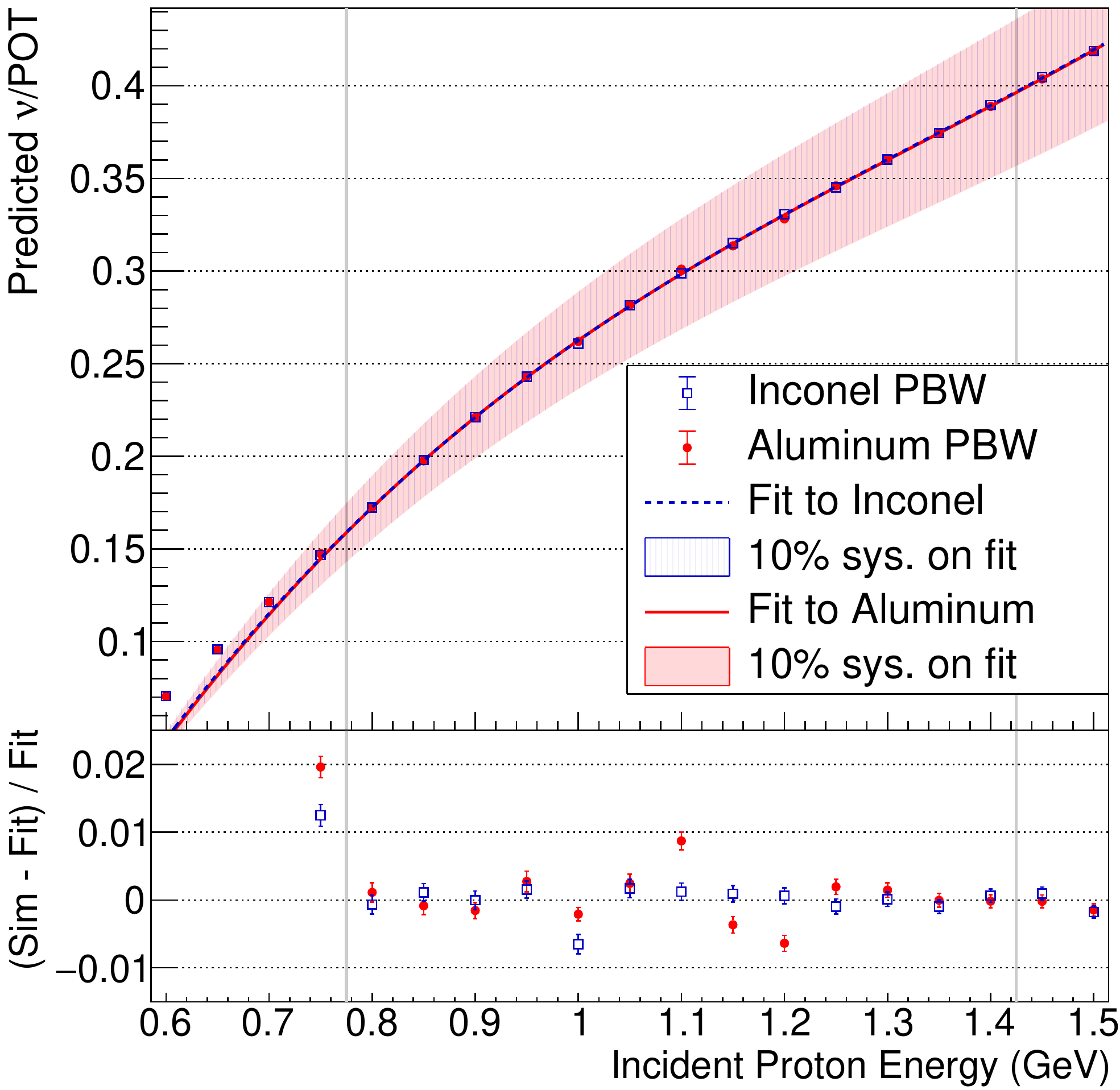}
  \includegraphics[width = 0.45\textwidth]{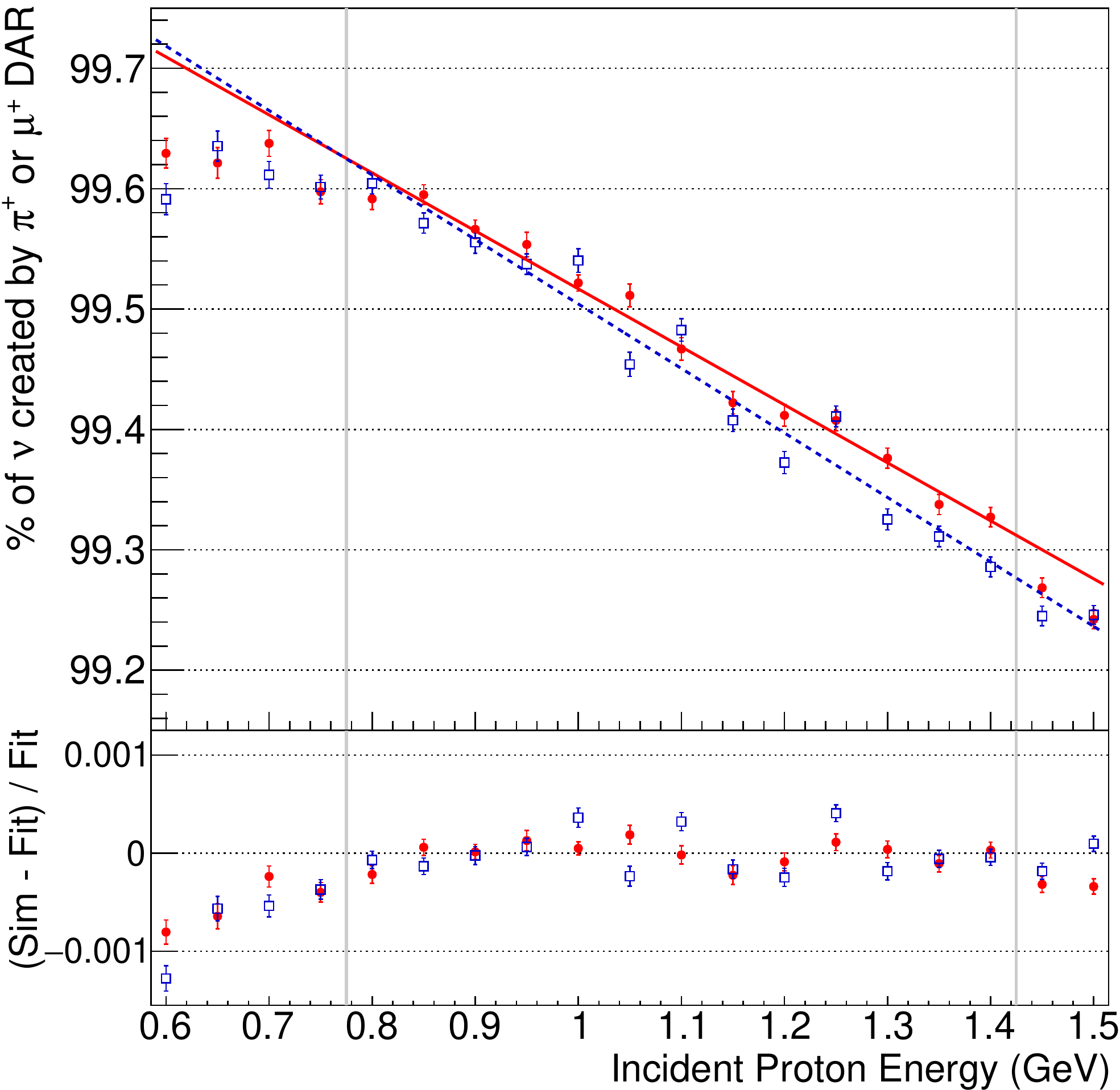}
  \caption{Dependence on proton energy of the predicted number of neutrinos produced per POT (left) and of the predicted percentage of neutrinos produced via decay-at-rest processes (right). The two colors indicate simulations with two different proton-beam window (PBW) materials; the bands on the left panel show the estimated 10\% systematic uncertainty on the neutrino-production simulation. Reproduced from Ref.~\cite{Akimov:2021geg}.}
  \label{fig:energy-dep}
\end{figure}

\begin{figure}[htbp] \centering
  \includegraphics[width = \textwidth]{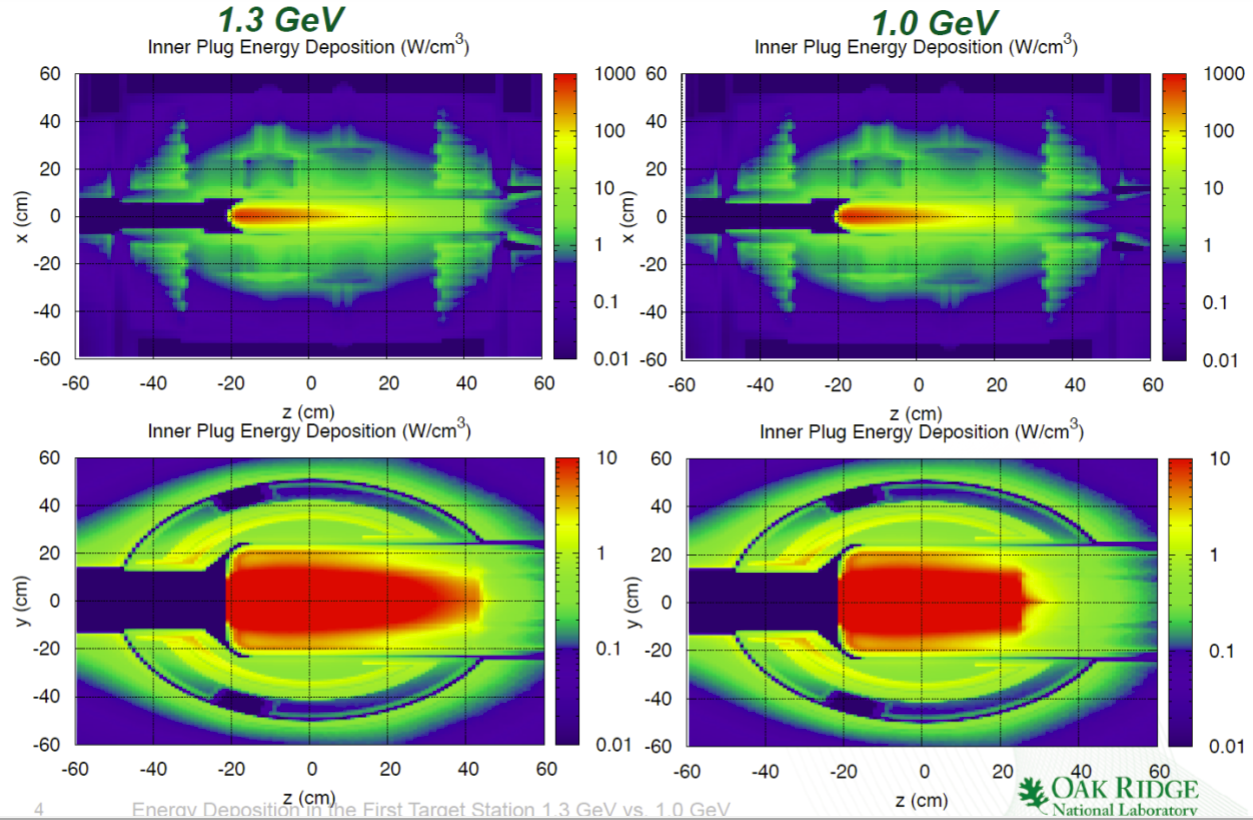}
  \caption{A comparison of the power deposition profiles at the SNS for operations after (left) and before (right) the PPU.  Reproduced from \cite{ppuDesign}.}
  \label{fig:ppuProtons}
\end{figure}

The planned Proton Power Upgrade (PPU) will increase the SNS proton power on target from 1.4 MW to 2.8 MW by 2028~\cite{ppuDesign}. This power increase entails a staged 50\% increase in the beam current, from 26 to 38~mA; neutrino production will scale linearly with this increase. The beam energy will also increase by 30\%, from 0.97 to 1.3~GeV with the addition of seven high-beta cryomodules to the superconducting linear accelerator.
The extensive testing of target gas injection during the ramp from 800 kW, and operation at 1.4 MW to mitigate cavitation damage, ensures that the target will accommodate the additional power when the SNS begins 2 MW operation in 2024.
The left panel of Fig.~\ref{fig:energy-dep} shows the predicted evolution of neutrino production with beam energy, based on our Geant4 simulation~\cite{Akimov:2021geg}. 

The right panel of Fig.~\ref{fig:energy-dep} shows a decreased fraction of decay-at-rest neutrinos resulting from increased penetration of the target, illustrated by Fig.~\ref{fig:ppuProtons}.  Pions and muons produced in the back of the target may not come to rest, increasing the average neutrino energy through boosts in the forward direction.  However, Fig.~\ref{fig:energy-dep} shows that this is a sub-percent effect at PPU energies; to a very good approximation, the neutrino source will simply increase in intensity for the PPU upgrade. 
To ensure control of systematics and backgrounds during the PPU staging process, it will be important for the COHERENT experiment to monitor both the neutrino flux~\cite{COHERENT:2021xhx} and the neutron background~\cite{COHERENT:2021qbu} in Neutrino Alley at as many energy steps as possible.

Anticipated in the mid-2030s, the Second Target Station (STS) will add an additional target site and experimental facility to the SNS. The STS offers the attractive possibility of literally building neutrino detection into the design: a dedicated, basement experimental hallway accommodating two 10-tonne neutrino detectors could be constructed for a moderate additional cost. Further information can be found in another Snowmass white paper~\cite{stsnuwp}.

\subsection{COHERENT detectors in Neutrino Alley}
The COHERENT Collaboration has deployed and is developing a number of subsystems with diverse target nuclei and detector technologies for the detection of neutrinos, neutrons, and possibly, dark matter particles from the SNS, as listed in Tab.~\ref{t:sub} and \ref{t:spd}.\footnote{The collaboration is adopting a more systematic naming scheme for its various subsystems, which starts with ``COH-'', followed by the target material, ``-Ge'' for example, and phase number, ``-2'', for example. The 50~kg HPGe subsystem listed in Fig.~\ref{t:sub} in this new naming scheme is called ``COH-Ge-2'', and the 10~kg CsI operated at 40~K is called ``COH-CryoCsI-1''. For historical reasons, the 750~kg liquid argon detector is called ``CENNS-750'' or ``COH-Ar-750'' instead of ``COH-Ar-2'', which may be subject to change in the future.}

\begin{figure}[htbp]\centering
\includegraphics[width=0.8\linewidth]{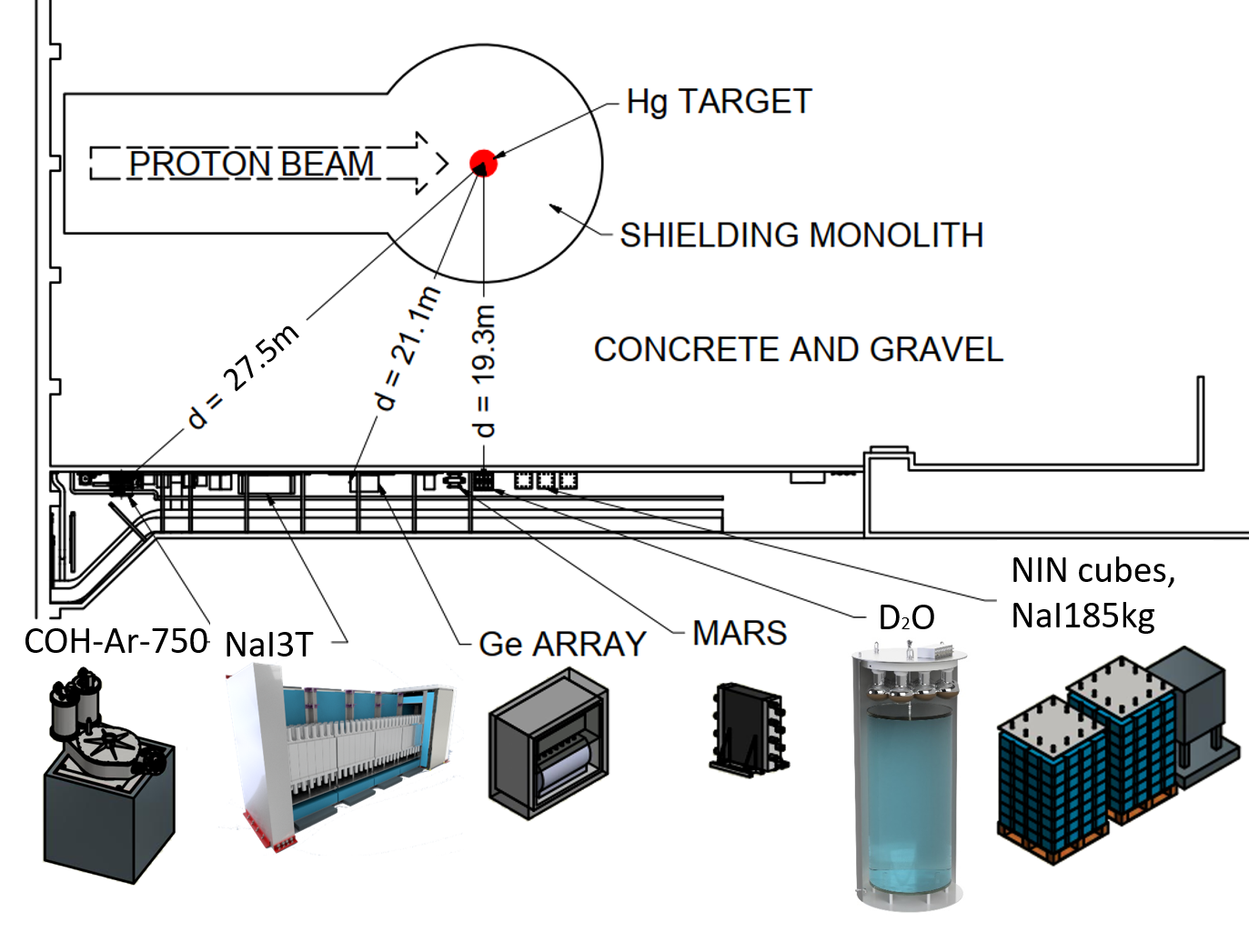}
\caption{Current and near-future detector subsystems in Neutrino Alley.}
\label{f:nuAlley}
\end{figure}

They are located 19$\sim$28~m from the Hg target in the basement of the SNS in what is known colloquially as ``Neutrino Alley'', as shown in Fig.~\ref{f:nuAlley}. With a 1~GeV proton beam, simulations~\cite{COHERENT:2021xhx} predict a flux of $4.7 \times 10^{7}$ neutrinos/(cm$^{2}$~s) 20~m away from the target.  Despite the proximity to the SNS beamline, beam-related neutrons are suppressed by a concrete fill between Neutrino Alley and the SNS target.  Cosmogenic background is reduced by an 8~m.w.e. (meter water equivalent) overburden. Due to the relatively large cross section of the CEvNS interaction for heavy nuclei and an intense neutrino source with excellent background rejection, detector mass can be reduced from the multiple kilotonne range down to the tens of kilogram scale. 

\begin{table}[htbp]\centering\small
\caption{Parameters of subsystems for CEvNS detection. \label{t:sub}}
\begin{tabular}{cccccc}
    \hline
    Nuclear & Detector & Target & Distance & Energy thresh-& Deployment\\
    target & Technology & Mass (kg) & from source & old (keV$^\dagger$)& dates \\\hline\rowcolor{lightgray}
    CsI[Na] & Scintillating crystal & 14 & 20 m & 5 & 2015-2019\\\rowcolor{lightgray}
    Ar & Single-phase LAr$\star$& 24 & 29 m & 20 & 2016-2021\\\rowcolor{lime}
    Ge & HPGe PPC$^\ddag$ & 18 & 22 m & $<$5 & 2022\\\rowcolor{lime}
    NaI[Tl] & Scintillating Crystal & 3500 & 22 m & 13 & 2022\\\rowcolor{lime}
    Ar & Single-phase LAr$\star$ & 750 & 29 m & 20 & 2025\\\rowcolor{lime}
    Ge & HPGe PPC$^\ddag$ & 50 & 22 m & $<$5 & 2025\\\rowcolor{lime}
    CsI & CsI+SiPM arrays at 40 K & 10$\sim$15 & 20 m & 1.4 & 2025\\\hline
\end{tabular}
\\\noindent \colorbox{lightgray}{Finished} \colorbox{lime}{Planned}, $^\star$liquid argon, $^\ddag p$-type point-contact, $^\dagger$nuclear recoil energy, approximate threshold
\end{table}

\begin{table}[htbp]\centering\small
\caption{Additional detectors that broaden the physics reach of COHERENT. \label{t:spd}}
\begin{tabular}{cllc}
    \hline
    Name & Detector Technology & Main purpose & Deployment dates \\\hline
    \multirow{2}{*}{NaIvE} & \multirow{2}{*}{185~kg NaI[Tl] crystals} & Measure $\nu_e+$ I CC cross section & 2016 - \\
    &  & \& beam-related backgrounds & present\\\hline
    \multirow{2}{*}{MARS} & scintillation panels inter- & Measure beam-related & 2017 -\\
    & leafed with Gd-painted foils & neutrons in Neutrino Alley & present\\\hline
    NIN  & liquid scintillator cells & Measure neutrino-induced & 2015 -\\
    cubes& in lead and iron shields  & neutrons (NIN) in lead \& iron & present\\\hline\rowcolor{lime}
     & heavy water & Measure neutrino flux precisely & \\\rowcolor{lime}
    \multirow{-2}{*}{D$_{2}$O} & Cherenkov detector & \& $\nu_e+$O inelastic cross section & \multirow{-2}{*}{2022}\\\hline\rowcolor{lime}
    LAr & liquid argon time- & Measure $\nu_e+$Ar inelastic & \\\rowcolor{lime}
    TPC & projection chamber &  cross section & \multirow{-2}{*}{2025}\\\hline
\end{tabular}
\\ Current \colorbox{lime}{Planned}
\end{table}

\section{Future Detectors}

In this section, we will describe COHERENT's future detector subsystems in approximate order of deployment in Neutrino Alley.  For some of these (heavy-water, COH-Ge-1 and COH-NaI-2), full resources have been secured and deployment is underway at the time of this writing.   Some of the farther future program is described in Ref.~\cite{stsnuwp}.

\subsection{Heavy-water Cherenkov detector}
\label{sec:d2o}

For the current COHERENT results~\cite{COHERENT:2017ipa, COHERENT:2020iec}, one of the dominant systematic uncertainties is due to the estimated 10\% uncertainty on the neutrino flux from the SNS target.
This uncertainty arises from comparisons between model predictions and the sparse available world data~\cite{Akimov:2021geg}.
Surface-based Cherenkov detectors have been successfully operated in the past, see e.g.~\cite{Willis:1980pj, MiniBooNE:2020pnu, ANNIE:2017nng}.
To that end, a heavy-water Cherenkov detector has been designed to operate in Neutrino Alley~\cite{COHERENT:2021xhx}.
This detector will make use of the well-understood $\nu_{e} + d \rightarrow p + p + e$ interaction cross section~\cite{Mosconi:2007tz, Ando:2019yum, Acharya:2019fij, Adelberger:2010qa} to greatly reduce the uncertainty on the SNS neutrino flux.

\begin{figure}[htbp]\centering
  \includegraphics[width=0.215\linewidth]{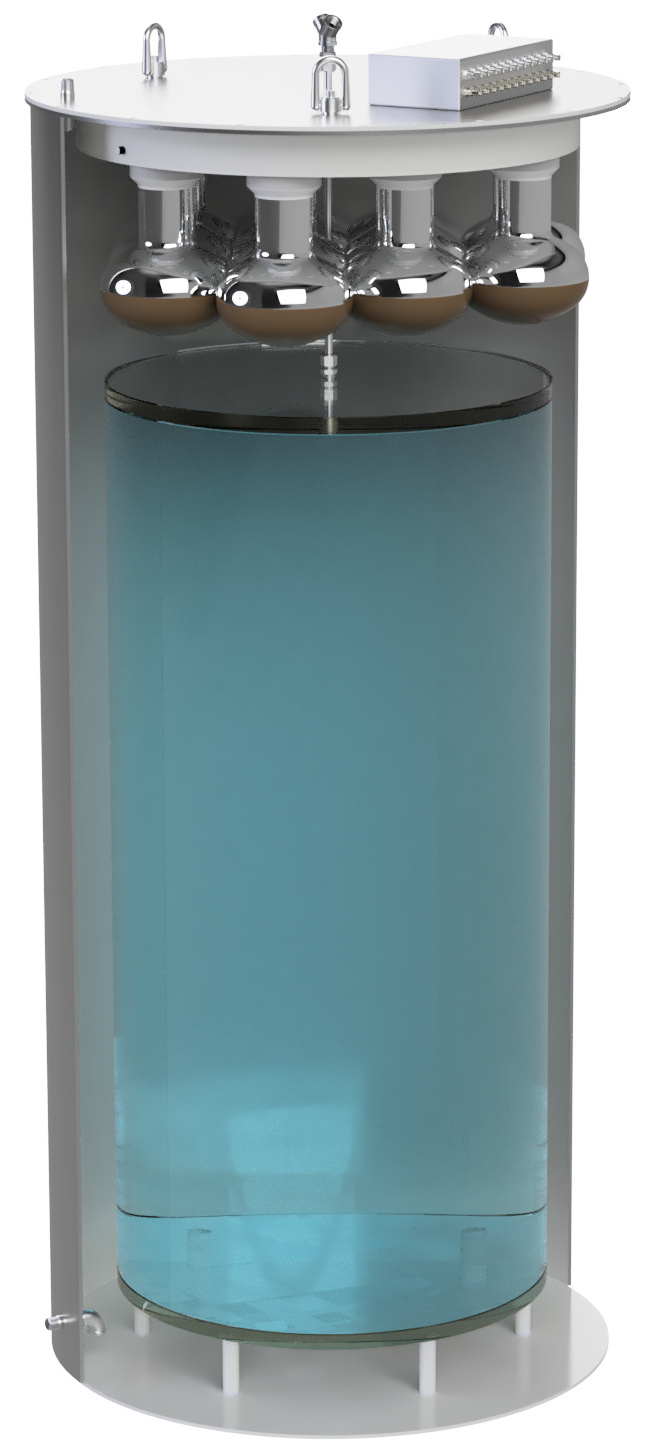}\hspace{2em}
  \includegraphics[width=0.65\linewidth]{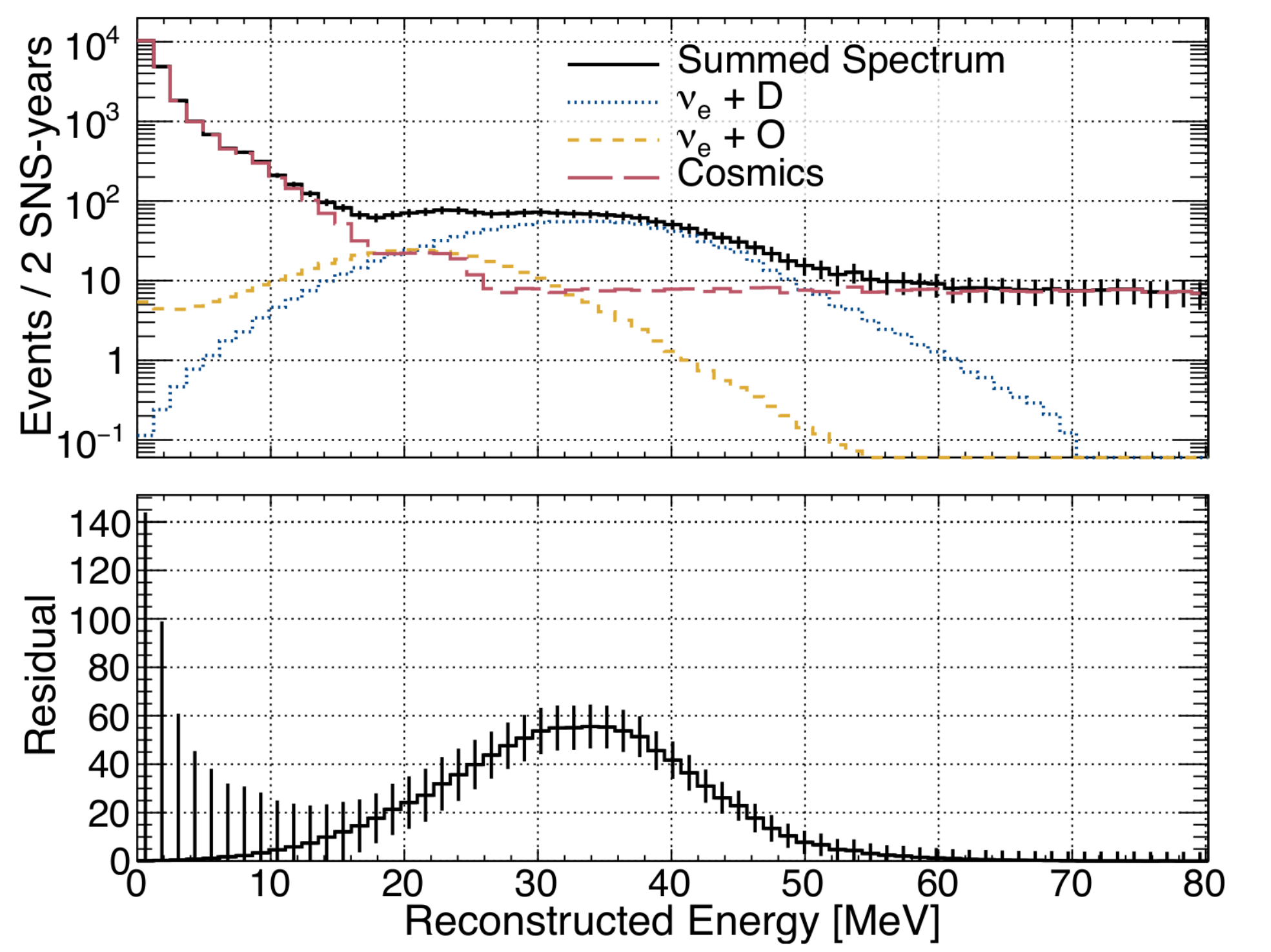}
  \caption{Left: Engineering drawing of the D$_2$O detector design, not showing shielding outside the water tank. Top right: Simulated signal and background energy spectra as reconstructed with the heavy-water detector. Bottom right: The background-subtracted $\nu_{e} + d$ spectrum with statistical errors, including smearing from imperfect energy resolution is included here. Both figures are reproduced from Ref.~\cite{COHERENT:2021xhx}. \label{d2ofigs}}
\end{figure}

As detailed in Ref.~\cite{COHERENT:2021xhx}, this D$_2$O detector is envisioned as a two-module system, with each module containing about 592~kg of D$_2$O and operating as an independent water-Cherenkov detector. It will be deployed roughly $90^\circ$ off the beam axis, about 20~m from the target, where the contribution from decay-in-flight neutrinos is minimal. The D$_2$O will be contained inside a transparent, cylindrical acrylic tank, which itself is contained inside a cylindrical steel tank. The volume between these tanks is filled with H$_2$O as a tail-catcher. The inner surfaces of the steel tank are lined in reflective Teflon$^{\rm{TM}}$, and twelve photomultiplier tubes view the system from above (Fig.~\ref{d2ofigs}). The system will be calibrated with Michel electrons~\cite{Michel:1949qe,TWIST:2004hce} and with an LED flasher system mounted within the steel tank. Figure~\ref{d2ofigs} shows the simulated signal and background spectra from a single module after two SNS-years of operation, resulting in an anticipated 4.7\% statistical uncertainty; we estimate that the statistical uncertainty will approach 2\% after 5 tonne-SNS-years of operation~\cite{COHERENT:2021xhx}. Continued neutrino-flux benchmarking with D$_2$O will be important during the proton power upgrade of the SNS, and during operations at the STS. In addition, this detector will measure for the first time charged-current neutrino reactions on oxygen in the energy range relevant fo supernova neutrinos in water Cherenkov detectors such as Super-Kamiokande and Hyper-Kamiokande.   

\subsection{High-Purity Germanium Detector Array}

The COH-GE-1 detector (also known as Ge-Mini) will be deployed in 2022.  This detector is based on well-understood technology including a number of experiments deployed for dark matter, neutrinoless double-beta decay and reactor CEvNS searches \cite{Aalseth:2008, Aalseth:2011, Aalseth:2011wp, Aalseth:2012if,Abdurashitov:2005tb,1748-0221-12-09-P09014,Abgrall:2014,Gerda:2017,Abgrall:2017syy,Zhao:2016dak,TEXONO:2014eky,CONUS:2020skt}.
COH-GE-1 consists of eight p-type point-contact high-purity Ge detectors of a bit more than 2 kg each, adding up to nearly 18 kg of natural germanium target.  Germanium offers several advantages for CEvNS:  it offers an intermediate value of $N$ as well as excellent energy resolution and thresholds in the few keVr range.  Signal timing resolution, while slower than scintillation light, does allow exploitation of the SNS pulsed beam for both background rejection and flavor separation.  
COH-Ge-1 detectors will be deployed in a multiport dewar with layers of copper, polyethylene and lead shielding.  The detector design incorporates also scintillator muon-veto panels.

The modular nature of the High-Purity Germanium (HPGe) detector array presently being deployed by COHERENT enables relatively straightforward and incremental increases of detector mass and sensitivity.
A detector array with a target mass in excess of 50 kg (COH-Ge-2) could be deployed in Neutrino Alley by making use of existing available ports in the current deployment arrangement (COH-Ge-1, at present the array is only 2/3 populated), and replicating the array and shielding assembly.
Preliminary studies with the array under construction will enable evaluation of the physics reach of a potential larger germanium detector array, for neutron form factor measurements, sterile oscillation searches, and searches for neutrino magnetic moments (see Sec.~\ref{s:pr}).

\subsection{COH-NaI-2 tonne-scale NaI[Tl] detector}
\label{sec:nai}
\begin{figure}[htbp]\centering
  \includegraphics[width = 0.415\textwidth]{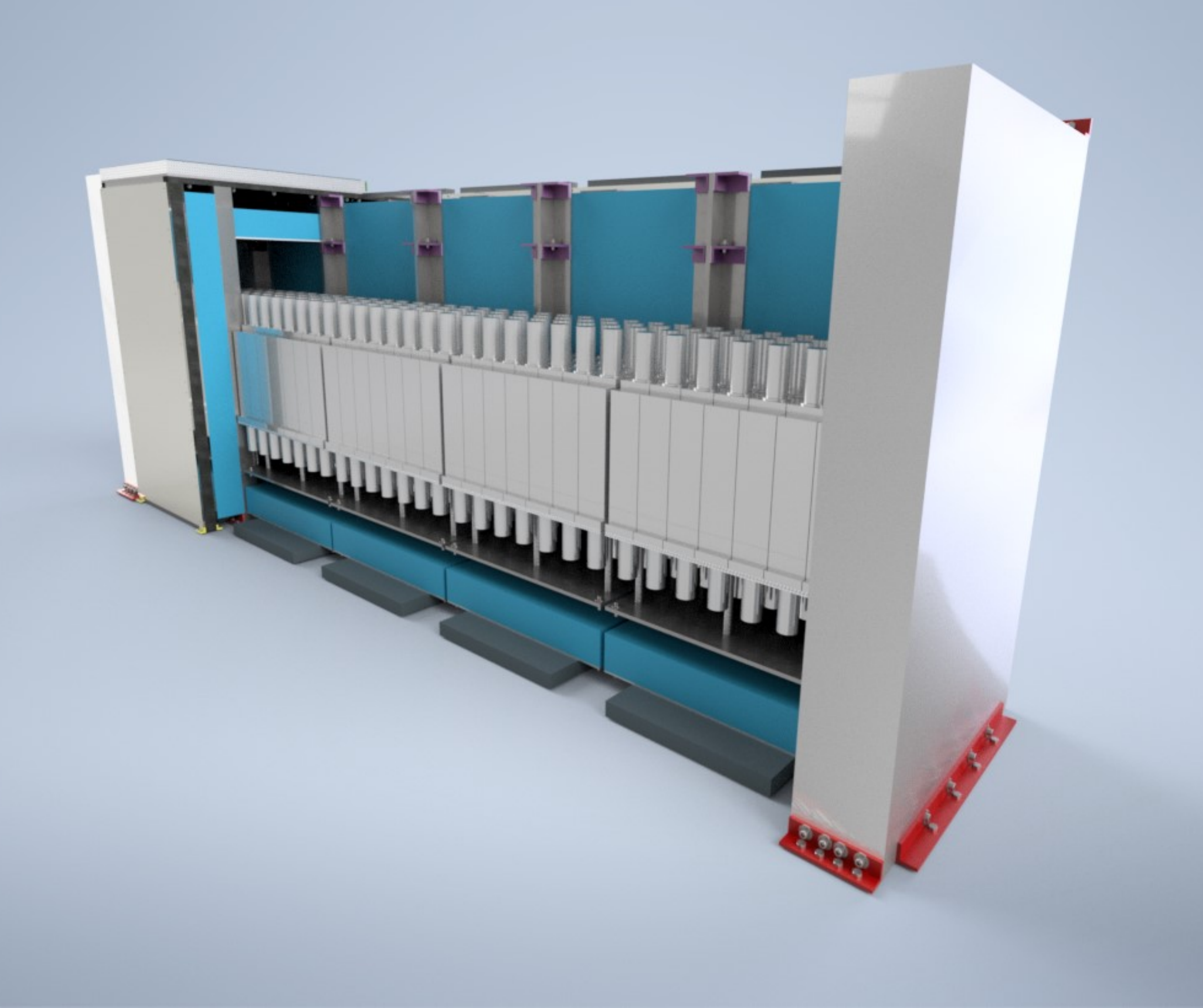}
  \includegraphics[width = 0.575\textwidth,trim={34 31 24 42},clip]{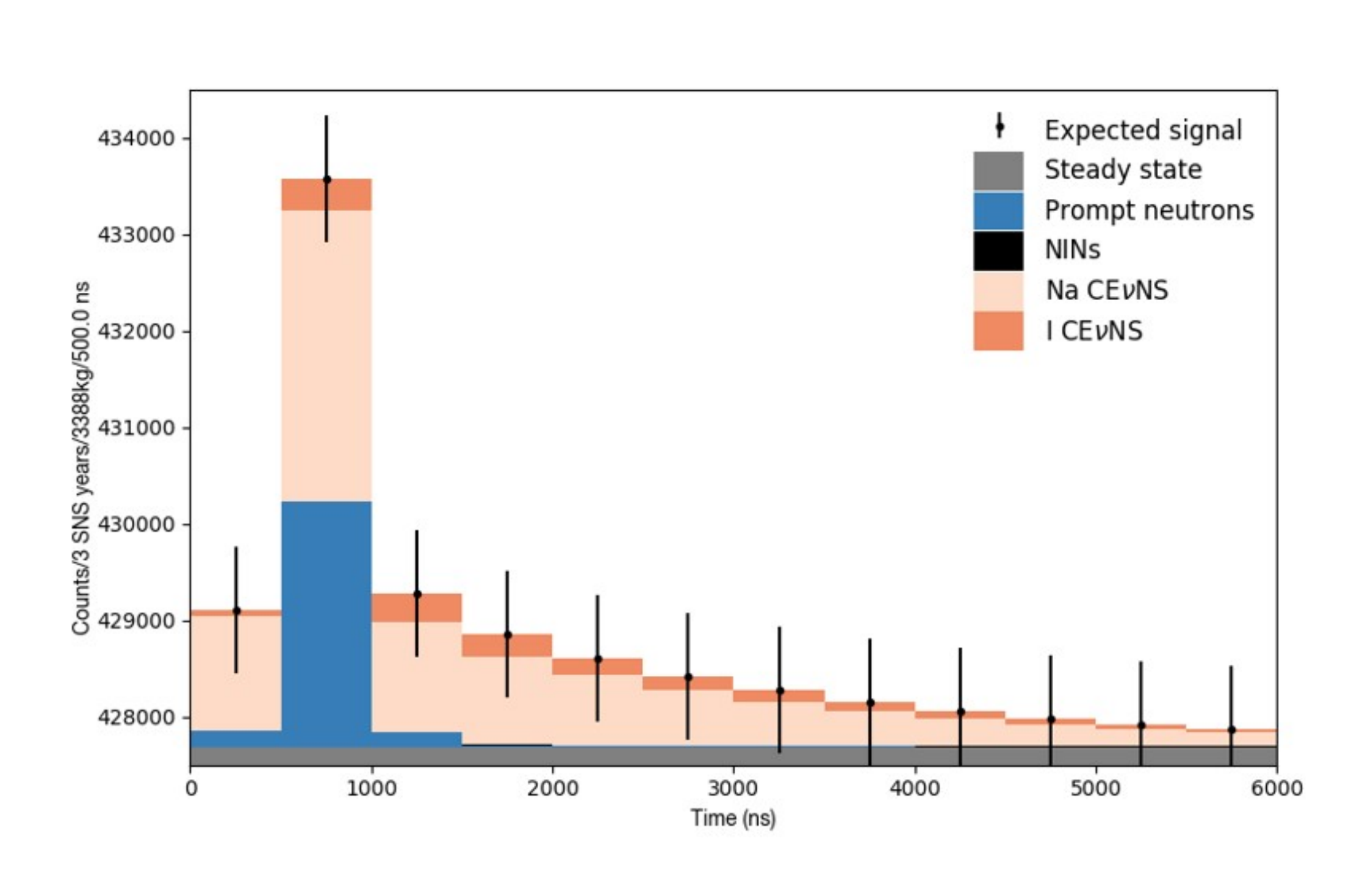}
  \caption{Left: NaIvETe modular detector array showing the NaI crystals surrounded by water bricks, steel and lead shielding in 5 modules.  The muon veto is not shown.  Right: Arrival time structure of the CEvNS signal and background simulation of the 7 module NaIvETe detector.  Errors shown are statistical.}
  \label{fig:nai-tonscale}
\end{figure}

A measurement of CEvNS on $^{23}$Na provides the lowest neutron-number data on the cross-section plot shown in Fig.~\ref{f:nuxs}.  The collaboration is currently constructing a multi-tonne modular array of re-purposed thallium-doped, 7.7~kg NaI crystals.  Each module of 63 crystals provides 485~kg of detector mass.  The current phase consists of the sectional deployment of 5 modules for a 2.4~tonne detector designed using dual-gain bases on the photomultiplier tubes to measure the low-energy CEvNS signal ($\sim$3-25~keVee) simultaneously with the high-energy ($\sim$10-50~MeV) charged-current signal on $^{137}$I.  Background studies with the NaIvE-185 detector array of 24 crystals deployed in Neutrino Alley since 2016, indicates that environmental and intrinsic backgrounds are sufficiently low for a successful CEvNS measurement.  Recent quenching-factor measurements and a calibration scheme will address  nonlinearity issues for low-energy signals. Initial charged-current studies with the NaIvE-185 detector will inform the analysis of the NaI Tonne-scale Experiment (NaIvETe) neutrino scattering measurement on $^{127}$I.  The future deployment of an additional two modules will bring the NaIvETe mass total to 3.4~tonnes.  Figure~\ref{fig:nai-tonscale} shows the modular design of the NaIvETe detector along with simulations of the 3.4~tonne configuration showing a 3$\sigma$ per year significance for a counting experiment.  The calculations assume a 13 keV-nr threshold, intrinsic backgrounds as measured in NaIvE-185, prompt neutron and neutrino-induced neutron backgrounds from CsI[Na] results and MCNP simulations, and a constant quenching factor in the energy region of interest.

\subsection{COH-Ar-750 liquid argon scintillation calorimeter}
\subsubsection{Overview}

\begin{figure}
    \centering
    \includegraphics[width=0.59\textwidth,trim={5 25 10 5},clip]{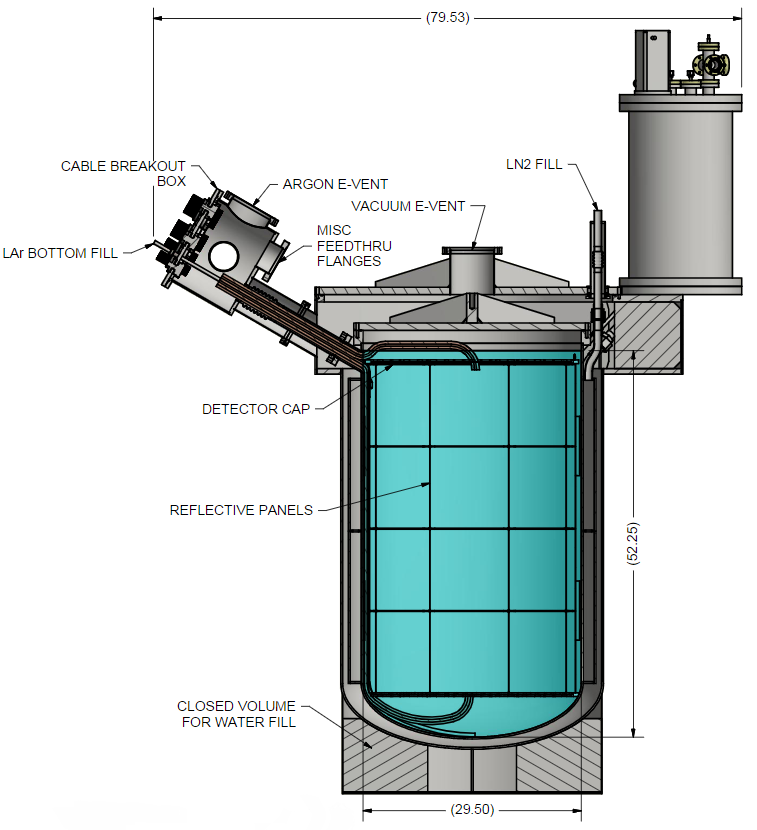}
    \includegraphics[width=0.39\textwidth,trim={10 5 15 10},clip]{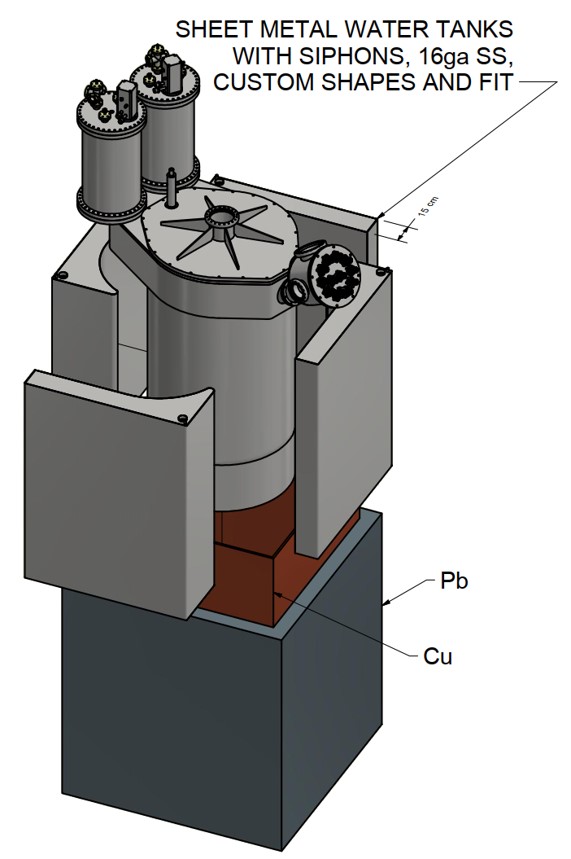}
    \caption{(left) Section view of COH-Ar-750, showing the internal cryostat geometry and active light-collection volume. (right) A rendering of COH-Ar-750 showing the compact shielding design. Layers of lead and copper shield from electromagnetic sources of steady-state background, and water shields from beam-related fast neutrons.}
    \label{fig:OneTonLAr}
\end{figure}

COH-Ar-750 (Fig.~\ref{fig:OneTonLAr}) represents the next phase of COHERENT's liquid argon CEvNS detection program. It uses a custom large-volume liquid argon cryostat with a small footprint to provide space for surrounding the apparatus with lead and water shielding to suppress background events. A cylindrical assembly of PMTs and wavelength-shifting (WLS) panels will view 610~kg of the volume. Current-generation cryogen-compatible PMTs have achieved remarkably high single-photon detection efficiency, thanks to ongoing WIMP searches, and we expect a 20-keVnr (nuclear recoil energy in keV) threshold to be readily achievable. We will design and construct a veto detector to identify and reject cosmic ray events, which are the dominant source of background for higher-energy charged-current events. The LAr must continually be chemically purified to assure that scintillation light can reach the photodetectors. We will implement a circulation and filtering system capable of maintaining the requisite part-per-million level chemical purity based upon the COH-Ar-10 design. To precisely measure the nuclear recoil distribution, we must continually calibrate the light-detection efficiency of the detector. This will be accomplished using a combination of radioactive sources and pulsed laser excitation, drawing from and improving upon the calibration methods used for COH-Ar-10. 

LAr-based detectors have been successfully deployed for WIMP searches~\cite{DarkSide:2018kuk}~\cite{DEAP:2019yzn} as well as neutrino detection~\cite{MicroBooNE:2018neo}. The scintillation properties of LAr, knowledge of the quenching of LAr nuclear recoil scintillation, and operational experience of LAr detectors are well documented in the literature (e.g. Ref.~\cite{DEAP:2019yzn}). The LAr detector must have a $\sim20$~keV or lower nuclear recoil energy threshold and a design that optimizes background reduction.

The COH-Ar-10 detector was installed in late 2016 and later upgraded for $\sim4.5$ photoelectron (PE) per keV electron-equivalent (keVee) light yield, and its performance closely informs the design of COH-Ar-750. It has observed CEvNS and verified the $N^2$ dependence of the CEvNS cross section~\cite{COHERENT:2020iec}. The active volume in COH-Ar-10 is formed with a 21-cm-diameter tetraphenyl butadiene (TPB)-coated Teflon$^{\rm{TM}}$ cylinder viewed by two TPB-coated 8-inch Hamamatsu R5912-02MOD PMTs with quantum efficiency of 18\%. Measurements with an internal $^{83m}$Kr calibration source showed that sufficient light yield for a nuclear recoil energy threshold down to 20 keVnr along with sufficient energy resolution and linearity for CEvNS detection~\cite{COHERENT:2020xlk}. Experience with the current detector also showed that liquid argon can discriminate between nuclear and electronic recoils via scintillation from singlet/triplet states with different time constants, reducing electronic recoil backgrounds by an order of magnitude.

\subsubsection{Cryostat design}

The 750~kg LAr volume will be contained in a cryostat precooled with LN$_{2}$ utilizing a separate plumbing loop welded to the outside of the inner cryostat. Liquid argon will be filled through contaminant filters from dewars into the inner (detector) cryostat. Utilizing both LN$_{2}$ and pulse-tube cooling schemes assures stable temperature control and adequate cooling power to compensate the heat load from internal instrumentation. A custom-designed liquefier and heat exchanger coupled to PT90 pulse tube refrigerators in a closed loop will re-liquefy the evaporated argon gas. This loop will include getters for continual chemical purification of the liquid to provide high, stable scintillation light yield.

Initial running will proceed with inexpensive natural ``atmospheric'' argon (AAr). This alleviates the need for recovery schemes requiring gas capture and re-compression, or long-term argon storage in custom-built dewars. For the further future, we are investigating the use of ``underground'' argon (UAr)~\cite{Back:2022maq}, which greatly reduces the fraction of $^{39}$Ar compared to AAr, almost entirely suppressing the steady-state background from $^{39}$Ar $\beta$-decay. The decay has a 565~keV endpoint, and thus a significant component in the energy range of interest for CEvNS. We have established a detector design that is compatible with UAr and are investigating the necessary recovery and storage schemes, and envision a future upgrade to the AAr-based detector on a timescale to match the availability of the UAr.

\subsubsection{Light detection and readout}

A TPB-coated Teflon$^{\rm{TM}}$ cylinder converts the short-wavelength primary scintillation light to visible light and conducts visible light to TPB-lined endcaps viewed by photodetectors. COH-Ar-10 data shows this is a reliable means for light transport. The detection volume consists of a skeletal aluminum support structure with curved, TPB-coated Teflon$^{\rm{TM}}$ panels fastened to the support structure to form the barrel of the cylinder. Flat TPB-coated mounting plates with holes to accept the array of PMTs are fastened to either end. The face of each PMT will also be coated with TPB.

\begin{figure}
    \centering
    \includegraphics[width=0.49\textwidth,trim={5 5 5 20},clip]{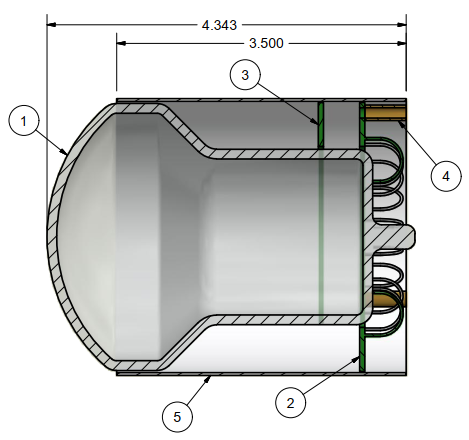}
    \includegraphics[width=0.37\textwidth,trim={10 10 10 10},clip]{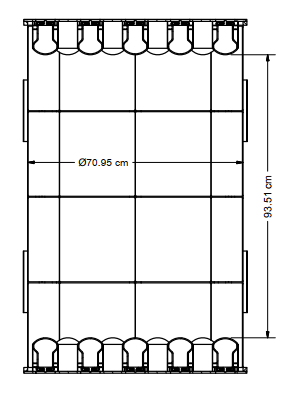}
    \caption{(left) A section view of the proposed PMT assembly: (1) The 3-inch Hamamatsu PMT housing (2-3) Compact custom signal readout and divider board, respectively (4) standoff for mounting to the light-collection volume (5) Teflon$^{\rm{TM}}$ PMT shroud. (right) The active light-collection volume in the COH-Ar-750 cryostat. Curved TPB-coated panels fasten to a cylindrical mount, and two panels (on top and bottom) support arrays of PMTs viewing the volume.}
    \label{fig:PMTassembly}
\end{figure}

The PMT array is a 2$\times$58 array of 3-inch Hamamatsu R14374-Y004 PMTs (Fig.~\ref{fig:PMTassembly}). The PMTs are compact and cryogenically compatible, with a nominal quantum efficiency approaching 20\%. We have confirmed experimentally that the dark rate and afterpulsing probability are consistent with Hamamatsu's specifications at LAr temperature. The PMT faces will be coated with TPB and read out by high channel-density digitizers.

We have simulated COH-Ar-750's response to nuclear recoils using Geant4, incorporating the LAr quenching factor, PMT quantum efficiency, and photon transport. The simulation uses a data-driven optical model matched to COH-Ar-10 calibration measurements. These simulations demonstrate the desired light-collection efficiency and $20$~keVnr threshould due in part to COH-Ar-750's wider aspect ratio, even though the fractional area of PMT coverage is slightly smaller than that for COH-Ar-10. We continue to perform Monte Carlo studies of the proposed detector to optimize event triggering and acceptance and study the pulse-shape response of nuclear and electronic recoil events to mitigate backgrounds.

\subsubsection{Calibration}

Comprehensive measurements of the detector response to low-energy recoil events is critical for precision CEvNS measurements and for maximizing DM sensitivity. Our controlled injection of $^{83m}$Kr used in COH-Ar-10 to perform low-energy electron recoil energy calibrations establishes the light yield and linearity of the detector~\cite{COHERENT:2020xlk}. For COH-Ar-750 we will combine this technique with direct light calibration using a UV fiber laser optically coupled to the light-collection volume. 

We are exploring additional calibration techniques such as \textit{in situ} radioactive point sources. Tagged fission sources are a potential means of performing \textit{in situ} measurements of the nuclear recoil (NR) response of the detector \cite{LUX:2016ezw}. Collimated, monoenergetic 14~MeV neutrons from a time-tagged $d$-$d$ or $d$-$t$ fusion source can be scattered from the COH-Ar-750 volume and detected at known angles using liquid scintillator detectors. We are currently modeling this and other NR calibration techniques for COH-Ar-750.

\subsubsection{Shielding}

Beam-unrelated steady-state backgrounds in Neutrino Alley come from environmental $\gamma$-ray backgrounds, which can be suppressed with lead and copper shielding, and from the radioactive $^{39}$Ar present in atmospheric argon. Both background sources are measured \textit{in situ} and subtracted, and these backgrounds only affect the statistical sensitivity of the measured signal.

\begin{figure}
    \centering
    \includegraphics[width=0.6\textwidth,trim={0 20 0 20},clip]{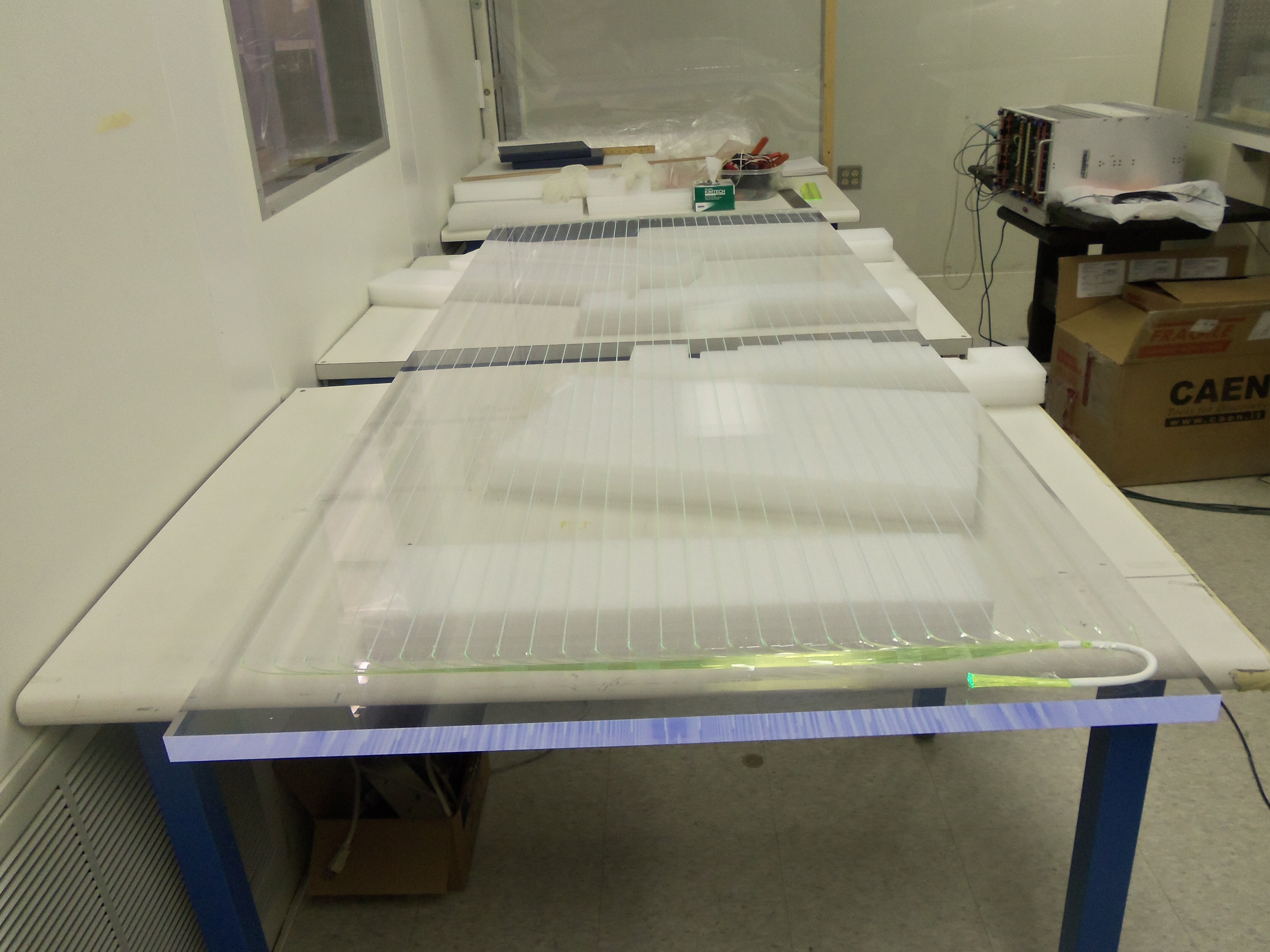} \\
    \caption{Veto panels for use in COH-Ar-750. Wavelength shifting fibers weave through the grooves of plastic scintillator panels, with SiPMs detecting light at the end of the fiber bundle.}
    \label{fig:veto_photo}
\end{figure}

Beam-related backgrounds come from fast neutrons from the spallation target, which can scatter within the active volume and make nuclear recoils. The rate of these events competes with the CEvNS rate and must be quantified. In COH-Ar-10 this is done by running with the fast neutron water tank shields alternately filled and drained, which changes the fast neutron background by a factor of 10. A Geant4 simulation modeling these backgrounds is tuned to agree with the empty-shield data and then generates a background prediction for the filled shields with associated uncertainties.

These measurements with COH-Ar-10 confirm our understanding of the sources of and mitigation strategies for the dominant backgrounds to CEvNS in Neutrino Alley~\cite{COHERENT:2020iec}. The relatively compact footprint of the COH-Ar-750 cryostat allows 10~cm of Pb for environmental $\gamma$-ray shielding and 15~cm H$_{2}$O for beam-related neutrons (Fig.~\ref{fig:OneTonLAr}) as is employed by the current detector.

\subsubsection{Muon veto} Cosmic ray muons must be vetoed to access the higher energy, lower rate charged-current neutrino events in COH-Ar-750. We will implement an organic scintillator-based veto detector above COH-Ar-750 based both on our ongoing Geant4 simulation effort and on current designs for other COHERENT detectors. The design and construction uses conventional technology with wavelength shifting fibers woven into the scintillator and coupled to SiPMs (Fig.~\ref{fig:veto_photo}), based upon the MAJORANA demonstrator design~\cite{Bugg:2013ica}. The final geometry will be optimized for high muon tagging efficiency and minimum dead time from the $\gamma$-ray backgrounds in Neutrino Alley.

\subsubsection{R\&D using COH-Ar-10}

We will perform small-scale tests of COH-Ar-750 components using COH-Ar-10. COH-Ar-10 will soon be modified to accept 3-inch PMTs and other candidate photodetectors to be used in COH-Ar-750 to measure the detectors' response to LAr scintillation light, analyze pulse-shape discrimination, and develop optimal readout and digitization schemes. The COH-Ar-10 cryostat will also be used to measure light-collection efficiencies and other optical properties of TPB-coated materials using improved techniques. This work will mitigate risk for the proposed design.

\subsection{COH-CryoCsI-1 cryogenic undoped CsI scintillating crystal}
\subsubsection{Motivation}
In 2017, the collaboration observed CEvNS using a 14.6~kg CsI[Na] detector~\cite{COHERENT:2017ipa}. A PMT was used as the light sensor in that detector. A serious background limiting its sensitivity was the Cherenkov radiation emitted from the PMT quartz window by charged particles. A switch from PMTs to SiPM arrays can be used to eliminate this background since SiPM arrays do not have a quartz window. In order to reduce the high dark count rate of SiPMs at room temperature, they need to be cooled to cryogenic temperatures~\cite{Acerbi:2016ikf}. The cryogenic operation calls for the switch from doped CsI crystals to undoped ones, since the latter at 40~Kelvin have about twice as high light yield as the former at 300~Kelvin~\cite{Bonanomi52, Hahn53a, Hahn53, Sciver56, Beghian58, Sciver58, Sciver60, Hsu66, Fontana68, West70, Fontana70, Emkey76, persyk80, Woody:1990hq, Williams90, Nishimura95, Wear96, Amsler:2002sz, Moszynski03, Moszynski03a, Moszynski05, Moszynski:2009zz, Sibczynski10, Sibczynski:2012hr, Sailer:2012ua}.  Compared to the original CsI[Na] detector, the combination of
\begin{itemize}
    \item the elimination of Cherenkov radiation,
    \item the higher photon-detection efficiency of SiPMs than PMTs,
    \item the low dark count rate of SiPMs, and
    \item the high light yield of undoped crystals at or below 77~Kelvin,
\end{itemize}
leads to at least an order of magnitude increase of the detectable CEvNS events, given a similar exposure. The physics reach of a $\sim$10 kg (COH-CryoCsI-1) and a $\sim$700 kg (COH-CryoCsI-2) cryogenic CsI detector at the first and the second target stations is detailed in Sec.~\ref{s:pr}. Detailed in this section are working principles, design considerations and projected performance of the proposed detectors.

\subsubsection{Light yields and nuclear quenching factors of undoped crystals}
\begin{figure}[htbp]\centering
  \includegraphics[width=0.51\linewidth]{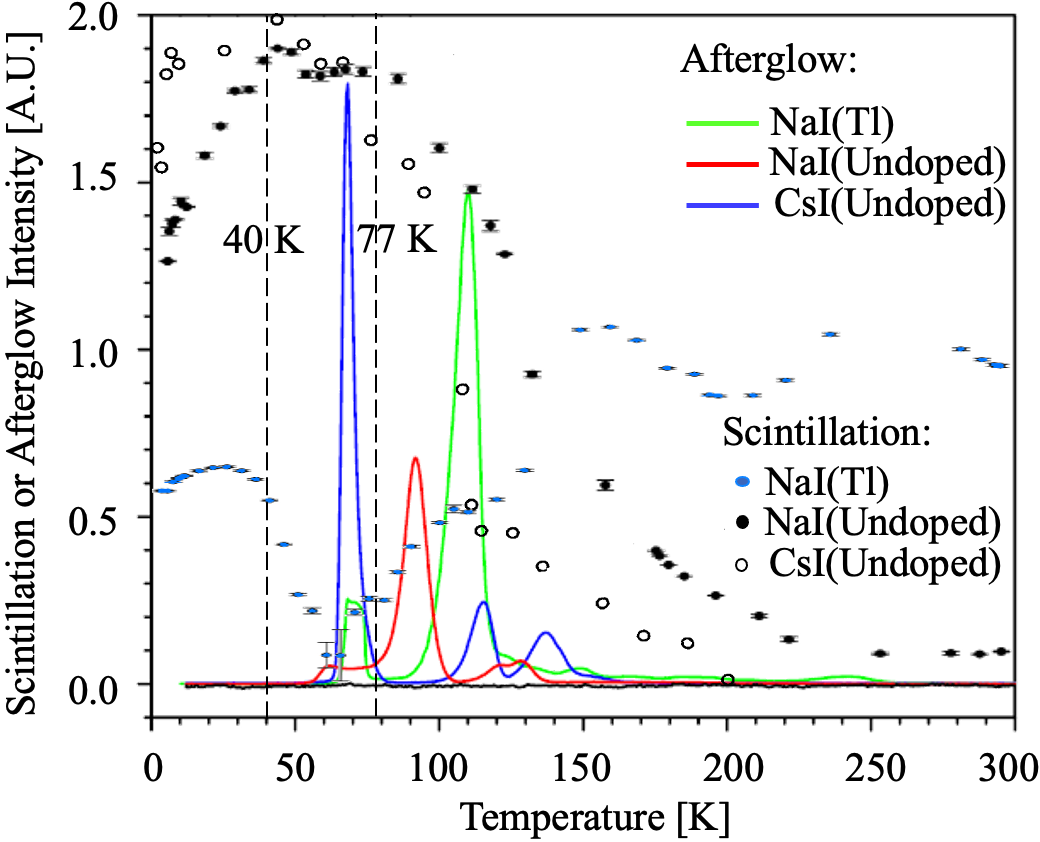}\hspace{2em}
  \includegraphics[width=0.30\linewidth]{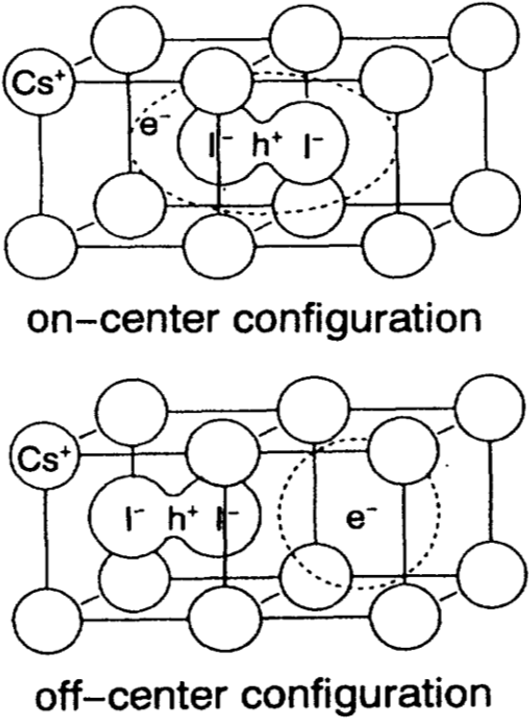}
  \caption{Left: relative scintillation yields~\cite{Nishimura95, Amsler:2002sz, Sailer:2012ua} and afterglow rates~\cite{Derenzo:2018plr} of various crystals versus temperature. Right: Two types of self-trapped excitons identified in undoped CsI (taken from Ref.~\cite{Nishimura95}).} \label{f:ly}
\end{figure}

As shown in the left graph in Fig.~\ref{f:ly}, light yields of undoped NaI/CsI increase rapidly when temperature goes down, and peak around 40~K~\cite{Bonanomi52, Hahn53a, Hahn53, Sciver56, Beghian58, Sciver58, Sciver60, Hsu66, Fontana68, West70, Fontana70, Emkey76, persyk80, Woody:1990hq, Williams90, Nishimura95, Wear96, Amsler:2002sz, Moszynski03, Moszynski03a, Moszynski05, Moszynski:2009zz, Sibczynski10, Sibczynski:2012hr, Sailer:2012ua}. Around liquid nitrogen (LN$_2$) temperature (77 K), the yields are about twice higher than those of NaI/CsI[Tl] at room temperature~\cite{Bonanomi52, Sciver56, Moszynski03, Moszynski05}. Note that simply cooling down existing doped crystals does not give the same benefit since their light yields go down when cooled~\cite{Sciver56, Sibczynski:2012hr, Sailer:2012ua, swiderski18}.

Scintillation centers in undoped NaI/CsI are understood to be self-trapped excitons instead of those trapped by doped impurities~\cite{pelant12}. Two types of excitons were observed in undoped CsI as shown in the right of Fig.~\ref{f:ly}~\cite{Nishimura95}. The energy dispersion among phonons and the two types of excitons dictates the temperature dependence of the light yield~\cite{Nishimura95, Mikhailik:2014wfa}.  Due to this completely different scintillation mechanism, scintillation wavelengths and decay times of undoped NaI/CsI are quite different from those of NaI/CsI[Tl], as summarized in Tab.~\ref{t:wd}. As the scintillation wavelengths of undoped NaI/CsI at 77~K are slightly below 420~nm, where detection efficiencies of most light sensors peak, a coating of wavelength-shifting material on crystals or light sensors can be used to increase the light detection efficiency.

\begin{table}[htbp] \centering\footnotesize
  \caption{Scintillation wavelength $\lambda$ and decay time $\tau$ of various crystals.}
  \label{t:wd}
  \begin{tabular}{ccccc}\hline
    Crystal & $\tau$ at RT [ns] & $\tau$ at 77~K [ns] & $\lambda$ at RT [nm] & $\lambda$ at 77~K [nm]\\\hline
    NaI[Tl] & $230\sim250$~\cite{robertson61, eby54, schweitzer83} & 736~\cite{Sibczynski:2012hr} & $420\sim430$~\cite{Sciver56,Sibczynski:2012hr} & $420\sim430$~\cite{Sciver56,Sibczynski:2012hr} \\
    CsI[Tl] & 600~\cite{Bonanomi52} & no data & 550~\cite{Kubota88} & no data \\
    undoped NaI & $10\sim15$~\cite{Sciver56,Sciver58,Beghian58} & 30~\cite{Sciver58, Beghian58}& 375~\cite{West70,Fontana70}& 303~\cite{Sciver56,Sibczynski:2012hr}\\
    undoped CsI & $6\sim36$~\cite{Kubota88,Schotanus:1990ha,Amsler:2002sz} & 1000~\cite{Nishimura95,Amsler:2002sz,Liu:2016iqv} & $305\sim310$~\cite{Kubota88,Woody:1990hq,Amsler:2002sz} & 340~\cite{Nishimura95,Woody:1990hq,Amsler:2002sz}\\\hline
  \end{tabular}
\end{table}

It is expected~\cite{birks1964} and observed~\cite{trefilova02, Park:2002jr} that dopant concentration affects light yield in inorganic scintillators. Microscopically, dopant is not uniformly distributed within a crystal. Charge carriers around short tracks of nuclear recoils have significantly less chance to meet doped trapping centers to de-excite through scintillation, resulting in quenching of scintillation of nuclear recoils compared to electronic recoils.

However, this is not the case in undoped crystals, where holes can be self-trapped anywhere, resulting in scintillation emission~\cite{Nishimura95}. Track lengths may not have the same effect on determining the scintillation efficiency as in doped crystals. Indeed, several early measurements at 77~K with $\alpha$-particles reported 85\% to 100\% quenching factor in undoped CsI~\cite{Hahn53, Hahn53a, Sciver56}. It was observed in a recent study~\cite{Clark:2017zaq} that the quenching factor of $\alpha$-particles changes with the crystal temperature. Around 77~K, the quenching factor is even larger than one. However, reported in another measurement at 108~K~\cite{Lewis:2021cjv} is a quenching factor very similar to that of CsI[Na] at room temperature. Possible causes of the discrepancy include different recoiling nuclei (Cs or I versus $\alpha$), measurement temperatures or origins of crystals. A series of measurements at the Triangle Universities Nuclear Laboratory (TUNL) will be performed by COHERENT collaborators to verify the quenching factor of CsI down to 40 K. To be conservative, a quenching factor of 5\% is assumed in sensitivity calculations presented in Sec.~\ref{s:pr}.

\subsubsection{SiPM arrays and readout electronics}
PMTs are great light sensors.  However, charged particles from natural radiation and cosmic rays can generate Cherenkov radiation when they pass through a PMT quartz (or fused silica) window. Given enough energy, a Cherenkov event can be easily distinguished from a scintillation event, since the former happens in a shorter time window, the current pulse of which is sharper than that of a scintillation event. However, close to the threshold, there are only a few detectable photons, which create a few single-PE pulses virtually identical in shape. The efficiency of pulse-shape discrimination becomes lower and lower as the energy goes down. Note that if two PMTs  are used on the two end surfaces of a cylindrical crystal, a requirement on coincident light detection in both of them will not help to remove Cherenkov events since the Cherenkov light created in one PMT can easily propagate to the other.
      
Two alternatives that do not generate Cherenkov radiation are avalanche photodiodes (APDs) and SiPMs, as they are made from thin silicon wafers, and no thick quartz window is attached to them. APDs have very high photon detection efficiencies (PDE, $\sim$80\%)~\cite{Baxter:2019mcx}. However, their gains are much less than those of PMTs and SiPMs, and they cannot be triggered at single-PE level. On the other   hand, a SiPM, which is basically an array of small APDs (micro cells) working in Geiger mode, is sensitive down to a single PE in each of its micro cells. The size of its micro cells has to be sufficiently small to avoid pileup. Spaces in between micro cells are not sensitive to photons. The peak PDE of a SiPM (up to 56\% at this moment) is hence smaller than that of an APD, but is typically higher than the peak quantum efficiency of a PMT~\cite{jac14}.
    
Since covering a large area with a monolithic SiPM die is not possible mainly due to the production yield, a compromise solution is to tile several dies tightly together to form an array. Given the same active area, a SiPM array uses less material, occupies less space, and can be made more radio-pure than a PMT. Table~\ref{t:sipm} lists a few SiPM arrays that are already available in the market. All have a PDE that is high enough for the proposed research. Their gains are also very close to that of a typical PMT, which makes the signal readout much easier than that for an APD. More importantly, most of them have been tested in liquid argon (LAr) or LN$_2$ temperature (for example, Ref.~\cite{Lightfoot:2008im, Lightfoot:2008ig, Rossi:2016jjt, Catalanotti:2015mna, Johnson:2018gyo} for SensL, Ref.~\cite{Otono:2006zz, akiba09, Igarashi:2015cma} for Hamamatsu, and Ref.~\cite{Csathy:2016wdy} for KETEK SiPMs).  FBK SiPMs  seem to be the only ones proven working down to 40~K with a good performance~\cite{DarkSide:2017zys,Acerbi:2016ikf,giovanetti17}. It is hence a major task of the detector R\&D to verify the performance of SiPM arrays from more manufacturers from 77~K down to around 40~K.

\begin{table}[htbp]\centering
  \caption{SiPM arrays available in the market possibly suitable for the proposed detector.}\label{t:sipm}
  \begin{tabular}{cccccc}\hline
    Company & SiPM & microcell size & PDE$^\dagger$ & Largest array size & Gain$^\diamond$ \\\hline
    SensL & J-series & 35~$\mu$m & 50\% & $50.4\times50.4$~mm$^2$ & $6.3\times10^6$\\
    SensL & C-series & 35~$\mu$m & 40\% & $57.4\times57.4$~mm$^2$ & $5.6\times10^6$\\
    Hamamatsu & S141xx &  50~$\mu$m & 50\% & $25.8\times25.8$~mm$^2$ & $4.7\times10^6$\\
    Hamamatsu & S133xx &  50~$\mu$m & 40\% & $25.0\times25.0$~mm$^2$ & $2.8\times10^6$\\
    KETEK & PM3325 & 25~$\mu$m & 43\% & $26.8\times26.8$~mm$^2$ & $1.7\times10^6$ \\\hline
  \end{tabular}
  \\\vspace{-1em}\flushleft\hspace{1.5cm}$^\dagger$ @ $420\sim 450$~nm. \hspace{1cm} $\diamond$ @ 5 volt over-voltage.
\end{table}

One major drawback of a SiPM array compared to a PMT is its high dark count rate at room temperature ($\sim$ hundred kHz). Fortunately, this rate drops quickly with temperature, and can be as low as 0.2~Hz/mm$^2$ below 77~K~\cite{DarkSide:2017zys}, while the PDE does not change much with temperature~\cite{Otono:2006zz, Lightfoot:2008im, akiba09, Janicsko-Csathy:2010uif}. However, a SiPM array that has an active area similar to a 3-inch PMT would still have about 100~Hz dark count rate at 77~K. A simple toy MC reveals that a 10-ns coincident window between two such arrays coupled to the same crystal results in a trigger rate of about $10^{-5}$~Hz. A further time coincidence with the SNS beam pulses would make the rate negligible.  Other potential problems of a SiPM include afterpulses~\cite{sipm} and optical crosstalk~\cite{Otono:2006zz, Lightfoot:2008im, akiba09, Janicsko-Csathy:2010uif}. Mitigation methods have been discussed in detail in a recent publications~\cite{Ding:2020uxu}.

Typically, $10\sim100\times$ amplification is still needed for a digitizer to record SiPM pulses down to the single PE level. Both nEXO~\cite{nEXO:2018ylp} and DarkSide~\cite{DIncecco:2017qta} chose cryogenic front-end electronics to amplify signals and reduce the amount of readout channels. Compared to these detectors, the proposed one is much smaller in its outer surface. Grouping channels is hence not a must, which makes room-temperature front-ends a viable option. This option dramatically reduces the complexity of the system, the amount of materials to be cooled, and the difficulty of maintenance.

\begin{figure}[htbp]\centering
  \includegraphics[width=0.48\linewidth]{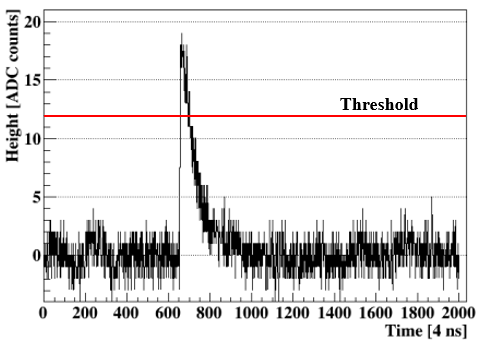}
  \includegraphics[width=0.51\linewidth]{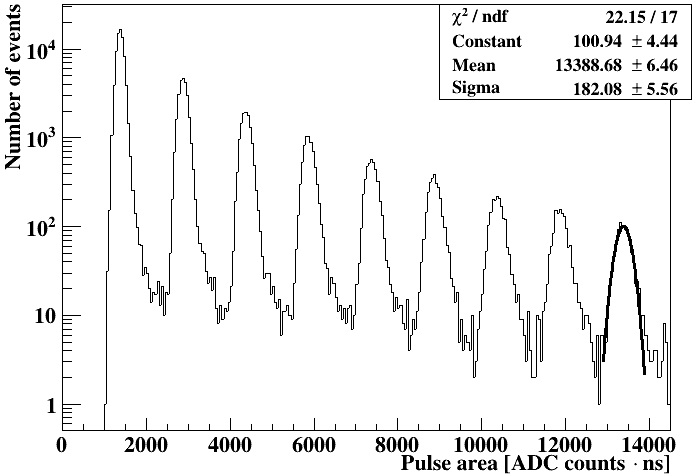}
  \caption{Typical single-PE pulse (left) and PE distribution (right) recorded by a cold SiPM and a warm amplifier with a long cable in between. Figure from Ref.~\cite{Ding:2022jjm}.} \label{f:1pe}
\end{figure}

CAEN has recently developed a CMOS-based ASIC front-end system for large detector arrays~\cite{a5202}. It features a standalone unit, A5202, that contains two WEEROC CITIROC chips, each providing a multiplexed output of 32 SiPM channels~\cite{citiroc1a}. It also features a flexible micro coaxial  cable bundle, A5260B, connecting a remote SiPM array with the A5202. Excellent single PE resolutions can be achieved with a cable length up to 3 m. Given the relatively small size of the proposed detector, that length is sufficient to bridge cold SiPM arrays and warm ASIC front-ends. The performance of such a setup has been confirmed by CAEN and COHERENT collaborators (see Fig.~\ref{f:1pe}) independently.

\subsubsection{Cryostat}
A pulse-tube-refrigerator-based cryostat is foreseen to be used for COH-CryoCsI-1 ($\sim10$ kg). Compared to a liquid-nitrogen-based one, it requires much less maintenance, and provides the possibility to run the detector at various temperatures. Less maintenance is not only for convenience, but also an essential feature of a neutrino detector that can run for a long time without periodic change of its temperature. Since many properties of the detector change with temperature. It is important to have the possibility to optimize its operating temperature. Flexible micro coaxial cables that contact cold SiPM arrays with warm front-end electronics will be thermalized through the cold head of the refrigerator. COH-CryoCsI-2 ($\sim700$ kg) will have a modularized design. A module is similar to COH-CryoCsI-1 but with a larger target mass. ASIC-based front-end + digital electronics at cryogenic temperature may be considered to reduce the amount of readout channels and feedthroughs.

\subsubsection{Current status}
\begin{figure}[htbp]\centering
    \includegraphics[width=0.49\linewidth]{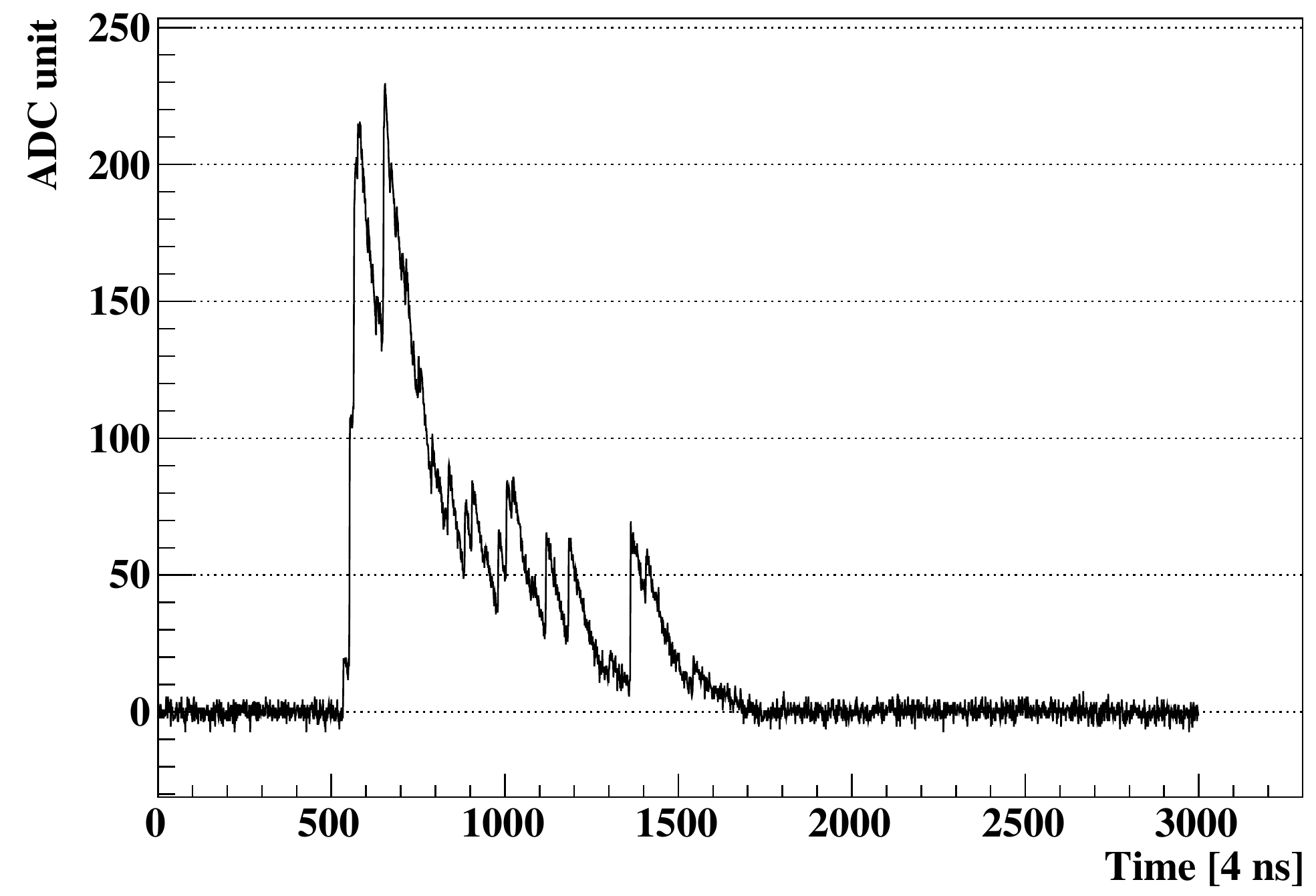}
    \includegraphics[width=0.49\linewidth]{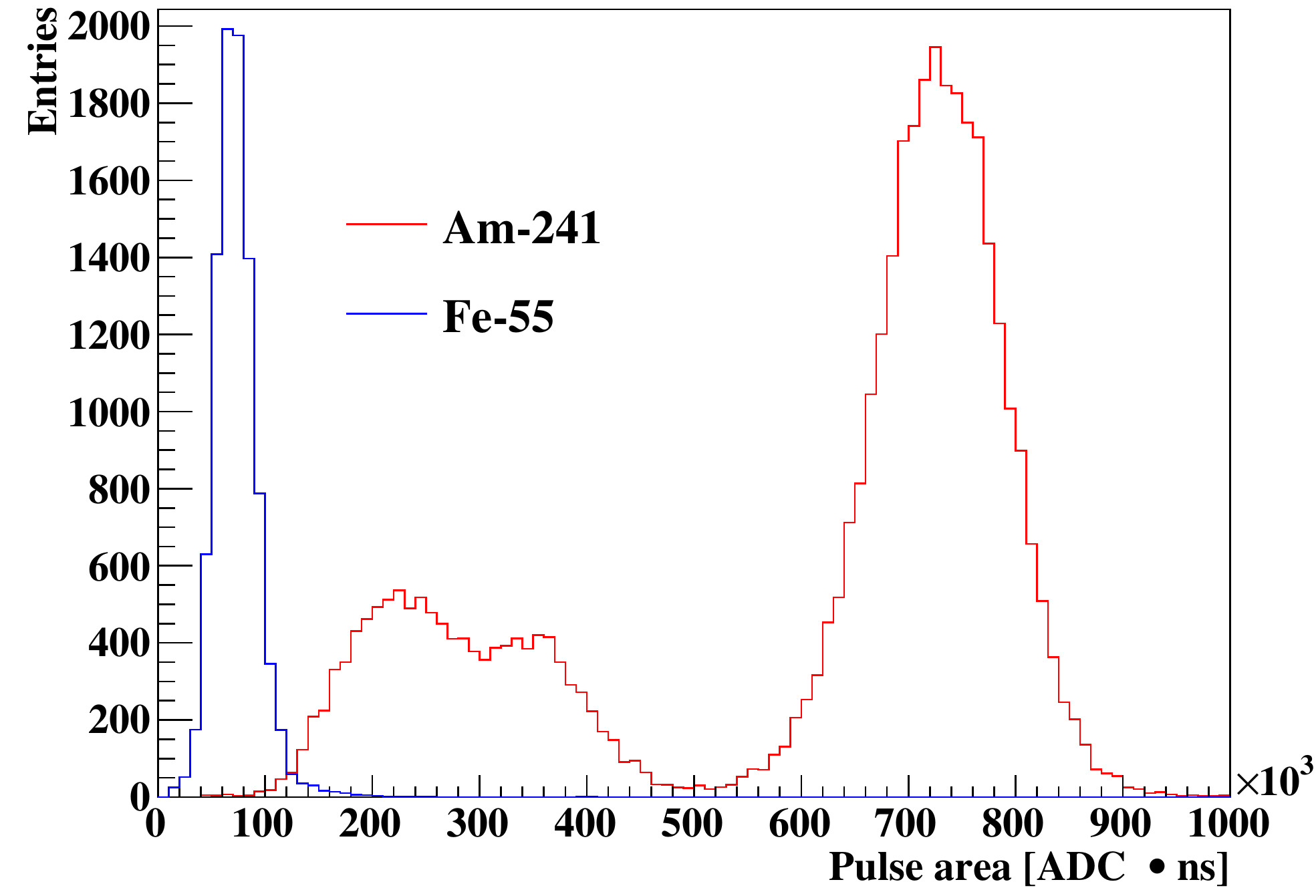}
    \caption{Left: a random Fe-55 pulse in a CsI collected by a SiPM at 77~K. Right: Energy spectra of Fe-55 and Am-241 recorded by a CsI + SiPMs at 77~K.}
    \label{f:csp}
\end{figure}

The light yield of undoped CsI has been steadily improved by COHERENT collaborators from 20 to 50 PE/keVee over the past few years.  By coupling a small undoped CsI crystal to an R8778MODAY(AR) PMT from Hamamatsu at 77~K, a yield of $\sim20$ PE/keVee was achieved in 2016~\cite{Liu:2016iqv}. To prove that this idea was applicable to larger crystals as well, two R11065 PMTs were used to detect light from a crystal of a diameter of 3~in and a height of 5~cm, and a yield of $\sim$26 PE/keVee~\cite{Chernyak:2020lhu} was achieved. Both measurements were done using $\gamma$-ray lines from 662~keV to 2.6~MeV. The light yield in a lower energy range was measured using an Am-241 source inside the cryostat. An average yield of $\sim$33~PE/keVee from 13 to 60~keVee was observed in this measurement~\cite{Ding:2020uxu}. Recently, two SensL SiPMs were used to read out a small cubic CsI crystal. SiPM signals were amplified outside of the cryostat. Nonetheless, the signal-to-noise ratio was great and individual PEs could be clearly distinguished (see Fig.~\ref{f:1pe}). The light yield observed in this setup was $38.9\pm0.7$~PE/keVee~\cite{Ding:2022jjm}. Figure~\ref{f:csp} shows a typical Fe-55 pulse (left) and the energy spectra (right) of Fe-55 and Am-241 recorded by one SiPM in such a setup. Many PEs can be clearly seen at an energy as low as 5.9 keV. The SiPMs were then coated with 1,1,4,4-Tetraphenyl-1,3-butadiene (TPB) to shift the 340~nm scintillation light from CsI to around 420~nm, where a typical SiPM's PDE reaches its maximum. The light yield was boosted to $50\pm1$~PE/keVee. The results of all measurements are summarized in Tab.~\ref{t:ly}. Even higher light yields may be achievable when the detector is operated at 40~K with SiPMs that have higher PDE.

\begin{table}[htbp]\centering
  \caption{Achieved~\cite{Liu:2016iqv, Chernyak:2020lhu, Ding:2020uxu} and predicted light yields of undoped CsI compared to that of the COHERENT CsI[Na] detector~\cite{COHERENT:2017ipa}.}\label{t:ly}
  \begin{tabular}{rcl} \hline
    Experiments &Type of crystals &Light yield [PE/keVee]\\\hline
    COHERENT 2017 & CsI[Na] & 13.5 $\pm$ 0.1~\cite{COHERENT:2017ipa}\\
    PMT+small crystal & undoped CsI & 20.4 $\pm$ 0.8~\cite{Liu:2016iqv}\\
    PMTs+large crystal & undoped CsI & 26.0 $\pm$ 0.4~\cite{Chernyak:2020lhu}\\
    Improved light collection & undoped CsI & 33.5 $\pm$ 0.7~\cite{Ding:2020uxu}\\
    PMT $\rightarrow$ SiPMs & undoped CsI & 38.9 $\pm$ 0.7~\cite{Ding:2022jjm}\\
    WLS coating on SiPMs & undoped CsI & 50.0 $\pm$ 1.0\\
    77 $\rightarrow$ 40 K, \& SiPMs with 50\% PDE & undoped CsI & 60 (projected)\\ \hline
  \end{tabular}
\end{table}

\subsection{Liquid argon Time Projection Chamber}

\subsubsection{Motivation}
\label{sec:LArTPC_Motivation}

Core-collapse supernova explosions and the everyday Sun nuclear cycle
both produce copious quantities of neutrinos in the few to tens of MeV range. The 
the detection of these neutrinos will provide unique information about internal  processes of these astrophysical objects~\cite{https://doi.org/10.48550/arxiv.2002.03005,Abi:2021vq,PhysRevLett.123.131803}.
This information, in turn, can improve our understanding of astrophysics
and constrain physics beyond the Standard Model~\cite{Thompson_2003,RAFFELT19901,DeRocco:2019aa,PhysRevD.99.121305,PhysRevD.100.075018}.
Argon has a relatively low threshold for charged-current (CC) electron
neutrino ($\nu_e$) interactions, which makes liquid-argon detectors uniquely
sensitive to MeV-scale $\nu_e$ interactions, including neutrinos produced
in supernova explosions and in the Sun~\cite{PhysRevLett.123.131803}.
Further,
liquid-argon time-projection chambers (LArTPCs) uniquely allow reconstruction
of both directionality and calorimetry for MeV-scale particles,
based on their millimeter resolution and potential on calorimetric capabilities~\cite{Bueno:kq}.
The Deep Underground Neutrino Experiment (DUNE), with its 40-kilotonne
underground LArTPC far detector,
will therefore offer unique opportunities for physics in the MeV regime,
including complementary measurements of supernova neutrinos to other underground,
massive neutrino experiments~\cite{https://doi.org/10.48550/arxiv.2002.03005,Abi:2021vq,PhysRevLett.123.131803}.

As the detected neutrino fluxes are effectively convolved with neutrino cross
sections and detector resolution,
it is crucial to disentangle each contribution.
However, neutrino-argon interactions at this energy range have never been measured,
and the systematic uncertainty originating from theoretical models dominates the
precision of supernova neutrino measurements~\cite{https://doi.org/10.48550/arxiv.2002.03005,Abi:2021vq}.
It is important to measure neutrino-argon cross sections,
thereby improving supernova and solar neutrino studies.
Moreover, cross sections as function of energy and direction of the outgoing charged
lepton (i.e., double-differential cross sections) are required for
further investigation.  For example, they will be needed to unfold the energy spectrum of supernova neutrinos, needed for understanding of
the explosion mechanism.

The SNS produces neutrinos from pions decaying at rest,
and has a highly beneficial properties in terms of energy, power and background rejection,
for neutrino cross-section measurements in the MeV to tens of MeV regime.
It is of great interest to conduct such measurements with a LArTPC detector,
which can directly translate the results for supernova and solar neutrino
studies in DUNE.
In addition, double-differential cross section measurements require a detector
with a scale of ten tonnes,
and including such a detector in the proposed Second Target Station of SNS
will allow the measurements and benefit the DUNE physics program.

\subsubsection{Detector description}
\label{sec:LArTPC_Detector}

In a time-projection chamber (TPC), an electric field of a few hundred volts
per centimeter is applied.
The charged particles produced in a neutrino-argon interaction ionize the argon atoms,
and the ionization electrons drift to the anode along the electric field with
a time scale of milliseconds.
The drifting electrons are eventually detected by the charge collection system,
where the millimeter-scale pitches determine the spatial resolution.
In parallel, the scintillation light produced from the de-excitation of argon
is collected within 10-100 nanoseconds by the light collection system,
determining the time the event occurs ($t_0$).
As the electrons drift with a constant velocity,
$t_0$ determines the position of the event along the drift direction.

LArTPCs located at the surface,
such as the experiments in the Short Baseline Neutrino (SBN) Program,
suffer from a high rate of cosmic rays traversing
the detector volumes within the millisecond readout time scale.
Hence, TPC readouts are usually accompanied by scintillator panels and preferably have some overburden in order to
to, respectively, identify cosmic muons and mitigate secondary particles
induced by cosmic rays~\cite{MicroBooNE:2019lta,Machado:2019oxb}.

The 8~m.w.e overburden of Neutrino Alley and background studies using COH-Ar-10 will benefit the background characterization for a new LArTPC detector deployment.

\subsubsection{Requirements for MeV physics}
\label{sec:LArTPC_Req}

Non-coherent interactions of MeV-scale $\nu_e$s typically leave a short
electron track surrounded by a number of blips of charge,
e.g. from bremsstrahlung photons later interacting by Compton scattering.
MeV-scale photons are emitted from inelastic scattering of neutrons
some distance away,
as well as de-excitation of argon nuclei, and can also be revealed as blips
of charge via Compton scattering~\cite{PhysRevD.99.012002,1748-0221-12-09-P09014}.
A modular LArTPC design with pixelated charge collection system has an advantage
for tackling the challenges of reconstructing short tracks and blips of charge
in a high-multiplicity environment with sizable cosmic ray and beam-induced
background~\cite{instruments5040031}.
The pixelated charge collection system natively provides a 3D event topology,
while a module confines the scintillation light signals from an interaction
and separates the interaction from the others~\cite{Dwyer_2018}.
However, the opaque pixelated charge collection system makes it impossible
to keep the conventional TPC design,
in which the light detectors are mounted behind the wire-based charge readout planes.
A few R\&D projects are currently progressing, pursuing several percent-level
light detection efficiency and nanosecond-scale time resolution~\cite{instruments5010004,instruments2010003,Anfimov_2020}.

A test stand LArTPC in Neutrino Alley, utilizing the facility and the studies of the SNS beam and the background will be an important first step to evaluate the opportunities and sensitivities of MeV physics with SNS.

\section{Physics reach of the COHERENT Program}
\label{s:pr}

Precision measurements of the CEvNS cross section and recoil spectrum using a variety of nuclei will evolve our understanding of the frontier of particle physics.  Initial data from small, first-light COHERENT detectors have already imposed world-leading constraints on BSM physics~\cite{Coloma:2017ncl,Liao:2017uzy,Miranda:2020tif,Papoulias:2017qdn,Dutta:2020vop} as well as improve standard-model nuclear and particle physics measurements~\cite{Cadeddu:2018rlm,Cadeddu:2018dux,Cadeddu:2021ijh,Cadeddu:2018izq}.  The next generation of CEvNS detectors will increase the signal event rate by over an order of magnitude, increasing potential to discover new physics.   Some selected physics topics are summarized below.

\subsection{Testing non-standard interactions with CEvNS}
\label{s:nsi}

Non-standard interactions (NSI) between neutrinos and quarks would modify the CEvNS cross section.  Such interactions are described generally by a matrix of vector couplings, $\varepsilon_{\alpha\beta}^{q}$,~\cite{Barranco:2005yy} with a Lagrangian
\begin{equation}
  \mathcal{L}=\sum_{q,\alpha,\beta}2\sqrt{2}G_F\varepsilon_{\alpha\beta}^{q}
  \left(\bar{\nu}_\alpha\gamma^\mu(1-\gamma^5)\nu_\beta \right) \left(\bar{q}\gamma_\mu(1-\gamma^5)q \right),
\end{equation}
assuming the mediator of these new interactions is heavy, $>\sqrt{Q^2}=\sqrt{2m_NE_r}=50$~MeV.  The terms $\varepsilon_{ee}^{q}$, $\varepsilon_{\mu\mu}^{q}$, and $\varepsilon_{\tau\tau}^{q}$ interfere with standard-model CEvNS and break lepton universality predicted for CEvNS at tree level, while $\varepsilon_{e\mu}^{q}$, $\varepsilon_{e\tau}^{q}$, and $\varepsilon_{\mu\tau}^{q}$ allow for flavor-changing transitions.  Non-zero values of $\varepsilon_{\alpha\beta}^{q}$ would change the CEvNS cross section by modifying the weak charge,
\begin{equation}
  \begin{split}
    Q_{\alpha}^2=\left[Z\left(g_p^V+2\varepsilon_{\alpha\alpha}^u+\varepsilon_{\alpha\alpha}^d\right)+N\left(g_n^V+\varepsilon_{\alpha\alpha}^u+2\varepsilon_{\alpha\alpha}^d\right)\right]^2 + \\
    \sum_{\alpha\neq\beta}\left[Z\left(2\varepsilon_{\alpha\beta}^{u}+\varepsilon_{\alpha\beta}^d\right)+N\left(\varepsilon_{\alpha\beta}^{u}+2\varepsilon_{\alpha\beta}^d\right)\right]^2.
  \end{split}
  \label{eqn:WeakCharge}
\end{equation}
Development of precise CEvNS measurements by COHERENT offers a novel strategy for constraining these effective parameters, many of which were previously only constrained to $\sim$~unity.  Understanding these parameters is critical for proper interpretation of neutrino oscillation data since various choices of $\varepsilon_{\alpha\beta}^q$ can bias experimental determination of neutrino mixing parameters such as $\Delta m^2_{32}$\cite{Coloma:2019mbs}, $\delta_{CP}$\cite{Denton:2020uda}, and $\theta_{12}$\cite{Coloma:2017ncl}.  

\begin{figure}[!tb]\centering
  \includegraphics[width=0.51\linewidth]{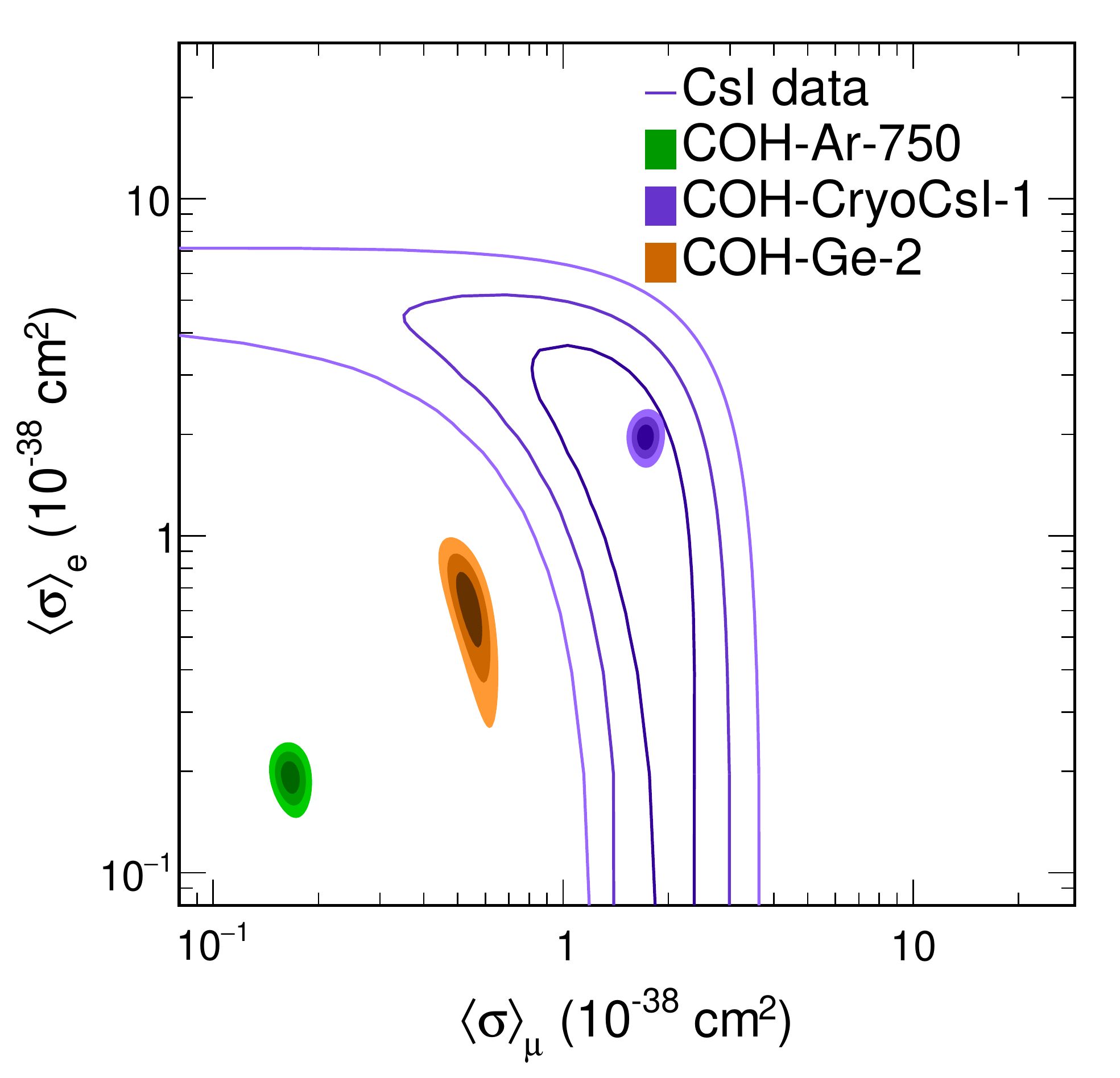}
  \caption{Sensitivity showing expected 1/2/3 $\sigma$ expected CEvNS cross section measurements for $\nu_\mu$ and $\nu_e$ separately using future detectors at the SNS assuming three years of exposure in each detector compared to current CsI[Na] data.}
  \label{f:FlavoredCEvNS}
\end{figure}

As NSI parameters will generally make the CEvNS cross section different for different flavors of neutrino, a source with multiple flavors is ideal.  The SNS offers a prompt $\nu_\mu$ flux separated in time from delayed $\bar{\nu}_\mu/\nu_e$ allowing us to test the flavored CEvNS cross sections, $\langle\sigma\rangle_\mu$ and $\langle\sigma\rangle_e$ as described in~\cite{Akimov:2021dab}.  A deviation of $\langle\sigma\rangle_\mu/\langle\sigma\rangle_e$ from the standard model expectation would be a sensitive probe of BSM physics, including NSI, as most systematic uncertainties in the ratio correlate.  The sensitivity of future COHERENT detectors to measuring these quantities is shown in Fig.~\ref{f:FlavoredCEvNS} compared to the measurement based on the full CsI[Na] dataset~\cite{Akimov:2021dab}.  Large CEvNS detectors will significantly improve on our current understanding of the flavored cross sections which will subsequently reduce viable NSI parameter space.

According to in Eq.~\ref{eqn:WeakCharge}, there are values of $\varepsilon_{\alpha\beta}^q$ that will keep $Q_\alpha^2$ equal to the standard-model expectation even if there are non-zero NSI parameters.  Fortunately, these values depend on $N/Z$, so a joint fit of multiple CEvNS datasets will break these degeneracies.  We focus on sensitivity of the future COH-Ar-750 and COH-CryoCsI-1 detectors to study how CEvNS data may limit NSI parameter space.  Data from COH-Ge-2 are not considered as its $N/Z$ is intermediate offering comparatively little leverage for relieving these degeneracies and PPC detectors have poor timing resolution leading to poorer $\nu_\mu/\nu_e$ separation as illustrated by Fig.~\ref{f:FlavoredCEvNS}.  

\begin{figure}[htbp]\centering
  \includegraphics[width=0.4\linewidth]{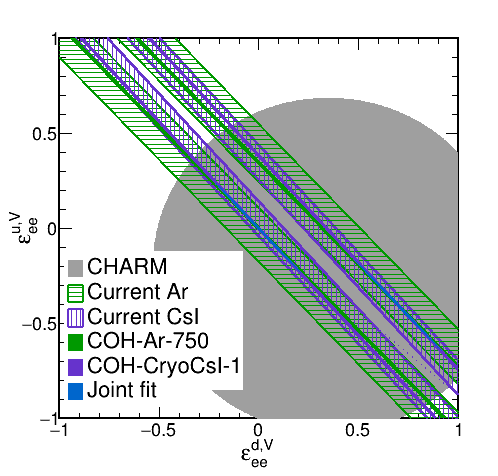}
  \includegraphics[width=0.4\linewidth]{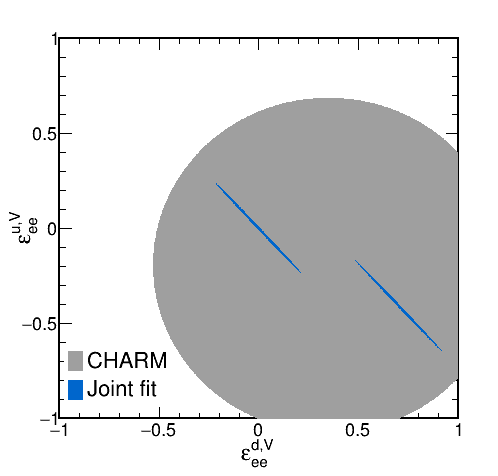}
  \includegraphics[width=0.49\linewidth]{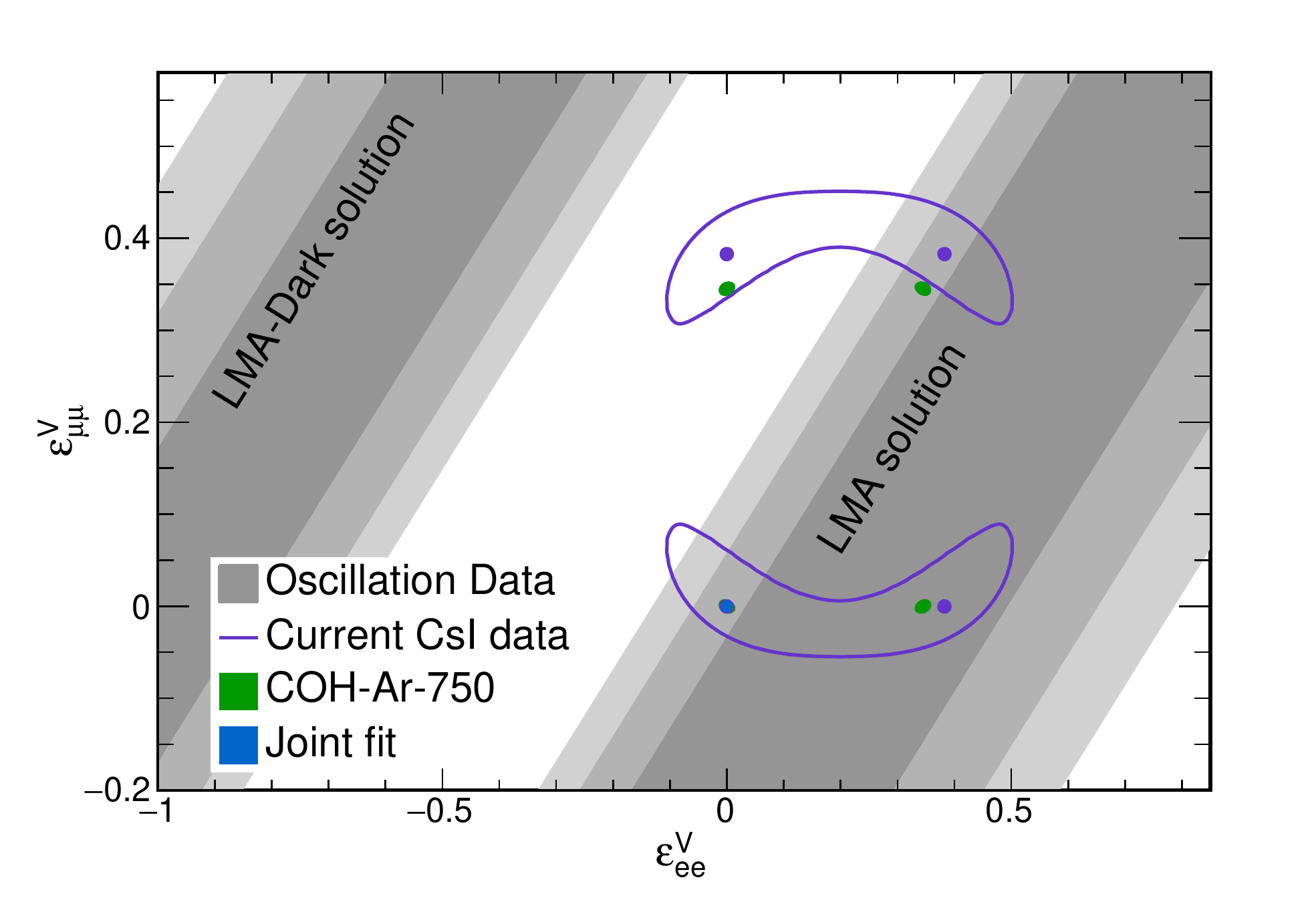}
  \includegraphics[width=0.49\linewidth]{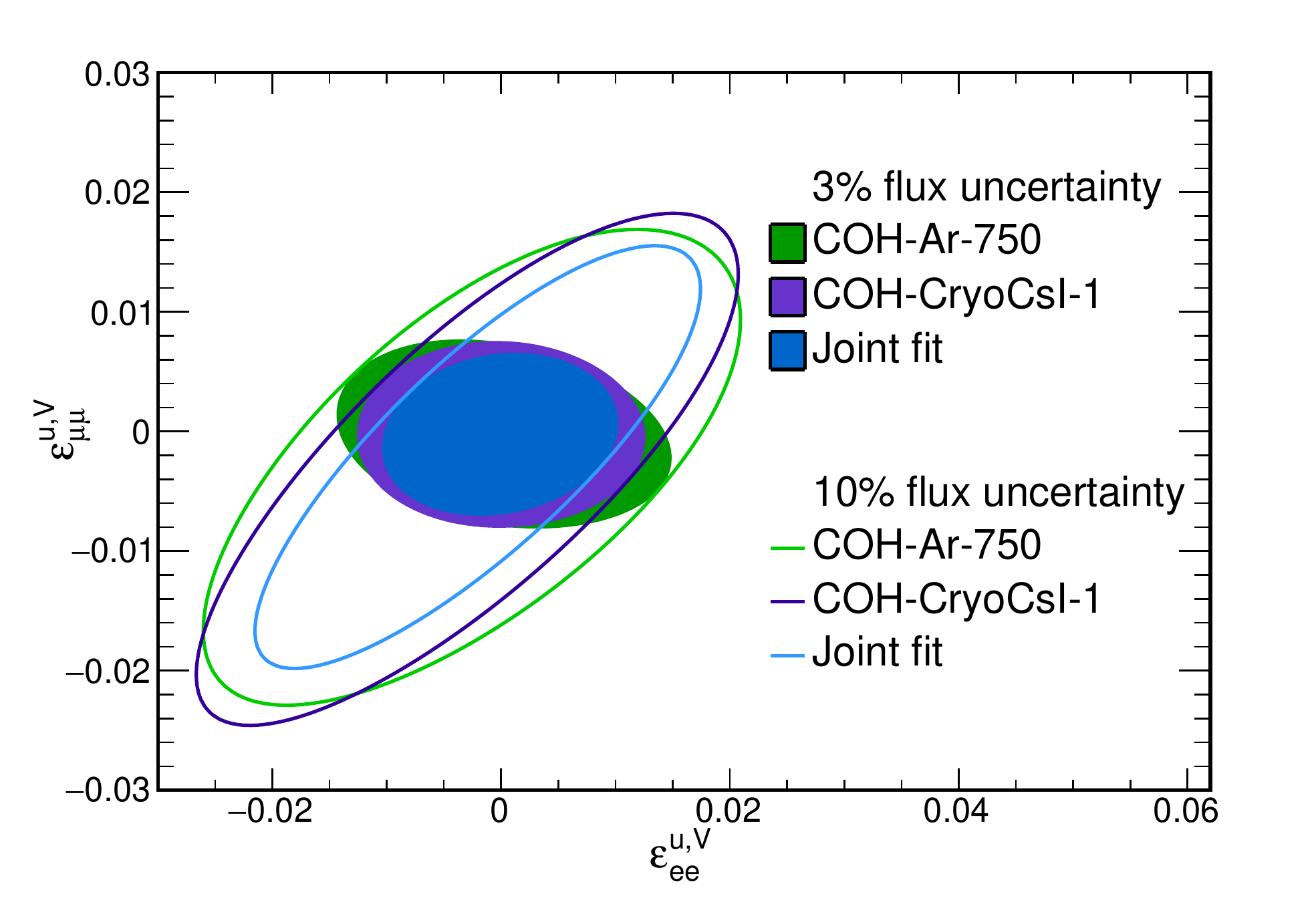}
  \caption{COHERENT future sensitivity to NSI scenarios after three years. We show expected constraints on $\varepsilon_{ee}^{u,V}$ and $\varepsilon_{ee}^{d,V}$ compared to current constraints (top) with just the projected joint fit plotted (top right) for clarity.  We also consider constraints on $\varepsilon_{ee}^{u,V}$ and $\varepsilon_{\mu\mu}^{u,V}$ compared to parameter space favored by the LMA and LMA-Dark solar oscillation parameters (bottom) with a comparison of constraints assuming both a 3$\%$ and 10$\%$ uncertainty on the neutrino flux (bottom right).}
  \label{f:NSIsens}
\end{figure}

In general, the flavor-changing couplings are better constrained and can be directly measured with oscillation measurements.  We thus fix these parameters to 0 and determine sensitivity to the diagonal couplings accessible at the SNS, $\varepsilon_{ee}^q$ and $\varepsilon_{\mu\mu}^q$.  Our sensitivity to $\varepsilon_{ee}^u$ and $\varepsilon_{\mu\mu}^d$, assuming all other couplings are 0, is shown in Fig.~\ref{f:NSIsens} along with current constraints from CsI[Na]~\cite{Akimov:2021dab} and COH-Ar-10~\cite{COHERENT:2020iec}.  Each individual constraint from COH-Ar-750 and COH-CryoCsI-1 gives an infinite linear degeneracy in the parameter space with a slope $\varepsilon_{ee}^u/\varepsilon_{ee}^d=-(Z+2N)/(2Z+N)$.  However, a future joint fit will reduce the allowed parameter space to two closed ellipses.  

We also consider variations in $\varepsilon_{ee}^u$ and $\varepsilon_{\mu\mu}^u$ in Fig.~\ref{f:NSIsens}, again assuming other parameters are 0.  With a rate-only analysis, degeneracies in these parameters manifest as annuli in the parameter space.  However, the separation of the $\langle\sigma\rangle_\mu$ and $\langle\sigma\rangle_e$, afforded by timing at the SNS, reduces the parameter space to four ellipses for a single dataset.  Again, by combining CEvNS data from multiple nuclei, this is reduced to a single closed contour near the standard-model parameters.  We show sensitivity assuming both a 10$\%$ neutrino flux uncertainty, reflecting current understanding of neutrino production at the SNS, and a 3$\%$ uncertainty which can be achieved with a tonne-scale D$_2$O detector used to calibrate the neutrino flux.  The improved flux uncertainty would reduce the uncertainty on $\lvert\varepsilon_{ee}^u\rvert$ and $\lvert\varepsilon_{ee}^u\rvert$ from $<0.03$ to $<0.01$.  This would also constrain $\lvert\varepsilon_{\tau\tau}^u\rvert$ when combined with oscillation data sensitive to $\lvert\varepsilon_{\mu\mu}^u-\varepsilon_{\tau\tau}^u\rvert$~\cite{Coloma:2019mbs}.

\subsection{Weak Mixing Angle Measuremsnt}

\begin{figure}[!tb]\centering
\includegraphics[width=0.49\linewidth]{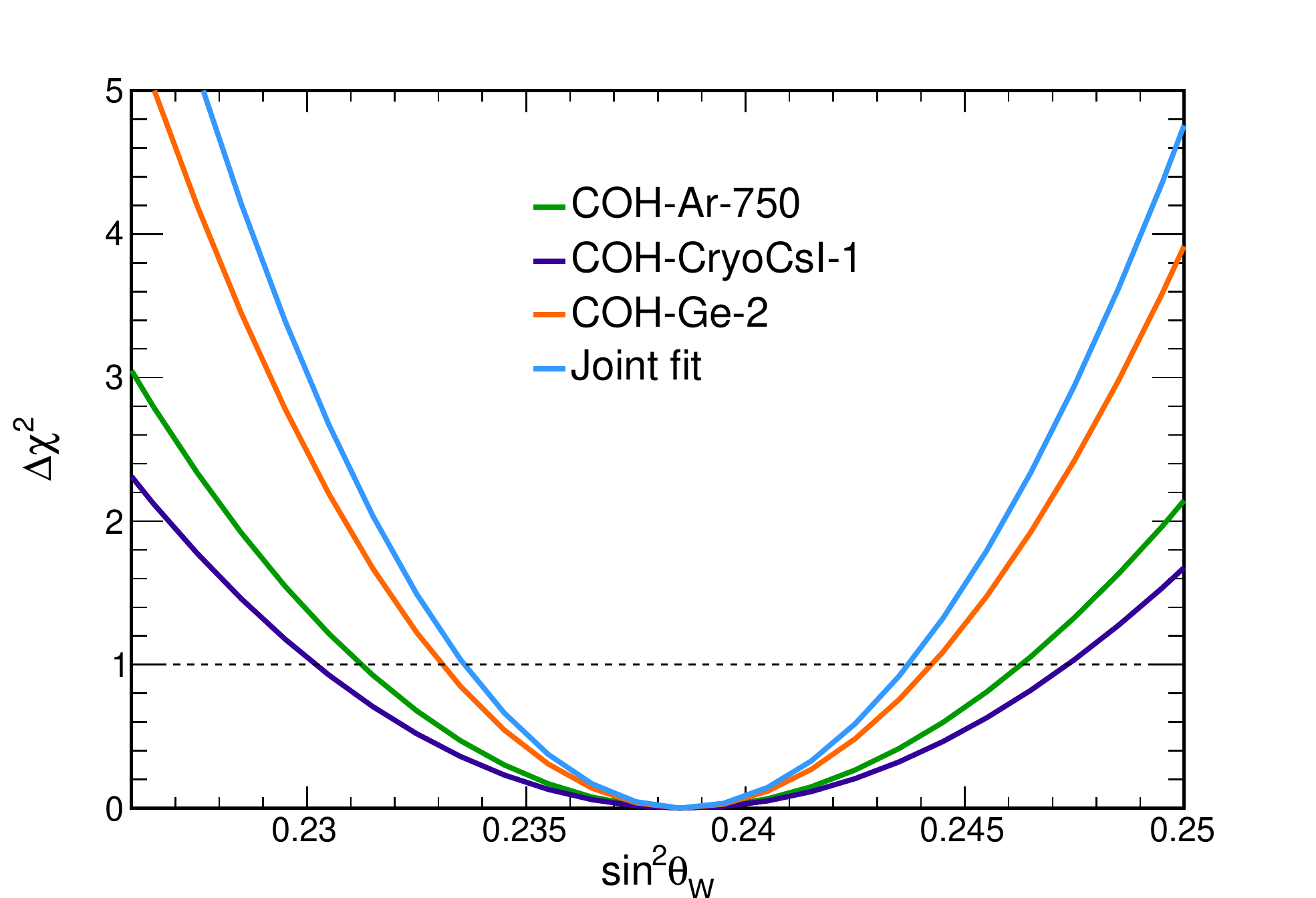}
\caption{COHERENT sensitivity to measuring the weak mixing angle using three years of CEvNS data from future Ar, CsI, and Ge detectors at 68$\%$ confidence.}
\label{f:EMsens}
\end{figure}

In addition to being a sensitive probe of BSM physics, CEvNS measurements will provide new understanding of electroweak parameters.  Among these is an opportunity to measure the weak mixing angle at $Q^2\sim50$~MeV~\cite{Canas:2018rng,Miranda:2019skf,Cadeddu:2021ijh}.  The CEvNS cross section relies on this parameter as $g_p^V=1/2-2\sin^2\theta_W$.  There are few measurements at low $Q^2$, so a test from COHERENT will test the standard model in an unexplored kinematic region.  Deviations of the standard-model expectation could identify new BSM physics in this $Q^2$ range.  The COHERENT sensitivity to the weak mixing angle is shown in Fig.~\ref{f:EMsens} with three years of data for COH-Ar-750, COH-CryoCsI-1, and COH-Ge-2.  A joint fit of all data will give a 2.1$\%$ measurement with the Ge detector dominating the sensitivity.  A precision test of the CEvNS cross-section ratio for $\nu_e$ to $\nu_\mu$ can also test for flavor-dependent effects on the cross section at loop level such as the neutrino ``charge radius'' at the sub-percent level.

\subsection{Accelerator-produced dark matter with CEvNS detectors}
\label{s:dm}

Detectors capable of detecting CEvNS can also directly detect accelerator-produced dark matter.  Dark matter particles produced at the SNS would also coherently scatter with nuclei inducing low-energy nuclear recoils similar to CEvNS \cite{deNiverville:2015mwa,Dutta:2020vop}.  Initial COHERENT data using a 14.6~kg, CsI[Na] scintillator detector placed leading constraints on sub-GeV dark matter~\cite{COHERENT:2021pvd} which will be improved with the next generation of COHERENT detectors.  

We are sensitive to sub-GeV dark matter accessible at the SNS beam energy.  Such light dark matter could not interact directly via the weak force~\cite{Lee:1977ua}.  Instead, a hidden sector dark matter particle, $\chi$, would interact with SM particles through a new mediator particle, $V$.  We use two dark matter models to benchmark our sensitivity.  In the first, a vector mediator kinetically mixes with the SM photon \cite{Fayet:2004bw,Boehm:2003hm,Pospelov:2007mp} with a coupling, $\varepsilon^2$.  A second coupling, $\alpha_D$ describes $V\rightarrow\bar{\chi}\chi$ decay.  We relate limits on this model in terms of $Y=\varepsilon^2\alpha_D(m_\chi/m_V)^4$~\cite{Izaguirre:2015yja} which is directly related to the relic dark matter abundance from thermal freeze-out.  We only consider $\alpha_D=0.5$ near the perturbative limit as lower values give tighter constraints.  

Leptophobic dark matter is also considered, where the mediator instead interacts with quarks.  This model is again determined by two couplings: $\alpha_B$ describing $V\bar{q}q$ interactions and $\alpha_\chi$ describing $V\bar{\chi}\chi$ vertices.  This model facilitates effective kinetic mixing between $V$ and the SM photon through a virtual $\bar{q}q$ loop which relates the relic abundance with $\alpha_B$.  We again conservatively choose $\alpha_\chi=0.5$.  This model is currently better constrained than the kinetically mixed model; however, significant parameter space remains unexplored for $m_\chi/m_V\sim2$.  In such cases, annihilation of dark matter during freeze-out proceeded resonantly which reduces the expected couplings required to match the thermal abundance~\cite{Feng:2017drg}.  This effect is parameterized by $\varepsilon_R=m_V^2/4m_\chi^2-1$.  We consider values of $\varepsilon_R$ down to $10^{-5}$.  At this point, a floor is reached for scalar dark matter beyond which lower values of $\varepsilon_R$ does not affect expectations for the thermal target.

CEvNS detectors are advantageous for these searches.  The scattering is coherent, which enhances the cross section by $Z^2$.  Thus, with only 14.6~kg of CsI[Na], we determined constraints competitive with $\sim$~100~t detectors sensitive to $\chi-e$ scattering.  Further, the $\chi-e$ channel is not sensitive to the leptophobic dark-matter model.  These first-light detectors can be upgraded to the tonne scale in Neutrino Alley or the 10-tonne scale at a future site around the STS, significantly improving on current results.  Additionally, accelerator-based searches are much less dependent on dark matter spin.  The scattering cross section for spin-\textonehalf, cold dark matter in the galaxy is suppressed by powers of $v/c$ up to 20 orders of magnitude compared to expected scalar dark matter scenarios.  Dark matter produced at an accelerator, however, would be relativistic, eliminating this strong suppression and making searches at accelerators the only viable option.  Among accelerator-based methods, CEvNS experiments also have powerful techniques for constraining systematic uncertainties on SM backgrounds.  Since dark matter is produced relativistically, it would arrive in our detectors coincident with the prompt $\nu_\mu$ flux.  Delayed CEvNS from the $\nu_e+\bar{\nu}_\mu$ flux can then be used as a background control sample to understand uncertainties on the neutrino flux, detector response, and neutrino interactions.  As these constraints improve with larger data sets, dark matter searches with CEvNS experiments remain limited by statistical uncertainty even after 100~tonne-yrs of exposure.  Lastly, if dark matter is detected at SNS, its flux would be boosted forward compared to the isotropic flux of the neutrino background.  Thus, with multiple detectors, we can confirm the dark matter nature of a signal by testing the angular dependence of the observed excess with respect to the beam direction.

\begin{figure}[!tb]\centering
  \includegraphics[width=0.49\linewidth]{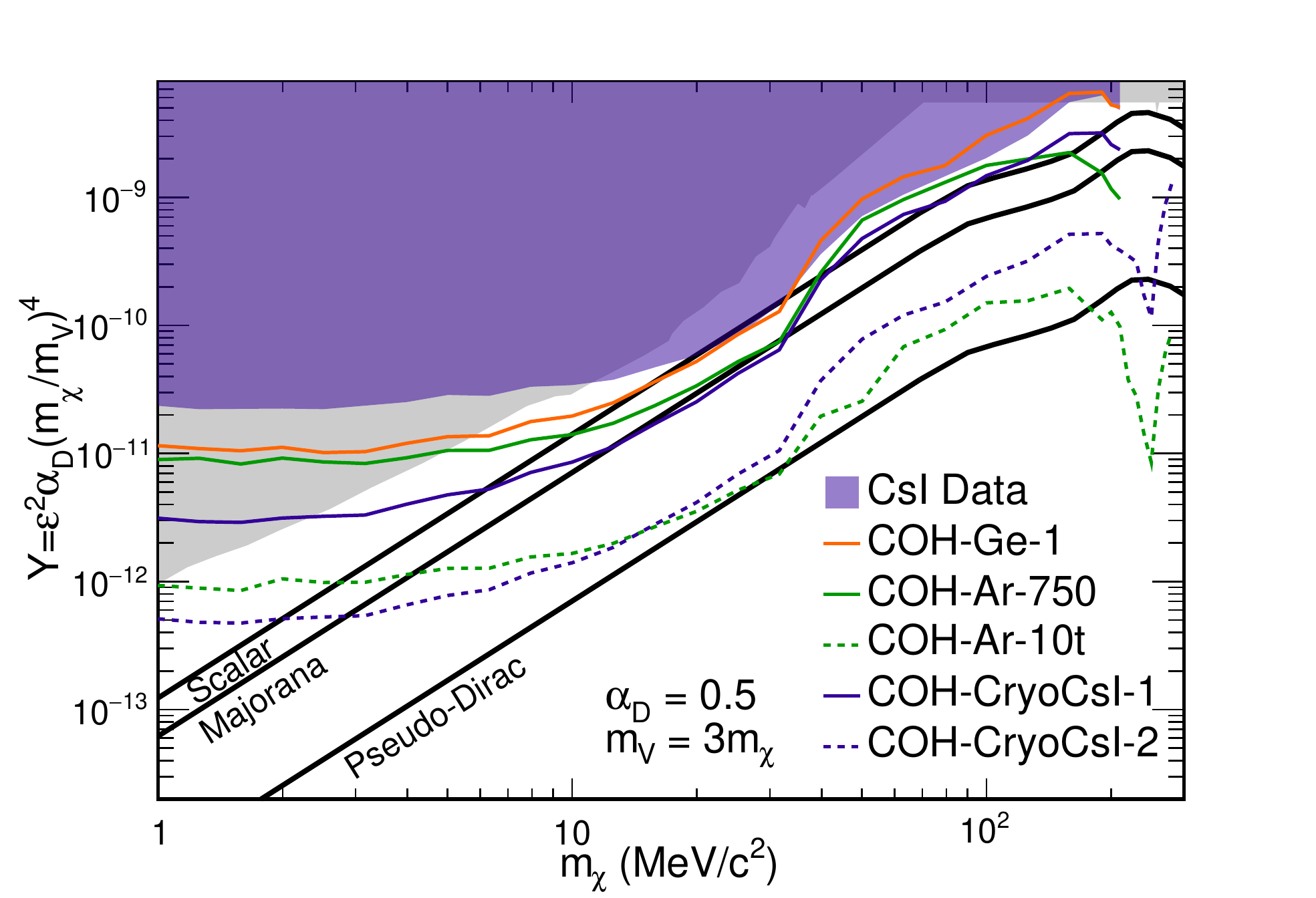}
  \includegraphics[width=0.49\linewidth]{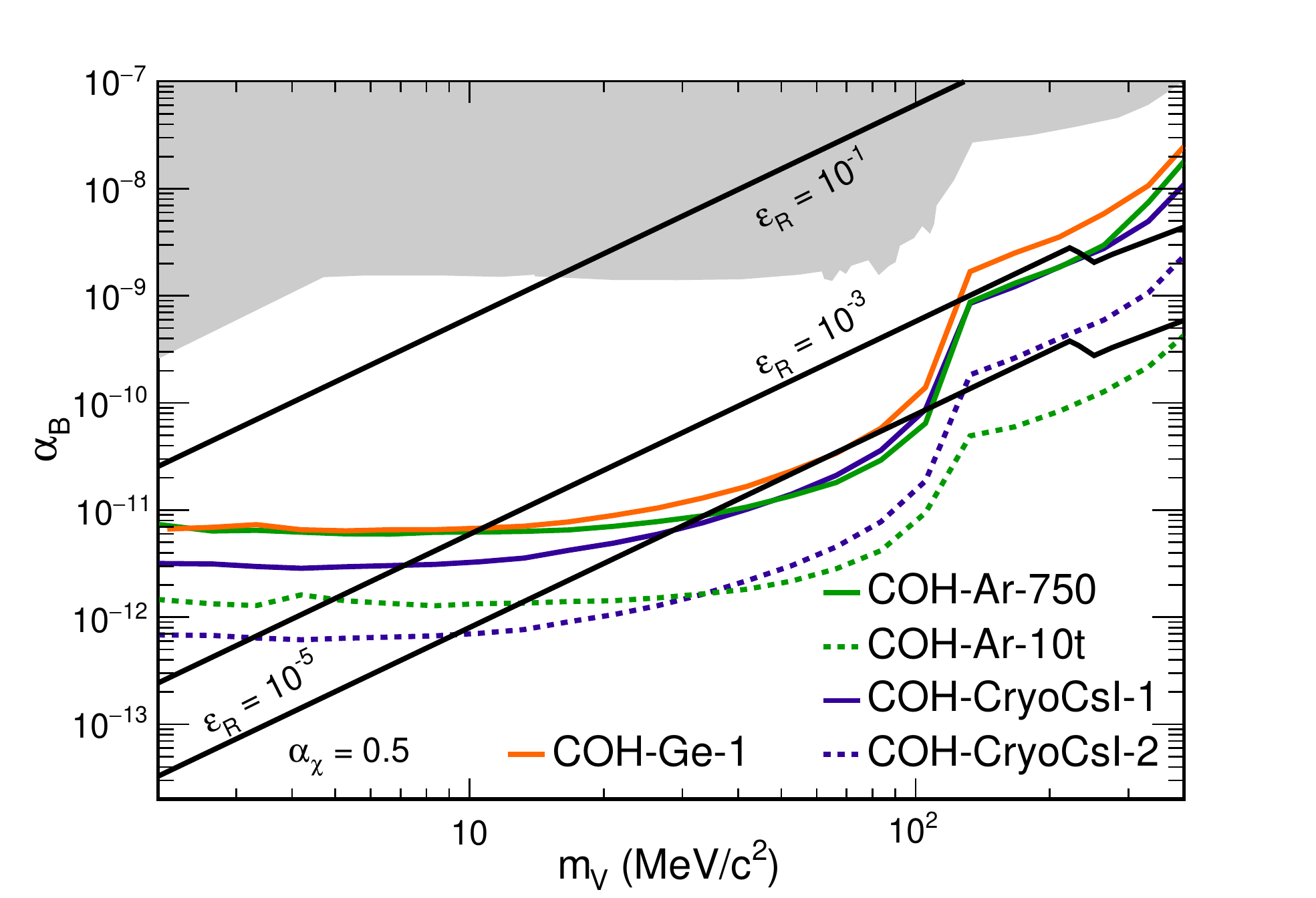}
  \caption{Future COHERENT sensitivity to kinetic mixing (left) and lepthophobic (right) dark-matter models.  Future detectors in Neutrino Alley (at the STS) are shown as solid (dashed) lines.  Solid black lines show the parameters consistent with the cosmological relic abundance of dark matter.}
  \label{f:DMsens}
\end{figure}

We determined the limits of our sensitivity to these models in Neutrino Alley, where we are limited by space in the hallway, as well as potential improvements at the STS which can accommodate detectors up to 10~tonnes.  Both argon and cryogenic scintillator detectors are adept for this type of search, through their low thresholds and favorable timing resolutions.  These two technologies would complement each other.  Cryogenic scintillator detectors give tighter constraints for low-mass dark matter due to its very low threshold.  Argon detectors can probe high-mass dark matter with larger detector masses.  Additionally, if the detectors are placed at different off-axis angles from the beam, the angular dependence of the dark matter flux can be studied.  Expected sensitivity after three years of running COH-Ar-750 and COH-CryoCsI-1 is shown in Fig.~\ref{f:DMsens}.  We also show projections for five years of running COH-Ar-10t and COH-CryoCsI-2 at the STS.  These detectors will not be constrained to Neutrino Alley and we assume both are placed 20~m from the target.  We assume COH-Ar-10t is placed within 20$^\circ$ of the STS beam to exploit the increased dark matter to CEvNS ratio due to the angular dependence of the dark matter flux.  COH-CryoCsI-2 is assumed to be orthogonal to the beam-line to study this angular dependence for any detected dark matter excess.

\subsection{Sterile neutrino search}

The LSND excess gave the first evidence that neutrino oscillation phenomenology extends beyond the three-flavor PMNS matrix~\cite{LSND:2001aii}.  Later, MiniBooNE~\cite{MiniBooNE:2020pnu}, gallium detectors~\cite{SAGE:1998fvr,Abdurashitov:2005tb,Kaether:2010ag}, and experiments at reactors~\cite{Mention:2011rk} saw similar evidence supporting the existence of a sterile neutrino.  Other searches designed to explore the LSND anomaly, however, have not found any evidence of a sterile neutrino~\cite{KARMEN:2002zcm,MINOS:2016viw,IceCube:2016rnb,MicroBooNE:2021rmx}.  A global fit to oscillation data suggests a sterile neutrino with $\Delta m^2_{41}=1.7$~eV$^2$~\cite{Gariazzo:2017fdh}, but more data is needed to discover a sterile neutrino.  

We detect CEvNS recoils from neutrinos with energies $\sim10$ to $53$~MeV.  Neutrino Alley extends from 19.3 to 28~m from the SNS target.  Thus, our detectors are optimally positioned to search for a sterile neutrino with $\Delta m^2_{41}$ between 0.4 and 3.4~eV$^2$, near the global best fit.  As CEvNS is a neutral-current process, we can search for neutral-current disappearance with approximate oscillation probabilities:
\begin{equation}
  \begin{split}
    1-P(\nu_e\rightarrow\nu_s)=1-\sin^22\theta_{14}\cos^2\theta_{24}\cos^2\theta_{34}\sin^2\frac{\Delta m^2_{41}L}{4E} \\
    1-P(\nu_\mu\rightarrow\nu_s)=1-\cos^4\theta_{14}\sin^22\theta_{24}\cos^2\theta_{34}\sin^2\frac{\Delta m^2_{41}L}{4E}
  \end{split}
\end{equation}
valid at short oscillation baselines with $\Delta m^2_{32}L/E\ll1$.  Neutral-current disappearance is sensitive to the three mixing angles $\theta_{14}$, $\theta_{24}$, and $\theta_{34}$.  However, all three can not be disentangled with only two channels.  We thus impose unitarity by including a Gaussian prior constraint on $\theta_{34}$ determined from three-flavor oscillation results~\cite{Denton:2021mso}.  Studying oscillations with $\nu_e$ charged-current events recorded in COH-Ar-750 events may also have competitive sensitivity to sterile mixing, though studies are preliminary.  If this can be achieved, the $\nu_e$ measurement from charged-current events will break the $\theta_{34}$ degeneracy in CEvNS channels, allowing a complete measurement of $\theta_{14}$, $\theta_{24}$, and $\theta_{34}$ using only COHERENT data.

COHERENT will operate CEvNS detectors throughout Neutrino Alley which will test the neutral-current disappearance at different baselines. This would map the $L/E$ oscillation dependence if the LSND anomaly is due to a sterile neutrino.  Having a neutrino flux with multiple flavors is also very beneficial, allowing for a simultaneous measurement of $\theta_{14}$ and $\theta_{24}$.  Further, though observed recoil energy in our detectors is not strongly correlated with the neutrino energy, the $\nu_\mu$ flux is monoenergetic and separated in time from the $\bar{\nu}_\mu/\nu_e$ flux.  Similarly, as the kinematic endpoint for CEvNS is $2E_\nu^2/m$, we can also study oscillations at the highest neutrino energies with narrow energy band, $E_\nu=48\pm4$~MeV, by studying oscillation effects at the highest recoil energies of our CEvNS spectrum.  Studying the $L/E$ dependence at these two precise energy bands would efficiently resolve $\Delta m^2_{41}$ if a sterile neutrino is observed.  Additionally, the CEvNS cross section is known very well, up to a few percent uncertainty from nuclear effects.  Thus, we can precisely predict the expected CEvNS spectrum in our detectors while some experiments searching for sterile neutrinos suffer from large neutrino interaction uncertainties.

\begin{figure}[!tb]\centering
  \includegraphics[width=0.4\linewidth]{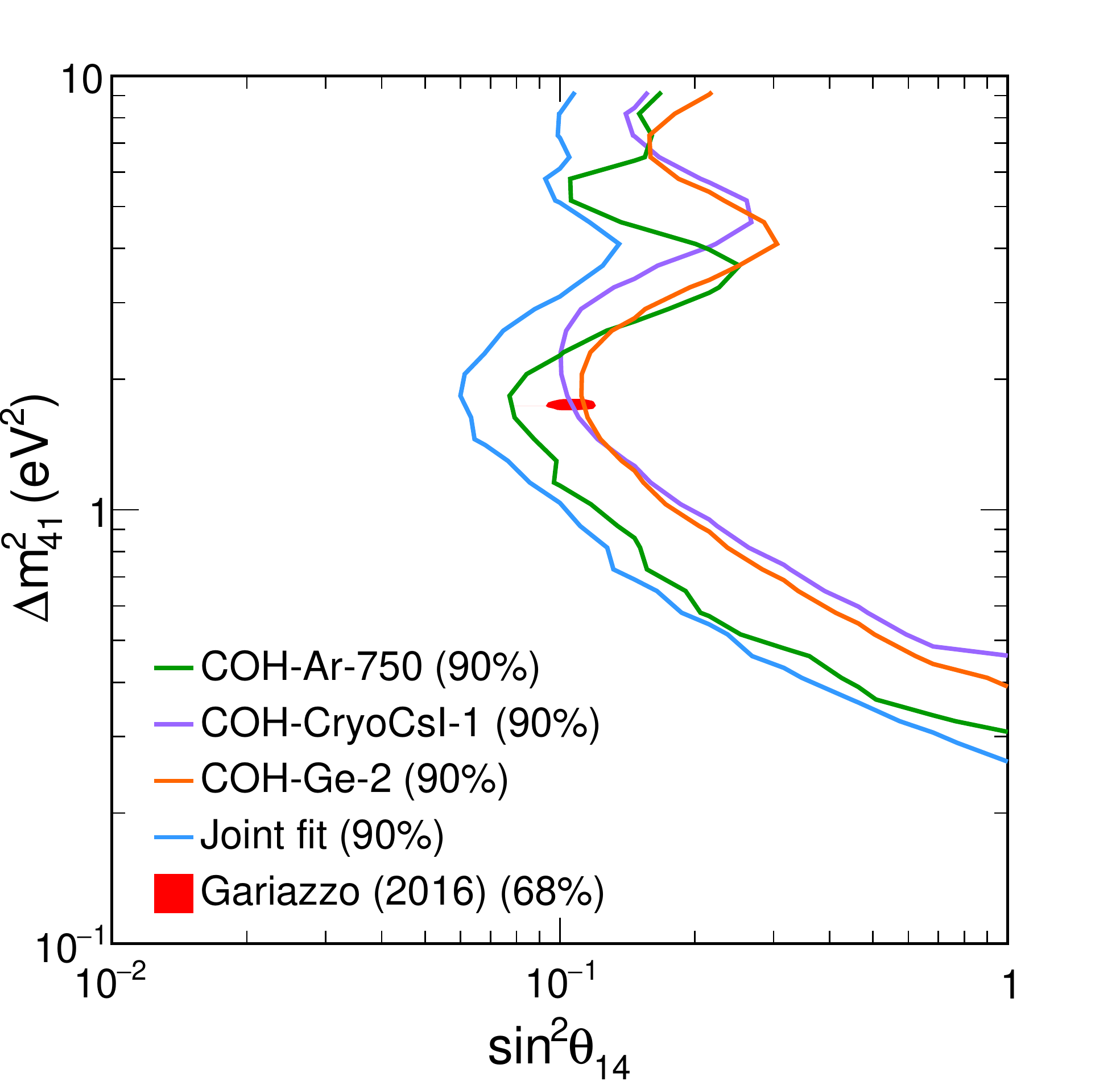}
  \includegraphics[width=0.4\linewidth]{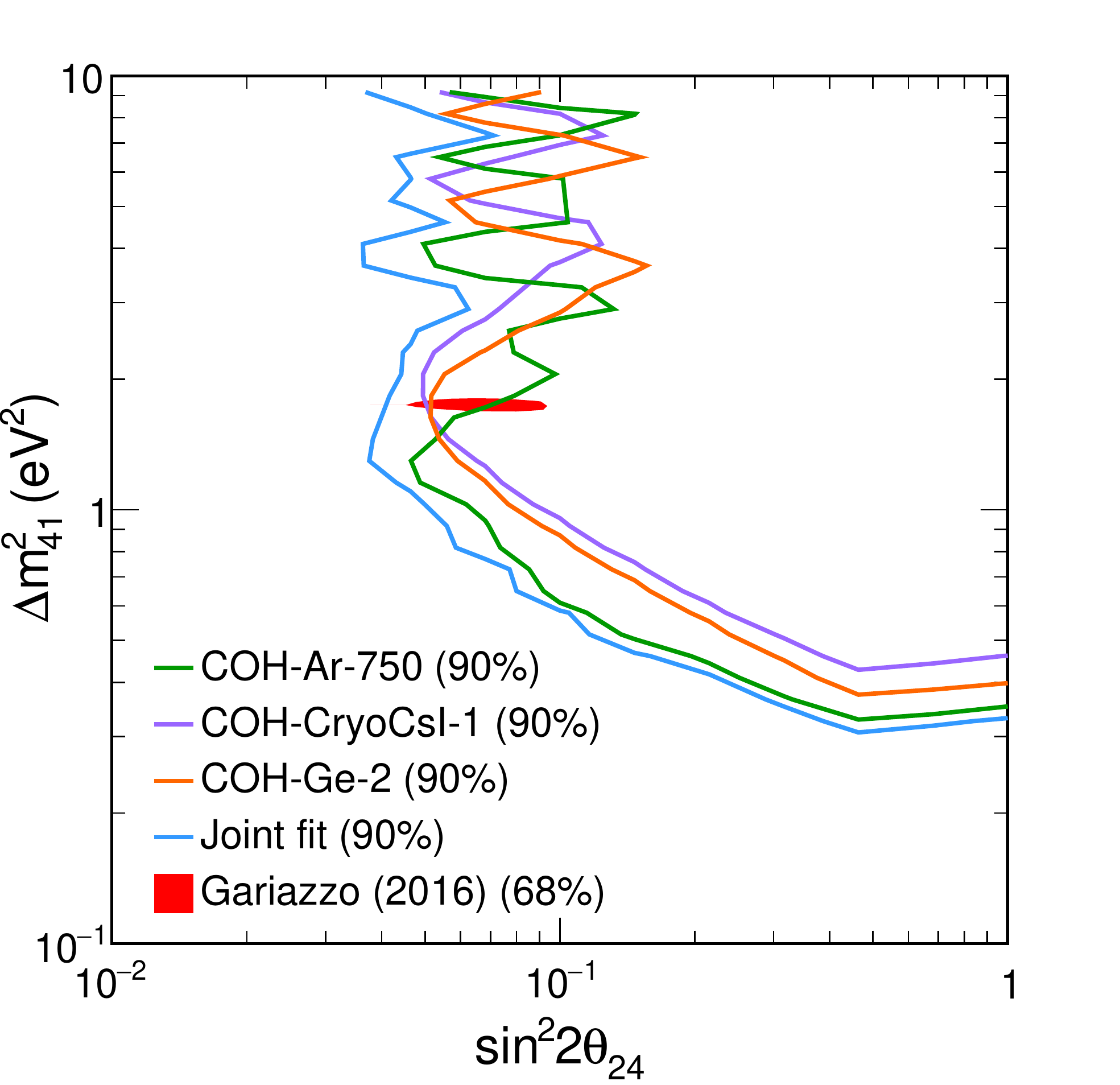}
  \includegraphics[width=0.49\linewidth]{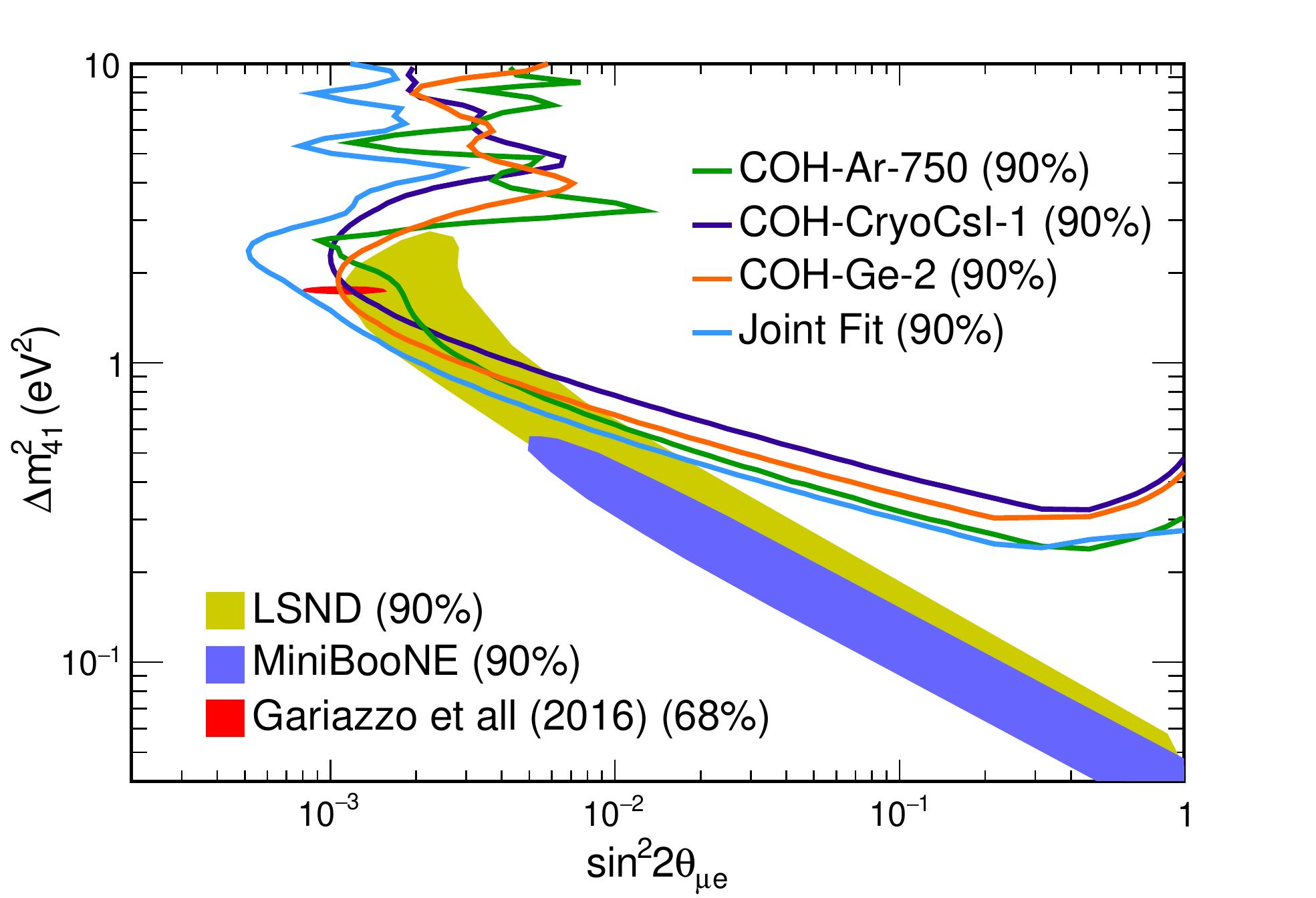}
  \caption{Sensitivity to sterile neutrinos for future COHERENT detectors at the FTS.  The mixing angle $\sin^22\theta_{14}$($\sin^22\theta_{24}$) is determined by neutral-current disappearance of $\nu_e$($\nu_\mu)$ in the SNS flux.  On the bottom, we fit in $\sin^22\theta_{\mu e}$ to compare to LSND and MiniBooNE allowed parameter space.}
  \label{f:SterileSens}
\end{figure}

As the key signature of a sterile search is the $L/E$ dependence, each CEvNS detector impacts limits on parameter space, with peak sensitivity to $\Delta m^2_{41}$ depending on the detector baseline: 19.3~m for COH-CryoCsI-1, 22~m for COH-Ge-1, and 28~m for COH-Ar-750.  We calculated expected sensitivity to identify oscillation effects from a single sterile neutrino mixing with the three active flavors, the $3+1$ scenario, for each detector after three years of exposure.  The mass splitting, $\Delta m^2_{41}$ and three mixing angles, $\theta_{14}$, $\theta_{24}$, and $\theta_{34}$ were allowed to float in the fits.  For each detector, the sensitivity was fit in recoil time and energy to separate populations of specific neutrino flavor and energy.  We also show the expectation for a joint fit which can reduce the influence of the dominant uncertainty on neutrino flux normalization due to its correlation across detectors.  Results of the calculation are shown in Fig.~\ref{f:SterileSens}.

We considered sensitivity both to $\sin^22\theta_{14}$ and $\sin^22\theta_{24}$ which is directly probed by neutral-current disappearance of the $\nu_e$ and $\nu_\mu$ flux, respectively.  At 90$\%$ confidence, we will probe parameter space preferred by a global fit~\cite{Gariazzo:2017fdh} for both mixing angles.  We will probe both angles to roughly the same degree, $\sin^22\theta>0.05$.  In each detector, sensitivity to $\sin^22\theta_{24}$ oscillates quickly at for $\Delta m^2_{41}>2$~eV$^2$ due to the monoenergetic $\nu_\mu$ component of the SNS flux.   There is significant evidence for $\nu_e/\bar{\nu}_e$ disappearance from gallium and reactor searches but none from sensitive $\nu_\mu$ disappearance experiments like MINOS and IceCube.  Thus, it is advantageous to search for both channels simultaneously with the same detector and technique which is allowed by the $\nu_\mu$ and $\nu_e$ components of the SNS flux that are separated in time.  

To compare to LSND and MiniBooNE, our sensitivity was also calculated in terms of $\sin^22\theta_{\mu e}=\sin^22\theta_{14}\sin^2\theta_{24}$ while profiling over $\sin^22\theta_{24}$ and $\sin^22\theta_{34}$.  The result is shown in Fig.~\ref{f:SterileSens}.  Each proposed detector could cover a fraction of the global preferred parameter space while the entire space can be explored with a joint fit.  Baselines allowed by Neutrino Alley are not long enough to cover parameter space at low mass splittings.  However, future COHERENT data will efficiently test the most favored $3+1$ oscillation parameters with a new, precisely predicted detection method.

\subsection{Measuring neutrino inelastic cross sections}

CEvNS measurements at the SNS will explore several new physics topics within and beyond the standard model, but the opportunities for neutrino physics extends yet beyond.  The $\pi$-DAR flux produced at the SNS is among the best beams for studying neutrinos in the tens-of-MeV range.  Though many experiments rely on understanding neutrino-nucleus scattering in these regions, there is a dearth of experimental measurements.  For example, the HALO experiment~\cite{Duba:2008zz} monitors for a flux of supernova neutrinos by detecting neutrino-induced neutrons emitted in Pb($\nu_e,en)$ interactions. This same interaction in lead shielding materials is responsible for a background in our CEvNS detectors, but there has not been a cross section measurement of this channel.  We have commissioned a detector to measure the rate of these emitted neutrons~\cite{COHERENT:2018gft}.  Additionally, we have collected data with 185~kg of NaI crystals suitable for measuring the $^{127}$I($\nu_e,e$) interactions exploring $g_A$ quenching whose understanding is vital for $0\nu\beta\beta$ searches~\cite{Singh:2018xch}.  With future detectors, we can expand the breadth of low-energy neutrino cross section measurements realized at the SNS.

The liquid argon program within COHERENT is of particular influence for the DUNE low-energy program~\cite{Caratelli:2022llt}, including measurement of supernova neutrinos~\cite{DUNE:2020zfm} and solar neutrino oscillations~\cite{Capozzi:2018dat} primarily through the charged-current $^{40}$Ar($\nu_e,e$) channel.  The low-energy regime of the atmospheric neutrino flux is also relevant as a background for searches for the diffuse supernova neutrino flux.  However, there is currently a wide spread in calculations of the charged-current cross section on argon, varying by as much as an order of magnitude.  Given this theoretical uncertainty, a measurement of this cross section is needed to properly interpret any DUNE low-energy neutrino data.  A large flux of neutrinos is produced at the SNS which covers the relevant kinematic region for these DUNE efforts, making COHERENT ideal for studying this scattering process.

\begin{figure}[!tb]\centering
  \includegraphics[width=0.49\linewidth]{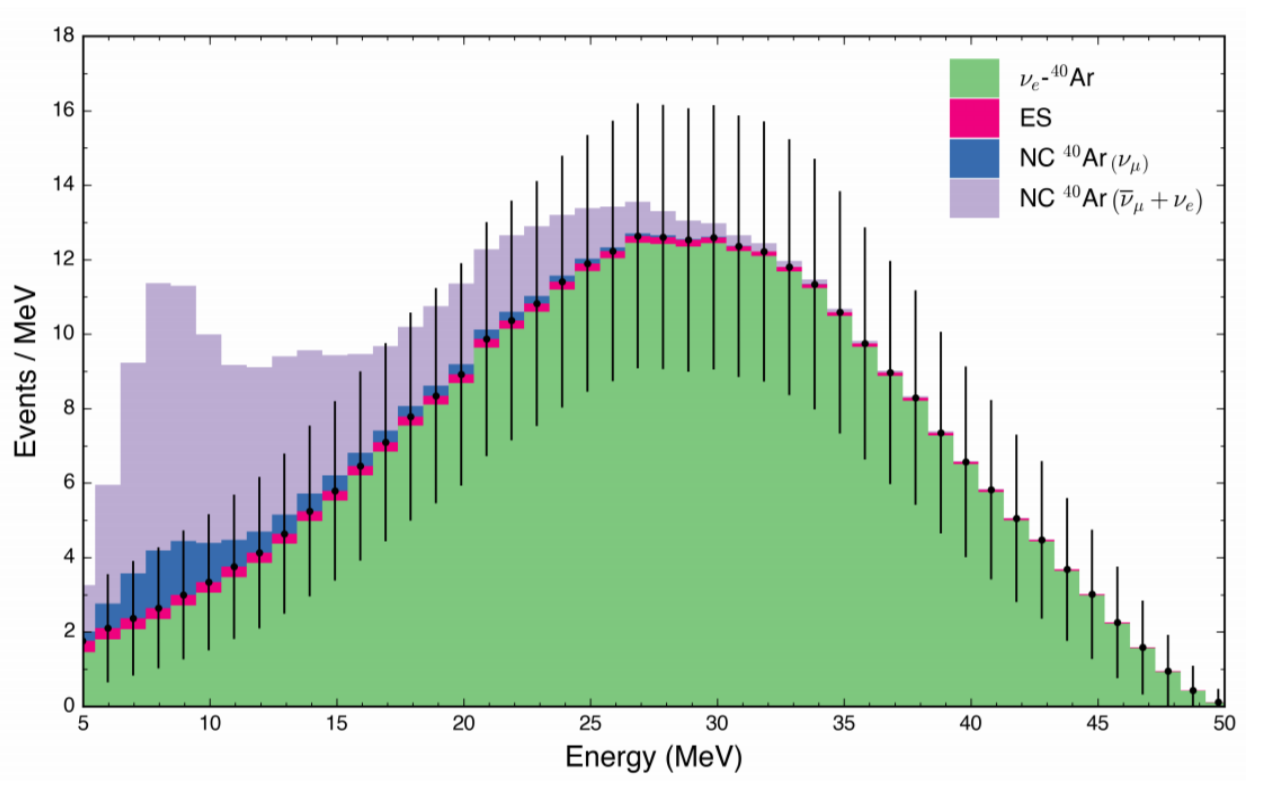}
  \caption{Expected inelastic neutrino events in COH-Ar-750 after one year of running.  The charged-current process, green, is the dominant interaction channel for DUNE low-energy physics objectives.}
  \label{f:LArInel}
\end{figure}

The COH-Ar-750 detector will record $\sim$~340 $\nu_e$ charged-current and $\sim$~100 neutral-current interactions of all flavors each SNS year, calculated by the MARLEY event generator~\cite{Gardiner:2021qfr}.  This would yield a $\sim$~5$\%$ measurement of the flux-averaged charged-current cross section, which would refine DUNE's understanding of its low-energy physics reach.  The expected inelastic spectrum in COH-Ar-750 is shown in Fig. \ref{f:LArInel}.  A liquid argon TPC at the STS would significantly improve this measurement.  The COH-ArTPC-1 detector will have sub-cm position resolution and favorable energy resolution.  With this we will reconstruct final state kinematics of the electron and deexcitation gammas on an event-by-event basis.  The large mass of the TPC also increases the expected rate with $\sim$~5500 events/yr expected.  This will allow precision differential cross section measurements.

Additionally, the COHERENT D$_2$O program is designed specifically to measure charged-current events from the SNS flux.  The primary purpose of this detector is to constrain systematic uncertainties on the neutrino flux normalization at the SNS.  The $d$($\nu_e,e$) interaction can be described theoretically to 2-3$\%$~\cite{Kozlov:1999ct,Nakamura:2002jg}.  Thus, a measurement of the rate of this interaction can be interpreted as a measurement of the SNS flux.  With 1.3~tonnes of D$_2$O instrumented at the SNS, we can achieve a 3$\%$ uncertainty on the SNS flux~\cite{COHERENT:2021xhx}.  A continued D$_2$O program at the SNS will continue to be necessary to monitor the neutrino flux as the SNS beam energy increases from 1~GeV to 1.3~GeV.  Coincidentally, understanding the $^{16}$O($\nu_e,e$)$^{16}$F$^*$ cross section is also important as this is a background for Super-Kamiokande (and subsequently Hyper-Kamiokande) searches for the diffuse supernova neutrino flux~\cite{Super-Kamiokande:2021jaq}; it is also important for interpretation of a supernova burst signal.  This cross section has not yet been measured and will constrain the Super-Kamiokande background after a measurement from COHERENT D$_2$O data.

\subsection{Probing nuclear physics}

The dominant uncertainty to the CEvNS cross section prediction comes from imperfect modeling of the spatial distribution of neutrons within the target nucleus, with the average radius of neutrons given by $R_n$.  The CEvNS cross section is coherent if $QR_n\ll1$ and becomes suppressed at higher $Q^2$.  This degree to which CEvNS is coherent is described by the nuclear form factor, $F_{nucl}(Q^2)$ and depends on $R_n$.  The best measurement of $R_n$ has recently been achieved using $^{208}$Pb~\cite{PREX:2021umo} which isolated the weak charge through parity violation, but data are needed on nuclei we use to measure CEvNS.  By measuring the suppression observed in the CEvNS recoil spectrum, we can infer estimates for $R_n$ on each nucleus we study.  This has been done using our CsI[Na] data, but errors are still large~\cite{Cadeddu:2018rlm} and larger-sample measurements are needed to precisely identify $R_n$.

Understanding $R_n$ is of particular interest in astrophysics.  An accurate determination of $R_n$, and the difference between the neutron and proton radii, is critical for understanding the equation of state for neutron stars~\cite{Horowitz:2000xj,Tsang:2012se} and graviational wave events~\cite{Baiotti:2019sew,Tsang:2019mlz,Fasano:2019zwm}.  The value of $R_n$ is related to density of matter within neutron stars and the mass range expected for these stars~\cite{Brown:2000pd}.

\begin{figure}[!tb]\centering
  \includegraphics[width=0.49\linewidth]{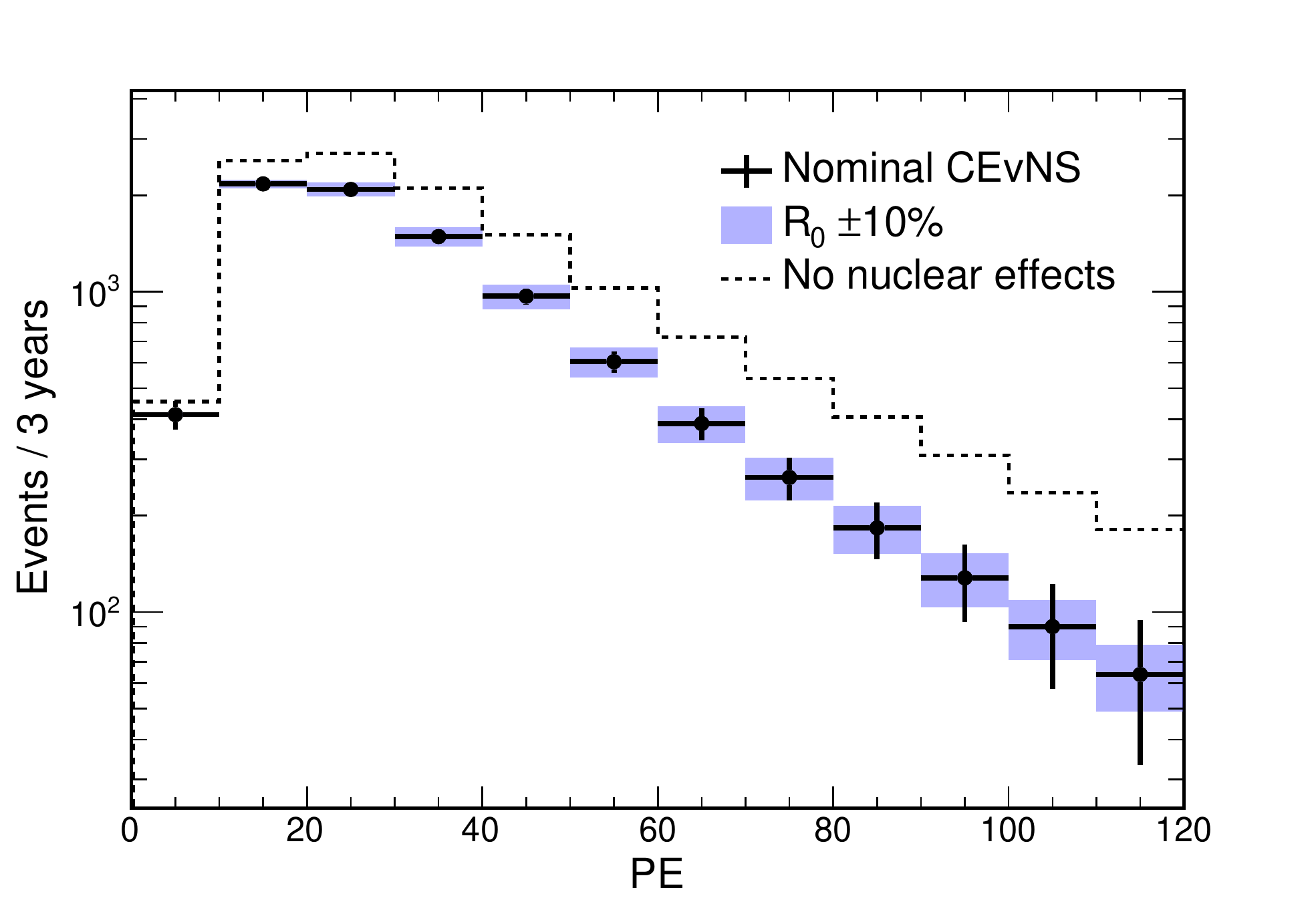}
  \includegraphics[width=0.49\linewidth]{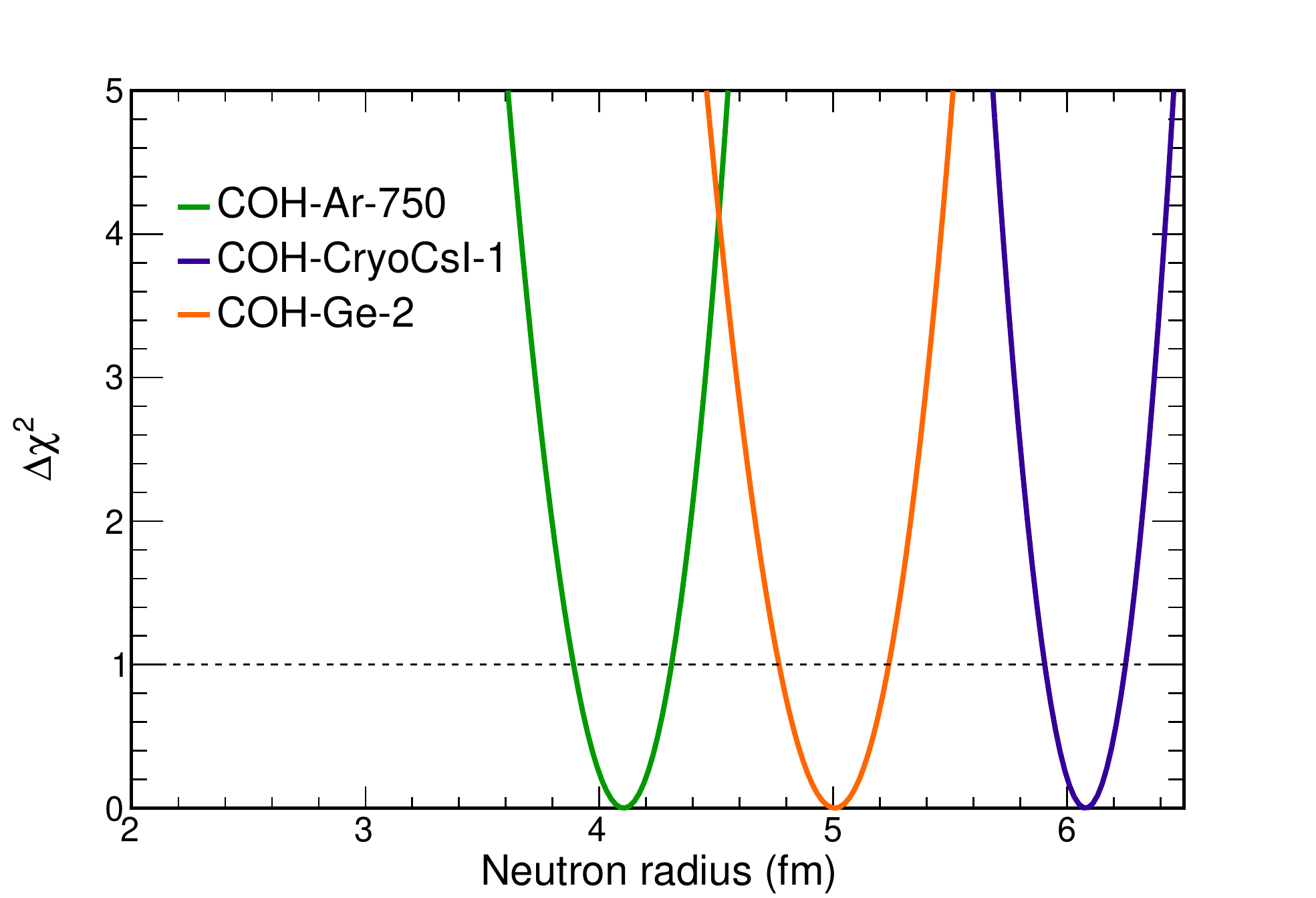}
  \caption{The influence of nuclear effects on CEvNS in COH-Ar-750 (left) along with our ability to determine $R_n$ in future Ar, Ge, and CsI detectors (right).}
  \label{f:NuclearRadius}
\end{figure}

The influence of these nuclear effects are shown in Fig.~\ref{f:NuclearRadius} for CEvNS events in COH-Ar-750 for nominal and adjusted choices of $R_n$.  We assumed the Klein form factor parameterization~\cite{Klein:1999qj} which predicts a characteristic $Q^2$ deformation with varying $R_n$.  We used this relation to fit a predicted uncertainty on $R_n$ from COH-Ar-750, COH-Ge-2, and COH-CryoCsI-1 datasets after three years of running, shown in Fig.~\ref{f:NuclearRadius}.  With these datasets, COHERENT will measure $R_n$ to 4.6, 5.1, and 2.9$\%$ in Ar, Ge, and CsI, respectively.  Given $R_p$ can be measured with high precision through electron scattering, COHERENT data will also similarly determine the neutron skin.

\subsection{Searching for axion-like particles at the SNS}

There is ongoing work within the CEvNS theory frontier to understand ability of CEvNS experiments at accelerators and nuclear reactors to detect axion-like particles (ALPs).  In particular, given the low threshold achieved by COHERENT detectors, CEvNS experiments can search for ALPs at lower masses than conventional beam dump experiments, covering an unexplored region of parameter space~\cite{CCM:2021lhc}.  Further, this new region of sensitivity is consistent with expectations for the QCD axion~\cite{Peccei:1977hh,Weinberg:1977ma}.  COH-Ar-750 can also identify ALPs produced at the SNS.  By identifying 10~keV to 100~MeV electron deposits in time with the arrival of the beam, we will constrain the ALP coupling to electrons, $g_{ae}$, and photons, $g_{a\gamma}$.  Understanding of COHERENT sensitivity is still preliminary as we develop understanding of the CENNS10 response to electron deposits $>1$~MeV, but a future argon detector shows promise with a tonne-scale of mass in the high-intensity, 1.4~MW SNS beam.
\section{Broad impact}

\subsection{Data sharing}

There is increasing discussion concerning the need for sharing experimental data, clear explanations of the data, and tools used to properly analyze the data~\cite{Chen:2018drk}. This is to assure reproducibility and to facilitate the broadest array possible of scientific studies using the data. To date, COHERENT has released data from our first CEvNS observation in CsI~\cite{COHERENT:2018imc} and our first detection of CEvNS in argon with COH-Ar-10~\cite{COHERENT:2020ybo}. Each release includes a technical note explaining the qualities and formatting of the data, as well as scripts to perform some basic treatment of the data. These data releases are hosted on \url{https://zenodo.org}, from where they obtained their digital object identifiers (DOI) for easy referencing. Up to now, these releases have approximately two thousand downloads and one hundred citations combined, and spurred multiple correspondences with members of the high energy physics community. With this positive response, the collaboration will continue timely data releases for all COHERENT scientific results.

In some cases, it may be beneficial to release larger datasets of order one terabyte to allow independent examination of raw data using alternative analysis techniques. The COHERENT data collected to determine the quenching factor for nuclear recoils in cesium iodide is a good example. The collaboration intends to share the raw digitized waveforms of these dedicated calibration runs which will approach ~100 gigabytes and plans to release these using the Oak Ridge Leadership Computing Facility's \href{https://www.olcf.ornl.gov/olcf-resources/rd-project/constellation-doi-framework-and-portal}{Constellation service}, which will archive the data, assign the data a DOI number, and present the data to the public via the \href{https://www.globus.org/data-transfer}{Globus interface}.

\subsection{COHERENT DEI and Outreach}

To constructively address diversity, equity, and inclusion (DEI) within the physics community, the COHERENT collaboration has initiated a set of specific actions.  A COHERENT DEI committee was formed in November 2020 along with a charge and set of guidelines for membership that ensures representation from people at different stages of the career pipeline.  The committee charge states:
\begin{quote}
 The Diversity, Equity, and Inclusion (DEI) Committee will: promote an environment within the
COHERENT collaboration that is respectful and inclusive of all who wish to participate; ensure
that the environments in which our personnel work and engage with fellow scientists are safe;
and grow the pool of talent from which future scientific efforts can draw by creating opportunities
for increased participation by scientists from groups underrepresented in physics.  
\end{quote}

Specifically, we will address these aims through the following actions:
\begin{enumerate}
    \item Periodically review the COHERENT Code of Conduct that every member agrees to, and make the Code, which includes a description for reporting violations, readily available to all collaborators.
    \item Facilitate open discussions of the research culture within the COHERENT Collaboration.
    \item Perform assessments of perceptions of our research culture.
    \item Evaluate the safety and inclusivity of the spaces in which Collaboration research is performed, including: ORNL, SNS, TUNL, Collaboration Institutions and meeting spaces.
    \item Generate opportunities for increased participation by students and researchers from underrepresented groups. This includes initiating contact with Minority Serving Institutions (MSIs) and establishing summer internships for students at COHERENT institutions. 
    \item Establish formal mentoring relationships between junior and senior scientific staff at all levels, outside of the formal student/faculty member relationship, to establish possible alternate lines of communication and guidance.
    \item Promote communication between early career student and post-doctoral research members and the senior faculty and research scientists through representation on the COHERENT collaboration board who presents any issues facing the ``junior membership'' and updates on resolving these issues at each meeting.
    \item Maintain records of responses of assessments and collaboration participation to evaluate changes of perception and representation over time.
    \item Provide input to the greater physics communities in which the collaboration members are constituents, including but not exclusive to: APS-DPF, APS-DNP, SESAPS, International Conference on Neutrino Physics.
\end{enumerate}

\begin{figure}[htbp]
\includegraphics[width=\linewidth]{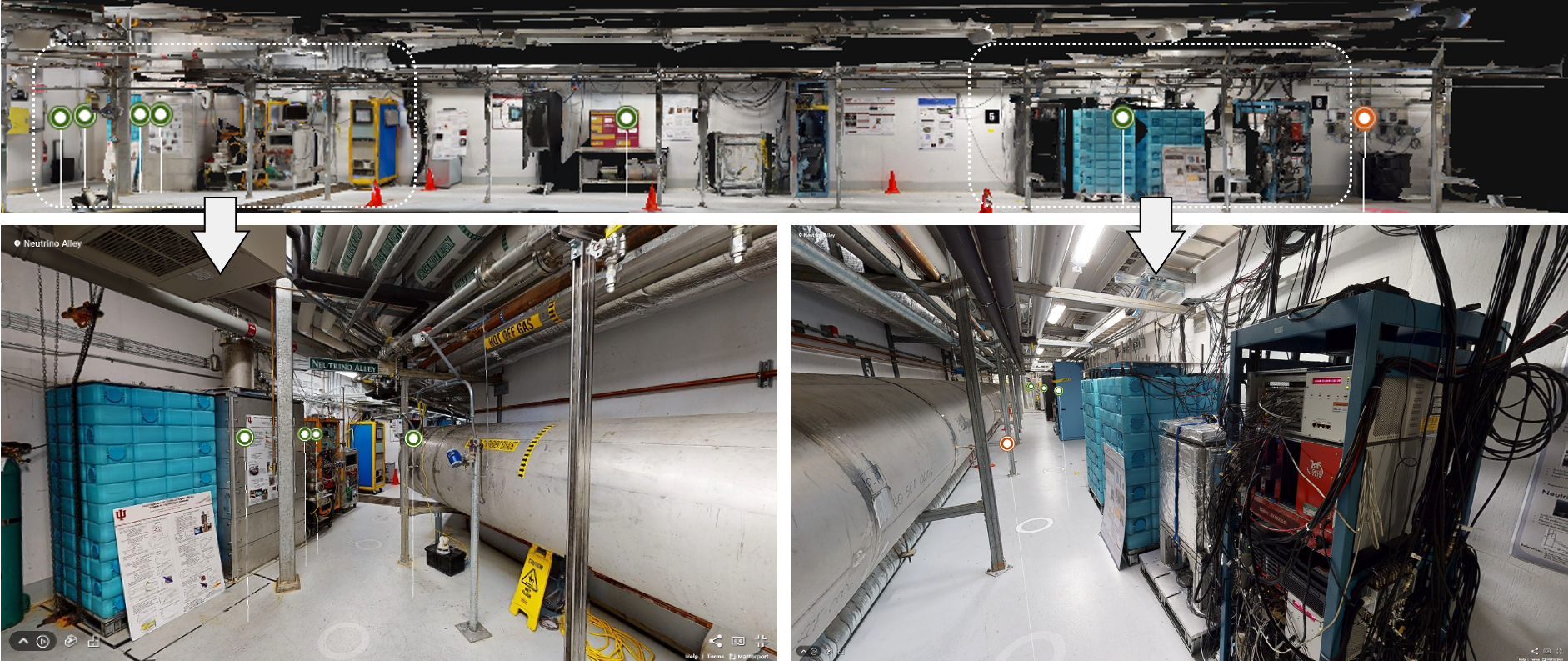}
\caption{Neutrino alley virtual tour at \url{https://my.matterport.com/show/?m=XYA19MBVdQS}.}
\label{f:nuAlley1}
\end{figure}

Although the pandemic has limited COHERENT collaborators from participating in hands-on, in-person outreach activities, we have presented virtual activities and talked to undergraduate institutions, specifically Minority Serving Institutions, and to middle and high-school classes.  Collaborators contributed contents to \href{https://en.wikipedia.org/wiki/Draft:COHERENT_Experiment}{Wikipedia} and a \href{https://my.matterport.com/show/?m=XYA19MBVdQS}{virtual tour of Neutrino Alley} hosted at ORNL, a screenshot of which is shown in Fig.~\ref{f:nuAlley1}.  To engage a broader audience, we are working on descriptions in multiple languages for people with varying backgrounds. 

\section{Acknowledgement}
The COHERENT collaboration acknowledges the resources generously provided by the Spallation Neutron Source, a DOE Office of Science User Facility operated by the Oak Ridge National Laboratory. This work was supported by the US Department of Energy (DOE), Office of Science, Office of High Energy Physics and Office of Nuclear Physics; the National Science Foundation; the Consortium for Nonproliferation Enabling Capabilities; the Institute for Basic Science (Korea, grant no. IBS-R017-G1-2019-a00); the Ministry of Science and Higher Education of the Russian Federation (Project ``Fundamental properties of elementary particles and cosmology'' No. 0723-2020-0041); and the US DOE Office of Science Graduate Student Research (SCGSR) program, administered for DOE by the Oak Ridge Institute for Science and Education which is in turn managed by Oak Ridge Associated Universities.  Sandia National Laboratories is a multi-mission laboratory managed and operated by National Technology and Engineering Solutions of Sandia LLC, a wholly owned subsidiary of Honeywell International Inc., for the U.S. Department of Energy’s National Nuclear Security Administration under contract DE-NA0003525. The Triangle Universities Nuclear Laboratory is supported by the U.S. Department of Energy under grant DE-FG02-97ER41033. Laboratory Directed Research and Development funds from Oak Ridge National Laboratory also supported this project. This research used the Oak Ridge Leadership Computing Facility, which is a DOE Office of Science User Facility.

\bibliographystyle{JHEP}
\bibliography{ref}

\end{document}